\documentclass{aastex63}
\usepackage{subfigure}

\received{???}
\revised{???}
\accepted{???}
\submitjournal{ApJ}

\shorttitle{Search for gravitational waves from young SNRs}
\shortauthors{Abbott et al.}

\begin{document}

\title{Searches for continuous gravitational waves from young supernova remnants \\in the early third observing run of Advanced LIGO and Virgo}

\author{R.~Abbott}
\affiliation{LIGO Laboratory, California Institute of Technology, Pasadena, CA 91125, USA}
\author{T.~D.~Abbott}
\affiliation{Louisiana State University, Baton Rouge, LA 70803, USA}
\author{S.~Abraham}
\affiliation{Inter-University Centre for Astronomy and Astrophysics, Pune 411007, India}
\author{F.~Acernese}
\affiliation{Dipartimento di Farmacia, Universit\`a di Salerno, I-84084 Fisciano, Salerno, Italy  }
\affiliation{INFN, Sezione di Napoli, Complesso Universitario di Monte S.Angelo, I-80126 Napoli, Italy  }
\author{K.~Ackley}
\affiliation{OzGrav, School of Physics \& Astronomy, Monash University, Clayton 3800, Victoria, Australia}
\author{A.~Adams}
\affiliation{Christopher Newport University, Newport News, VA 23606, USA}
\author{C.~Adams}
\affiliation{LIGO Livingston Observatory, Livingston, LA 70754, USA}
\author{R.~X.~Adhikari}
\affiliation{LIGO Laboratory, California Institute of Technology, Pasadena, CA 91125, USA}
\author{V.~B.~Adya}
\affiliation{OzGrav, Australian National University, Canberra, Australian Capital Territory 0200, Australia}
\author{C.~Affeldt}
\affiliation{Max Planck Institute for Gravitational Physics (Albert Einstein Institute), D-30167 Hannover, Germany}
\affiliation{Leibniz Universit\"at Hannover, D-30167 Hannover, Germany}
% ...



\begin{abstract}

We present results of three wide-band directed searches for continuous gravitational waves from 15 young supernova remnants in the first half of the third Advanced LIGO and Virgo observing run. 
We use three search pipelines with distinct signal models and methods of identifying noise artifacts.
Without ephemerides of these sources, the searches are conducted over a frequency band spanning from 10~Hz to 2~kHz. 
We find no evidence of continuous gravitational radiation from these sources.
We set upper limits on the intrinsic signal strain at 95\% confidence level in sample sub-bands, estimate the sensitivity in the full band, and derive the corresponding constraints on the fiducial neutron star ellipticity and $r$-mode amplitude.
The best 95\% confidence constraints placed on the signal strain are $7.7\times 10^{-26}$ and $7.8\times 10^{-26}$ near 200~Hz for the supernova remnants G39.2--0.3 and G65.7+1.2, respectively.
The most stringent constraints on the ellipticity and $r$-mode amplitude reach $\lesssim 10^{-7}$ and $ \lesssim 10^{-5}$, respectively, at frequencies above $\sim 400$~Hz for the closest supernova remnant G266.2--1.2/Vela Jr.

\end{abstract}


\section{Introduction}
\label{sec:intro}
Transient gravitational waves (GWs) from compact binary coalescences~\citep{LVC-catalog,abbott2020gwtc2} have been directly observed by the Advanced Laser Interferometer Gravitational-Wave Observatory (Advanced LIGO) detectors~\citep{LIGO2014} and the Advanced Virgo detector~\citep{Virgo2014}.
Continuous gravitational waves (CWs) have not yet been detected.
The most likely sources of CWs detectable by ground-based interferometers are non-axisymmetric, rapidly rotating neutron stars.
Searches for CWs have been carried out targeting various isolated sources, including known pulsars with electromagnetic ephemerides \citep{Abbott_2019_known_pulsar, abbott2020J0537}, neutron stars without ephemerides in the galactic center or in globular clusters \citep{DirectedBSD,aasi2013directed,PhysRevD.99.084048,LVC2017GC}, neutron stars in binary systems \citep{scox1o2,Zhang_2021,lmxb_other}, and young supernova remnants (SNRs) \citep{aasi2015searches,2016PhRvD..94h2004S,PhysRevD.100.024063,Abbott_2019,millhouse2020search,Lindblom_2020,Papa:2020vfz,Beniwal2021arXiv210206334B}. 
Searches have also been conducted over the whole sky for CWs instead of targeting at a particular direction \citep{allskyO2,Vladimir_allsky,Steltner2021,wette2021deep,Covas_allsky,abbott2020binaryallsky}. 
This work searches for CWs from SNRs in the first half of the third observing run (O3a), which commenced on April 1st, 2019 and ended on March 27th, 2020~\citep{O3detector,Virgo2014}.

Young neutron stars in SNRs are one potential source of continuous, quasi-monochromatic GWs. If pulsations are observed in electromagnetic emission from the neutron star, one can search for CWs guided by the ephemerides obtained from those observations, as in e.g., \cite{Abbott_2019_known_pulsar} and \cite{O3targetCW}.
Even so, there is no guarantee that the GW-emitting quadrupole is phase locked to the electromagnetic pulsations.
When there is no phase locking, search algorithms are needed that can track small (and possibly randomly varying) displacements between the gravitational and electromagnetic frequencies~\citep{Beniwal2021arXiv210206334B,abbott2019narrow}.
If the neutron star does not pulsate, it may be observed as an X-ray point source, known as a central compact object \citep{Gotthelf_2013}.
In the latter scenario, the maximum GW strain can be inferred from the age of the SNR \citep{wette2008searching,riles2013gravitational}, as has been done in recent GW searches \citep{millhouse2020search,Beniwal2021arXiv210206334B}.

A rotating, non-axisymmetric neutron star has a time-varying mass quadrupole (from the point of view of a distant observer) and emits GWs at a strain proportional to the stellar ellipticity, which is affected by the nuclear equation of state, the history of strain build up and diffusion in the crust, and the magnetic field configuration \citep{2018ASSL..457..673G}.
For an isolated star, young neutron stars may have larger non-axisymmetries than older ones and consequently may produce stronger GW emissions \citep{knispel2008blandford,riles2017recent}.
As the star ages, Ohmic \citep{haensel1990ohmic}, thermal \citep{gnedin2001thermal,potekhin2015neutron}, tectonic, or other relaxation processes work to reduce the asymmetries introduced in the birth process.
Young neutron stars are therefore promising targets for CW searches.
The GW frequency is proportional to the stellar spin frequency $f_\star$.
For thermoelastic \citep{ushomirsky2000deformations,johnson2013maximum} or magnetic \citep{cutler2002gravitational,mastrano2011gravitational,lasky2013tilted} mass quadrupoles, the predicted frequency is either $f_\star$ or $2f_\star$; $r$-mode current quadrupoles emit at $\sim 4f_\star/3$ \citep{1998PhRvD..58h4020O,1998ApJ...502..708A,caride2019search}, with minor equation-of-state dependent corrections; also, pinned superfluids in neutron stars may produce CWs at frequencies proportional to $f_\star$~\citep{Jones2010,2015ApJ...807..132M}.

In young, rapidly-rotating neutron stars, $f_\star$ evolves quickly under the action of gravitational and electromagnetic torques \citep{knispel2008blandford,riles2013gravitational}.
Rapid spin-down in young SNRs creates challenges for traditional CW search methods, especially over a long observation with duration $T_{\rm obs} \gtrsim 1\,{\rm yr}$. 
Most previous searches for SNRs have been restricted to short ($\sim 1$ month) stretches of data \citep[e.g.][]{abadie2010first,Abbott_2019}, limited parameter space \citep[e.g.][]{Lindblom_2020}, or have had a high associated computational cost \citep[e.g.][]{Papa:2020vfz,2016PhRvD..94h2004S}.
Accounting for spin down in a coherent search requires a very large number of templates, which increases computation cost beyond feasibility.
Furthermore, $f_\star$ may wander randomly, a phenomenon known as spin wandering or timing noise \citep{hobbs2010analysis,shannon2010assessing,price2012time,ashton2015effect,10.1093/mnras/stz2383,namkham2019diagnostics,10.1093/mnras/staa615}, due to unknown internal or magnetospheric processes~\citep{1981ApJ...245.1060C,melatos2014pulsar}. 
One computationally efficient alternative to a coherent search is a semi-coherent search in which the integration is calculated coherently on blocks of short duration $T_{\rm coh}$ and added incoherently over the full $T_{\rm obs}$.

We apply three semi-coherent methods to search for signals from 15 known young SNRs in the data collected in the first half (six months) of O3: the directed Band-Sampled-Data (BSD) pipeline~\citep{BSD}, based on the FrequencyHough (FH) transform~\citep{FrequencyHoughmethod,FrequencyHough}, and the single-harmonic Viterbi and dual-harmonic Viterbi pipelines, both based on a hidden Markov model (HMM) tracking scheme~\citep{sun2018hidden,Sun2019}.
The two Viterbi methods achieve a lower sensitivity compared to the BSD pipeline, but take into consideration the uncertainties associated with the star's stochastic spin evolution, with one of them tracking two harmonics of the star's spin frequency simultaneously~\citep{sun2018hidden,Sun2019}, making the three methods complementary to each other.

The structure of the paper is as follows. In Section~\ref{sec:targets}, we introduce the 15 young SNR targets, listing their location, estimated age and distance. In Section~\ref{sec:instrument}, we briefly describe the interferometric data analyzed.
In Section~\ref{sec:searches}, we review each of the three search methods and the parameter space covered. The strain upper limits, estimated sensitivity, and astrophysical interpretation are discussed in Section~\ref{sec:limits}. A conclusion is given in Section~\ref{sec:conclusion}. 
The postprocessing procedure applied to the candidates identified in each search is presented in Appendix~\ref{sec:postprocessing}. 
Technical details on the pipelines are described in Appendix~\ref{appendix:pipelines}.

\section{Targeted sources}
\label{sec:targets}
The target SNRs are selected from the Green supernova catalogue \citep{Green2019} and the SNRcat, an online catalogue of high-energy galactic SNRs hosted by the University of Manitoba \citep{SNRcatpaper,SNRcat}, as SNRs with X-ray point sources are likely to contain neutron stars.
Of the 15 SNRs in Table~\ref{tab:targets}, seven are searched using all three different pipelines, while the remaining eight are only searched by the single-harmonic Viterbi pipeline.
The characteristic ages of the neutron stars are inferred from the estimated supernova ages listed in the table.
In the three pipelines, we cover parts of different parameter spaces, corresponding to slightly different assumptions of the characteristic age of the star.
See Section~\ref{sec:searches} for details for each pipeline.

The 15 SNRs were previously searched in the earlier LIGO observing runs, but no CW signal was identified~\citep{Abbott_2019,millhouse2020search,Lindblom_2020,Papa:2020vfz}.
Additionally, \citet{Papa:2020vfz} performed a follow-up search for sub-threshold candidates obtained in the first observing run of Advanced LIGO (O1)~\citep{Ming2019} for three of the SNRs, Cassiopeia A (Cas A), Vela Jr. and G347.3--0.5, using data collected in the second observing run of Advanced LIGO (O2), and reported one possible CW candidate in G347.3--0.5. 
This fully coherent follow-up search uses two stretches of data in O2 ($T_{\rm coh} \sim 4$ months each). 
As indicated in Table~\ref{tab:targets}, only the single-harmonic Viterbi pipeline (which allows for stochastic spin wandering) searches G347.3--0.5 semi-coherently using a short $T_{\rm coh}$.
Since the signal-to-noise ratio roughly scales $\propto T_{\rm coh}^{1/2}$, the sensitivity presented in \citet{Papa:2020vfz} exceeds that presented here for G347.3--0.5, provided that the signal power leaked into adjacent frequency bins due to the spin down and spin wandering over the coherent duration is negligible. In addition, the candidate reported in \citet{Papa:2020vfz} was originally identified as a sub-threshold one.
Therefore it is not surprising that we do not find a possible candidate in G347.3--0.5.

\begin{table}[!tbh]
\centering
\setlength{\tabcolsep}{1.6pt}
\renewcommand\arraystretch{1.2}
\begin{tabular}{llllll}
\hline
Source & Age& Distance & Right ascension & Declination & References\\
           & (kyr) & (kpc) & (h:m:s) & ($^\circ$:$'$:$''$)& \\
\hline
G18.9--1.1 & 2.6--6.1     & 1.6--2.5    & 18:29:13.1 & $-$12:51:13     &  \citet{Ranasinghe2019,Shan2018}     \\ 
         &           &           &            &                 &    \citet{harrus2004x} \\
G39.2--0.3/3C 396 & 3--7.3 & 6.2--8.5  & 19:04:04.7 & 5:27:12         &   \citet{Shan2018,su2010molecular}     \\ 
         &           &           &            &                 &    \citet{Harrus_1999} \\
G65.7+1.2/DA 495  & 7--20   & 1--5  & 19:52:17.0 & 29:25:53        &   \citet{Karpova2015,kothes2008495}    \\
G93.3+6.9/DA 530  & 2.9--7    & 1.7--3.5  & 20:52:14.0 & 55:17:22        &    \citet{Straal2019, jiang2007chandra}    \\
         &           &           &            &                 &    \citet{Landecker_1999,foster2003new} \\
G189.1+3.0/IC 443 & 3--30  & 1.4--1.9  & 06:17:05.3 & 22:21:27        &    \citet{Ambrocio-Cruz2017, kargaltsev_2017}    \\ 
                 &           &           &            &                 &    \citet{swartz2015high,fesen1980spectrophotometry} \\
G266.2--1.2/Vela Jr. & 0.69--5.1 & 0.2--1 & 08:52:01.4 & $-$46:17:53&  \citet{allen2014expansion,liseau1992star}      \\
G353.6--0.7 & 10--40    & 3.2--6.1  & 17:32:03.3 & $-$34:45:18     &  \citet{Klochkov2015,Fukuda_2014}       \\ 
                 &           &           &            &                 &    \citet{2008ApJ...679L..85T}\\
\hline
G1.9+0.3               & 0.10--0.26    & 8.5--10   & 17:48:46.9  & $-$27:10:16       & \citet{reynolds2008youngest,2014IAUS..296..197R}  \\
G15.9+0.2              & 0.54--5.7    & 6.0--16.7 & 18:18:52.1  & $-$15:02:14       & \citet{reynolds2006new,2018MNRAS.479.3033S}      \\
G111.7--2.1/Cas A      & 0.28--0.35    & 3.3--3.4  & 23:23:27.9  & 58:48:42          & \citet{1972AA....18..169I, reed1995three};     \\
   &  &  & &       & \citet{1971ApJ...165..457V,fesen2006expansion}     \\
G291.0--0.1/MSH 11--62 & 1.2--10      & 3.0--10   & 11:11:48.6  & $-$60:39:26       & \citet{1986MNRAS.219..815R,moffett2001g291};       \\
   &  &  & &       & \citet{harrus2004x,slane2012broadband}     \\
G330.2+1.0             & 0.8--9.8     & 4.9--10   & 16:01:03.1 & $-$51:33:54      &    \citet{2001ApJ...551..394M,park2009nonthermal};    \\
   &  &  & &       & \citet{2018ApJ...868L..21B,2020ApJS..248...16L}     \\
G347.3--0.5            & 0.1--6.8     & 0.9--6.0  & 17:13:28.3 & $-$39:49:53      &   \citet{1999ApJ...525..357S,1997AA...318L..59W};     \\
   &  &  & &       & \citet{cassam2004xmm,2003ApJ...593L..27L}     \\
   &  &  & &       & \citet{2016PASJ...68..108T}     \\
G350.1--0.3            & 0.6--2.5     & 4.5--9.0  & 17:20:54.5 & $-$37:26:52     &     \citet{gaensler2008re,lovchinsky2011chandra}    \\
   &  &  & &       & \citet{2014PASJ...66...68Y,2020ApJS..248...16L}     \\
G354.4+0.0             & 0.1--0.5        & 5--8        & 17:31:27.5 & $-$33:34:12       &   \citet{roy2013discovery}      \\
\hline
\end{tabular}
\caption{\label{tab:targets} The 15 SNRs covered in this analysis. Sources in the upper half of the table are searched by all three pipelines described in Section~\ref{sec:searches}. Sources in the bottom half are searched by a single pipeline described in Section~\ref{sec:viterbiregular}. The ages and distances listed are consistent with the values used in the previous LIGO analysis~\citep{Abbott_2019}. 
}
\end{table}

\section{Instrumental overview and data}
\label{sec:instrument}
The O3 observing run started on April 1st, 2019 at 15:00 UTC and ended on March 27th, 2020 at 17:00 UTC. 
For the search, we use data collected by the two Advanced LIGO detectors in Hanford, Washington (H) and Livingston, Louisiana (L) and Advanced Virgo in the first half of O3, from the start until October 1st, 2019.
This time period is referred to as ``O3a''.
The data collected by the two LIGO detectors during the second half of O3 (O3b), starting from November 1st, 2019 until the end of O3, are used by the BSD pipeline (Section~\ref{sec:BSD}) and dual-harmonic Viterbi pipeline (Section~\ref{sec:viterbi2f}) to cross-check candidates.
Data collected by Virgo are only used by the BSD pipeline, which runs the initial search using individual detectors separately (Section~\ref{sec:BSD}). 
In the two Viterbi-based pipelines, the Virgo data are not used due to the detector's relatively lower sensitivity, and the two pipelines both operate on all detectors combined.
All three pipelines use data collected when the detectors are in the nominal low-noise observing mode~\citep{O2O3detchar}. The BSD pipeline (Section~\ref{sec:BSD}) uses low-latency calibrated data (C00 frames)~\citep{Sun_2020} for H and L detectors and the ``online" calibration version for Virgo, after a procedure of removing significant short-duration noise transients, known as ``glitches"~\citep{O2O3detchar}, in the Short Fourier Transform Database (SFDB)~\citep{FFTpeakmaps}. Tests show that the difference between the C00 data, after glitch removal in SFDB, and glitch gated C01 frames is negligible. The two Viterbi pipelines (Sections~\ref{sec:viterbiregular} and \ref{sec:viterbi2f}) use the high-latency calibrated data (C01 frames)~\citep{Sun_2020}, passed through a procedure of glitch gating~\citep{gating-T2000384}.

\section{Search methods}
\label{sec:searches}

\subsection{BSD}
\label{sec:BSD}
The BSD directed search pipeline is a hierarchical semi-coherent method based on the FH transform \citep{FrequencyHoughmethod,FrequencyHough}. A previous search using the BSD directed search pipeline, pointing to the Galactic Center in Advanced LIGO O2, was reported in \cite{DirectedBSD}. The pipeline descibed in this section is based on the BSD framework, i.e. a library of functions which allows the user to freely select a subset of the detector strain data (both in frequency and time domain), starting from a collection of basic files (BSD files) in a special data format.
All the properties of the framework are described in \cite{BSD}, and here we only remind the reader that the standard format of the BSD files, containing an opportunely down-sampled complex time series,  covers a 10-Hz frequency band and $\sim1$ month of data. For the purpose of this search, where the actual signal frequency is unknown, each BSD file is partially corrected for the Doppler modulation in each 1-Hz frequency sub-band using its central frequency (see \citealt{DirectedBSD} for more details). From this partially corrected time series, a collection of time-frequency peaks (called ``peakmaps") is obtained, by choosing all the local maxima above a given threshold from equalized spectra~\citep{FFTpeakmaps}. The equalization is given by the square modulus of the periodogram divided by the average spectrum. In this way also narrow peaks are kept. This peakmap is the input of the FH transform, which maps each time-frequency peak into the intrinsic source frequency and spin-down $(f_0,\dot{f}_0)$ plane at a given reference time. 
The resolution of a single FH map is the size of the bins in the template grid
\begin{eqnarray}
\delta f_{\rm FH} = \frac{1}{T_{\rm coh}K_{f}},
\label{eq:deltaf}
\\
\delta \dot{f}_{\rm FH}=\frac{1}{T_{\rm coh}T_{\rm obs}K_{\dot{f}}},
\label{eq:deltadf}
\end{eqnarray}
where $T_{\rm coh}$ is the coherence time, while $T_{\rm obs}$ is the observational time. The parameters $K_{f}$ and $K_{\dot{f}}$ are the over-resolution factors as described in \cite{FrequencyHoughmethod}, here chosen as $K_{f}=10$ and $K_{\dot{f}}=2$. 
The coherence time $T_{\rm coh}$ scales with the maximum frequency of the band as $1/\sqrt{f_{\rm max}}$, and hence the frequency and spin-down bin sizes in Eqs. (\ref{eq:deltaf}) and (\ref{eq:deltadf}) change for each 10-Hz band. For a source with age $t_{\rm age}$, the spin-down range is defined as  $-{f_{\rm max}}/{t_{\rm age}}\leq \dot{f} \leq 0.1 {f_{\rm max}}/{t_{\rm age}}$, where $f_{\rm max}$ is the maximum frequency in each 10-Hz band. In this analysis, the age of the source affects the parameter space investigated, with a wider spin-down range covered when the source is younger. When possible, we use the youngest age estimate available in the SNRcat catalog \citep{SNRcat,SNRcatpaper}. On the other hand, according to the age of the source, we can consider the effects of the second order spin down as negligible or not (a discussion is reported in Appendix~\ref{appendix:BSD}). In this search, we investigate a frequency band of $[10, 600]$~Hz for targets with assumed $t_{\rm age}\leq3$ kyr, and a wider range of $[10, 1000]$~Hz for older sources. We remind the reader of the subtle difference when talking about the source age estimates (which is most of the time inferred from the SNR age) and the characteristic age of the star (which is unknown because they have no observed electromagnetic pulsations). The maximum coherence time used is 17.8~hr for the frequency band $[10, 20]$~Hz and a minimum of 2.5~hr for $[990, 1000]$~Hz.
We search both positive and negative $\dot{f}$ to allow for the possibility of unexpected spin up.
A summary of the parameter space investigated for each source is shown in Table~\ref{tab:parabsd}.

\begin{table}[!tbh]
	\centering
	\setlength{\tabcolsep}{3pt}
	\renewcommand\arraystretch{1.2}
	\begin{tabular}{lcccc}
		\hline
		Source &  minimum $t_{\rm age}$ (kyr) & $T_{\rm coh}$ (hr)  & $f$ (Hz) & $\dot{f}$ $(\rm{Hz\,s^{-1}})$ \\
		       &                              & (@100 Hz)           &          &   (@100 Hz)\\
		\hline
		G65.7+1.2, G189.1+3.0, G266.2--1.2 & 3   & 8 & [10, 600]  & $[-1.06\times 10^{-9}, 1.06\times 10^{-10}]$  \\    
		G353.6--0.7 					   & 27  & 8 & [10, 1000] & $[-1.17\times 10^{-10}, 1.17\times 10^{-11}]$ \\
		G18.9--1.1 						   & 4.4 & 8 & [10, 1000] & $[-7.13\times 10^{-10}, 7.13\times 10^{-11}]$ \\
		G39.2--0.3 						   & 4.7 & 8 & [10, 1000] & $[-6.75\times 10^{-10}, 6.75\times 10^{-11}]$  \\
		G93.3+6.9 						   & 5   & 8 & [10, 1000] & $[-6.34\times 10^{-10}, 6.34\times 10^{-11}]$  \\
		\hline
	\end{tabular}
	\caption{Sources searched in the BSD analysis (Section~\ref{sec:BSD}) and the parameter space covered. The coherence time and the spin-down/up range scale with the maximum frequency in each 10-Hz frequency band. For each source, we report the $T_{\rm coh}$ and spin-down/up range used for the frequency band [90, 100] Hz where $f_{\rm max}=100$ Hz.}
	\label{tab:parabsd}
\end{table}

The first set of candidates is selected from a final FH map, which is the sum of all the single monthly-based FH maps spanning the same frequency and spin-down ranges. These candidates are independently selected in each detector, including Virgo, using the ranking procedure of \citet{FrequencyHoughmethod} where candidates with the highest FH number count are kept. At a later stage, coincidences are calculated between the candidate sets from the two LIGO detectors using a coincidence distance defined as 

\begin{equation}
d=\sqrt{\left(\frac{\Delta f}{\delta f_{\rm FH}}\right)^2+\left(\frac{\Delta \dot{f}}{\delta \dot{f}_{\rm FH}}\right)^2},
\label{Eq:distance}
\end{equation}
where $\Delta f$ and $\Delta \dot{f}$  are the differences between the candidate parameters in each data set.  A candidate is then selected when the coincidence distance is below a given threshold distance, $d_{\rm thr}$ in this search chosen equal to 4.  The choice of the window size has been widely discussed in \citet{FrequencyHoughmethod}, using injected simulated signals.

The coincidence step has been applied first to the pair of LIGO candidates. At a later stage, the same coincidence criterion has been applied between the HL coincident candidates and the most significant Virgo candidates.  Candidates found in triple coincidence were discarded after applying the post-processing methods described in Appendix~\ref{sec:postprocessing}.   
However, we cannot conclude with certainty that a pair of LIGO candidates are non-astrophysical if they have $d < d_{\rm thr}$ but are not seen in Virgo data, because Virgo is less sensitive than LIGO. For this reason we also postprocessed all the candidates found in coincidence between H and L only.

Surviving candidates are further investigated through a followup process described in Appendix~\ref{sec:postprocessing}.
Also, we apply a threshold to the Critical Ratio (CR) $\rho_{\rm CR}$, which measures the statistical significance of a candidate based on the number count associated with the pixel of the FH map where the candidate lies.
The threshold $\rho_{\rm CR,thr}$ is chosen as the mean $\rho_{\rm CR}$ plus one standard deviation of the CR distribution across the candidates excluding those due to known instrumental lines (Appendix~\ref{subsec:line_veto}) and with an inconsistent significance among the two detectors (Appendix~\ref{subsec:itf_veto}). 
For the targets G65.7+1.2, G189.1+3.0, and G266.2--1.2, we use $\rho_{\rm CR,thr}=4.7$; for G18.9--1.1 and G93.3+6.9, we use $\rho_{\rm CR,thr}=4.6$; and for G353.6--0.7 and G39.2--0.3, we use $\rho_{\rm CR,thr}=4.5$.
The threshold chosen here is less stringent than in \citet{DirectedBSD} where the threshold was $\approx 6.5$, corresponding to the probability of picking an average of one false candidate over the total number of points in the parameter space, under the assumption of Gaussian noise. For this work, a lower CR threshold is picked since we are using some new postprocessing methods, described in Appendix~\ref{sec:postprocessing}, which allow us to followup a higher number of candidates, given the low computational cost of each step.

\subsection{Single-harmonic Viterbi}
\label{sec:viterbiregular}
An HMM is an efficient search algorithm capable of handling both spin down and spin wandering.
Previous searches for young SNRs using an HMM \citep{sun2018hidden} were conducted in the Advanced LIGO O2 data, but no evidence for a GW signal was reported~\citep{millhouse2020search}. 

An HMM models a time-varying signal with underlying hidden (i.e. unobservable) parameters by treating the hidden parameters as links in a Markov chain, with each hidden parameter linked to an observable through a likelihood statistic.
Given an observed sequence, the goal is to infer the most probable hidden sequence.
For a set of $N_T$ observations at discrete times $\{t_0,t_1, ..., t_{N_T-1}\}$, the corresponding discrete states $\{q(t_0), q(t_1), ..., q(t_{N_T-1})\}$ (chosen from $N_Q$ possible hidden states $\{q_1, ..., q_{N_Q}\}$) form a Markov chain with transition probabilities from $t_k$ to $t_{k+1}$ defined by $A_{q_iq_j} = P[ q(t_{k+1}) = q_j | q(t_k) = q_i ]$.
For this search, we choose $A_{q_iq_i} = A_{q_{i\pm1}q_i} = 1/3$ and all other $A_{q_iq_j} = 0$, allowing the frequency to remain static or wander up or down one bin for each time step.
This allows us to track both spin down and stochastic spin wandering, which may cause spin up.
Strictly speaking, spin down is expected to be more rapid than spin up due to spin wandering, but the exact values of $A_{q_i q_j}$ have minimal effect on the performance of an HMM, provided they capture the behaviour of the signal in a broad sense \citep{quinn2001estimation,suvorova2016hidden}.
We assume a uniform prior over the initial state, i.e. $\Pi[q(t_0)] = N_Q^{-1}$.
The observations are denoted $\{o(t_0), o(t_1), ..., o(t_{N_T-1})\}$ and are connected to $q(t_k)$ through unknown parameters.
We call the probability of observing $o(t_k)$ given some state $q(t_k)$ the emission probability $L_{o(t_k) q(t_k)} = P[o(t_k) | q(t_k)]$.
Given some observed sequence $O$, we can then infer the most likely hidden sequence $Q^*$ by maximizing
\begin{equation} P(Q^*|O) =\Pi[q(t_0)]\prod_{k = 1}^{N_T-1} L_{o(t_k)q(t_k)}A_{q(t_k) q(t_{k-1})}.
\label{eqn:qstar}
\end{equation}
The Viterbi algorithm is an efficient implementation of the inference step, using dynamic programming to sample and discard unfavourable paths at each time-step \citep{ViterbiOriginal,suvorova2016hidden}. 

For our purposes, the hidden state is the true GW frequency and the observable is the value of the $\mathcal{F}$-statistic, calculated coherently over a block of duration $T_{\rm coh}$ and width (in the frequency domain) $(2T_{\rm coh})^{-1}$.
The $\mathcal{F}$-statistic is a maximum likelihood filter for a CW signal of frequency $f$ with time derivatives $\dot{f}$, $\ddot{f}$, etc. (for more details on the $\mathcal{F}$-statistic, please see \citealt{Jaranowski1998}).
In this search, we compute the $\mathcal{F}$-statistic as a function of $f$ only, and account for spin down by choosing $T_{\rm coh} \propto \left|\dot{f}_0^{\rm max}\right|^{-1/2}$ (as in \citealt{sun2018hidden}), where $\dot{f}_0^{\rm max}$ is the maximum $\dot{f}$ within $T_{\rm coh}$, such that the signal should wander by at most one frequency bin per time step. 

We choose our parameter space according to the detectability of a potential signal.
First, we estimate the maximum expected GW strain for a neutron star at distance $D$ with characteristic age $t_{\rm age}$ and a principle moment of inertia $I_{zz}$ using
\begin{equation}
h_0^{\rm age} = 2.27\times 10^{-24} \left( \frac{1 \, {\rm kpc}}{D}\right) \left(\frac{1 \, {\rm kyr}}{t_{\rm age}}\right)^{1/2} \left(\frac{I_{zz}}{10^{38} \, {\rm kg \, m}^2}\right)^{1/2}
\label{eqn:maxstrain}
\end{equation}
and assuming purely gravitational spin down \citep{wette2008searching}.
We also estimate the minimum detectable strain using an analytic estimate of the 95\% confidence sensitivity for a semi-coherent search, given by \citep{sun2018hidden,wette2008searching}
\begin{equation}
\label{eqn:sensitivity}
h_0^{\rm est}=\Theta S_n(f)^{1/2}\left(T_\mathrm{obs}T_{\rm coh}\right)^{-1/4},
\end{equation}
where $S_n(f)$ is the noise amplitude spectral density.
The statistical threshold $\Theta$ is defined by the location in parameter space and typically lies in the range $30 \lesssim \Theta \lesssim 40$.
Following previous studies for CWs with an HMM, we take $\Theta = 35$ \citep{sun2018hidden,wette2008searching}.
The frequency range for each source is defined by $h_0^{\rm est} < h_0^{\rm age}$.
The parameter space for each source, including $T_{\rm coh}$, is summarized in Table~\ref{tab:vit_targets}, and the process for defining the parameter space is described in Appendix~\ref{appendix:Viterbi}.

We split the data into $N_{\rm band}$ frequency sub-bands of width 2~Hz to ensure loud, non-Gaussian noise artifacts (e.g. lines) are confined to one sub-band and do not affect the whole analysis.
We overlap the frequency sub-bands by 0.57~Hz, ensuring that any signal corresponding to a rapidly spinning down neutron star can always be contained in a single sub-band.

For each sub-band, we apply the Viterbi algorithm outlined above and obtain $N_Q$ frequency paths ending in $N_Q$ different bins with associated likelihoods $\mathcal{L}$.
Alternative implementations of Viterbi (including \citealt{suvorova2016hidden} and \citealt{sun2018hidden}) used a Viterbi score as their detection statistic (see Section \ref{sec:viterbi2f}).
This statistic generally requires $N_T \ll N_Q$.
\citet{millhouse2020search} demonstrated that this statistic fails to identify an injected (or real) path for $N_T \sim N_Q$ because the score is calculated for the optimal path relative to other paths in the band.
If most of the paths overlap, the optimal path is similar to other paths in the band.
In this search, we have a minimum $T_{\rm coh} = 1$~hr ($N_T = 4391$, $N_Q = 14400$), which is sufficient for almost one third of  Viterbi paths to converge over $T_{\rm obs}$ and consequently lower the sensitivity of the Viterbi score.
To maintain the search sensitivity with $N_T \sim N_Q$, we use the log-likelihood $\mathcal{L}$ as our detection statistic. 
Using the process outlined in Appendix~\ref{appendix:Viterbi}, we determine the 1\% false alarm threshold for each source and denote the corresponding likelihood $\mathcal{L}_{\rm th}$. 
We follow up all unique frequency paths with $\mathcal{L} > \mathcal{L}_{\rm th}$ using the procedure described in Appendix \ref{sec:postprocessing} and find no CW candidates which cannot be described by non-astrophysical noise.

\begin{table}[!tbh]
\centering
\setlength{\tabcolsep}{3pt}
\renewcommand\arraystretch{1.2}
	\begin{tabular}{llllll}
		\hline
		Source     & Minimum $t_{\rm age}$ (kyr) & $D$ (kpc) &$T_{\rm coh}$ (hours) & \(f\) (Hz) &$\dot{f}$ (Hz/s)\\
		\hline         
        G1.9+0.3    & 0.10   &   8.5  & 1.0  & $[31.56, 121.7 ] $ & $ \left[-3.858\times 10^{-8}        ,   3.858\times 10^{-8}        \right]$   \\
		G15.9+0.2   & 0.54   &   8.5  & 1.0  & $[44.03, 657.1] $  & $ \left[-3.858\times 10^{-8}        ,   3.858\times 10^{-8}        \right]$   \\
		G18.9--1.1  & 4.4    &     2  & 1.9  & $[31.02, 1511 ] $  & $ \left[-1.507\times 10^{-8}        ,   1.507\times 10^{-8}        \right]$   \\
		G39.2--0.3  & 3.0    &   6.2  & 2.8  & $[62.02, 459.2] $  & $ \left[-1.968\times 10^{-8}        ,   1.968\times 10^{-8}        \right]$   \\
        G65.7+1.2   & 20     &   1.5  & 4.7  & $[35.10, 1128 ] $  & $ \left[-3.149\times 10^{-9}        ,   3.149\times 10^{-9}        \right]$   \\
		G93.3+6.9   & 5.0    &   1.7  & 1.9  & $[30.00, 1668 ] $  & $ \left[-1.335\times 10^{-8}        ,   1.335\times 10^{-8}        \right]$   \\
        G111.7--2.1 & 0.30   &   3.3  & 1.0  & $[25.71, 365.1] $  & $ \left[-3.858\times 10^{-8}        ,   3.858\times 10^{-8}        \right]$   \\
		G189.1+3.0  & 3.0    &   1.5  & 1.4  & $[26.13, 2000 ] $  & $ \left[-1.968\times 10^{-8}        ,   1.968\times 10^{-8}        \right]$   \\
		G266.2--1.2 & 0.69   &   0.2  & 1.0  & $[18.36, 839.6] $  & $ \left[-3.858\times 10^{-8}        ,   3.858\times 10^{-8}        \right]$   \\
		G291.0--0.1 & 1.2    &   3.5  & 1.0  & $[31.97, 1460 ] $  & $ \left[-3.858\times 10^{-8}        ,   3.858\times 10^{-8}        \right]$   \\
		G330.2+1.0  & 1.0    &     5  & 1.1  & $[36.57, 1039 ] $  & $ \left[-3.858\times 10^{-8}        ,   3.858\times 10^{-8}        \right]$   \\
		G347.3--0.5 & 1.6    &   0.9  & 1.0  & $[21.74, 1947 ] $  & $ \left[-3.858\times 10^{-8}        ,   3.858\times 10^{-8}        \right]$   \\
		G350.1--0.3 & 0.60   &   4.5  & 1.0  & $[31.96, 730.1] $  & $ \left[-3.858\times 10^{-8}        ,   3.858\times 10^{-8}        \right]$   \\
		G353.6--0.7 & 27     &   3.2  & 10   & $[77.86, 318.3] $  & $ \left[-2.295\times 10^{-9}        ,   2.295\times 10^{-9}        \right]$   \\
		G354.4+0.0  & 0.10   &     5  & 1.0  & $[25.72, 121.7] $  & $ \left[-3.858\times 10^{-8}        ,   3.858\times 10^{-8}        \right]$   \\
		\hline
	\end{tabular}
	\caption{\label{tab:vit_targets} Sources searched in the single-harmonic Viterbi analysis (Section~\ref{sec:viterbiregular}) and the parameter space covered. The parameter space for each of the 15 sources is derived using the age and distance estimates in columns two and three.} 
\end{table}

\subsection{Dual-harmonic Viterbi}
\label{sec:viterbi2f}
\newcommand{\fstar}{f_\star}
\newcommand{\Tcoh}{T_{\rm coh}}
\newcommand{\Tobs}{T_{\rm obs}}
\newcommand{\fdot}{\dot{f}_0}
\newcommand{\fdotmax}{\dot{f}_{\rm max}}

Methods in Sections~\ref{sec:BSD} and \ref{sec:viterbiregular} assume that the star rotates about one of its principal axes of the moment of inertia, and hence the GWs are emitted at $2 \fstar$.
This assumption is based on the fact that the phenomenon of free precession is not clearly observed in the population of known pulsars~\citep{Jones2010}.
However, the superfluid interior of a star pinned to the crust along an axis nonaligned with any of its principal axes could allow the star to emit GWs at both $\fstar$ and $2\fstar$, even without free precession~\citep{Jones2010,Bejger2014,2015ApJ...807..132M}. 
The dual-harmonic emission mechanism motivates searches combining the two frequency components of a signal to improve signal-to-noise ratio. 
The HMM tracking scheme described in Section~\ref{sec:viterbiregular} has been extended to track two frequency components simultaneously~\citep{Sun2019}. 
The signal model considered in this section consists of both $\fstar$ and $2 \fstar$ components, given by \citep{Jaranowski1998,Sun2019}
\begin{eqnarray}
\label{eqn:h2p} h_{2+} &=& \frac{1}{2}h_0(1+\cos^2\iota)\sin^2\theta\cos 2\Phi ,\\
h_{2\times} &=& h_0 \cos \iota \sin^2\theta \sin 2 \Phi, \\
h_{1+} &=&\frac{1}{8}h_0\sin 2\iota \sin 2\theta \sin \Phi,\\
\label{eqn:h1c} h_{1\times} &=& \frac{1}{4} h_0 \sin \iota \sin 2 \theta \cos \Phi,
\end{eqnarray}
where $\iota$ is the inclination angle of the source, $\theta$ is the wobble angle between the star’s rotation axis and its principal axis of the moment of inertia, and $\Phi$ is the GW signal phase observed at the detector.
In general, when precession and triaxiality of the star are included, emission occurs at other frequencies too~\citep{Zimmermann1979,VanDenBroeck2005,Lasky2013}.

In this analysis, the HMM formulation generally follows the description in Section~\ref{sec:viterbiregular}, with three major updates. 
First, 
two different coherent times of $\Tcoh = 12$~hr and 9~hr are selected for three sources with $t_{\rm age}\gtrsim 20$~kyr and four sources with $t_{\rm age}\lesssim 5$~kyr, respectively. 
Second,
two frequency components are tracked simultaneously.
The GW signal for each frequency component is assumed to be monochromatic over $\Tcoh$.
The signal power in each frequency bin is computed by the two-component $\mathcal{F}$-statistic, denoted by $\mathcal{F}_1(f_i) + \mathcal{F}_2(2f_i)$,
where $\mathcal{F}_1$ and $\mathcal{F}_2$ are the $\mathcal{F}$-statistic outputs computed in two separate frequency bands, and $f_i$ is the frequency value in the $i$th bin. We use $\Delta f = 1/(4 \Tcoh)$ and $2\Delta f = 1/(2 \Tcoh)$ as frequency bin sizes when computing $\mathcal{F}_1$ and $\mathcal{F}_2$, respectively, such that both the $\fstar$ and $2\fstar$ signal components stay in one bin for each time interval $\Tcoh$.
Third, 
we assume that the signal frequency evolution is dominated by secular spin down, and can be approximated by a negatively biased random walk. The unknown spin-down rate lies in the range between zero and the maximum estimated spin-down rate and can vary over time.
Hence we use a transition probability matrix 
$A_{q_{i-1} q_i} = A_{q_i q_i} = 1/2$,
with all other entries being zero.
The full frequency band is divided into 1-Hz and 1.5-Hz sub-bands for $\Tcoh = 12$~hr and $\Tcoh = 9$~hr, respectively, to parallelize computing. The detection statistic used in this analysis requires that the number of frequency bins in each sub-band (with bandwidth $B$) is significantly larger than the total number of tracking steps (i.e., $2B\Tcoh \gg \Tobs/\Tcoh$). Thus for $\Tcoh = 9$~hr, we choose a 0.5-Hz wider sub-band such that the requirement is satisfied.
More details are provided in Appendix~\ref{appendix:Viterbi2f}.

Seven sources in the top half of Table~\ref{tab:targets} with an assumed age of $t_{\rm age} \gtrsim 3$~kyr are searched using this method.
Due to the fact that two frequency bands are combined, this method is susceptible to noise features present in either band. Coherent times shorter than $\sim 5$~hr and correspondingly, wider $\Delta f$, can further degrade the sensitivity.
Hence we do not search the other eight sources with $t_{\rm age} \lesssim 3$~kyr that require a much shorter $\Tcoh$.
The parameter space covered for each source is listed in Table~\ref{tab:para}. 
The $\dot{f}_\star$ range covered in this analysis is hence $|\dot{f}_\star| \in [0, 1/(4\Tcoh^2)]$.
The frequency range is determined as follows. For all seven sources, we fix the minimum frequency at 50~Hz and 100~Hz for $\fstar$ and $2\fstar$, respectively. We do not search below 50~Hz because the number of instrumental lines in each 1-Hz band significantly increases at low frequencies and the optimal Viterbi paths would be dominated by noise artifacts. 
The maximum frequency is set by the assumed minimum characteristic age of the source, $t_{\rm age}$ (the second column in Table~\ref{tab:para}), assuming $|\dot{f}_\star| =  f_\star (n-1)^{-1} t_{\rm age}^{-1}$~\citep{sun2018hidden,Abbott_2019},
where $n=f_\star \ddot{f}_\star/\dot{f}_\star^2$ is the braking index with $\ddot{f}_\star$ being the second time derivative of $f_\star$. We assume the spin down of the star is dominated by gravitational radiation due to a non-zero ellipticity, i.e., $n=5$.

\begin{table}[!tbh]
	\centering
	\setlength{\tabcolsep}{5pt}
	\renewcommand\arraystretch{1.2}
	\begin{tabular}{lllll}
		\hline
		Source & Minimum $t_{\rm age}$ (kyr) & $T_{\rm coh}$ (hr) & $f_\star$ (Hz)& $\dot{f}_\star$ $(\rm{Hz\,s^{-1}})$ \\
		\hline
		G65.7+1.2 & 20 & 12 &  [50, 338]& [$-1.34\times 10^{-10}$, 0]  \\
		G189.1+3.0 & 20 &12 & [50, 338] & [$-1.34\times 10^{-10}$, 0]  \\
		G353.6--0.7 & 27 &12 & [50, 457]& [$-1.34\times 10^{-10}$, 0]  \\
		G18.9--1.1 & 4.4 & 9 & [50, 132] & [$-2.38 \times 10^{-10}$, 0] \\
		G39.2--0.3 & 3 & 9 & [50, 90] & [$-2.38 \times 10^{-10}$, 0] \\
		G93.3+6.9 & 5 & 9 & [50, 150]& [$-2.38 \times 10^{-10}$, 0]  \\
        G266.2--1.2 & 5.1 & 9 & [50, 153] & [$-2.38 \times 10^{-10}$, 0]  \\
		\hline
	\end{tabular}
	\caption{Sources searched in the dual-harmonic Viterbi analysis (Section~\ref{sec:viterbi2f}) and the parameter space covered.}
	\label{tab:para}
\end{table}

We use the Viterbi score $S$ as the detection statistic in the dual-harmonic search, which indicates the significance of the optimal Viterbi path obtained in each sub-band compared to all other paths in that band at the final step of the tracking. Given that the condition $N_T \ll N_Q$ is generally satisfied with the choices of $\Tcoh$ in this method, the issue described in Section~\ref{sec:viterbiregular} with short $\Tcoh \sim 1$~hr does not happen. 
The full mathematical definition of $S$ is given in \cite{Sun2019}. We determine a threshold corresponding to 1\% false alarm probability $S_{\rm th}=5.47$ and $S_{\rm th}=5.33$ for $\Tcoh = 12$~hr and $\Tcoh = 9$~hr, respectively, obtained from Monte-Carlo simulations in Gaussian noise and verified in real O3a data. The results obtained from simulations in O3a interferometric noise are consistent with the Gaussian noise thresholds.

\section{Sensitivity and constraints}
\label{sec:limits}

A total of 42464, 9236, and 477 first-stage candidates are identified across all SNRs in BSD, single-harmonic Viterbi, and dual-harmonic Viterbi pipelines. We apply a hierarchical veto procedure (Appendix~\ref{sec:veto}) to the full population and perform dedicated follow-up analyses on 35, 1, and 25 candidates for BSD, single-harmonic Viterbi, and dual-harmonic Viterbi, respectively (Appendix~\ref{subsec:further_followup}). No candidate survives from any pipeline. All are consistent with a non-astrophysical origin. In this section, we present the sensitivity of each pipeline and the constraints obtained from this analysis.

\subsection{BSD constraints}
\begin{figure*}
	\centering
		\subfigure[][G65.7+1.2]
	{
		\label{fig:ulBSD_G65_7}
		\scalebox{0.29}{\includegraphics{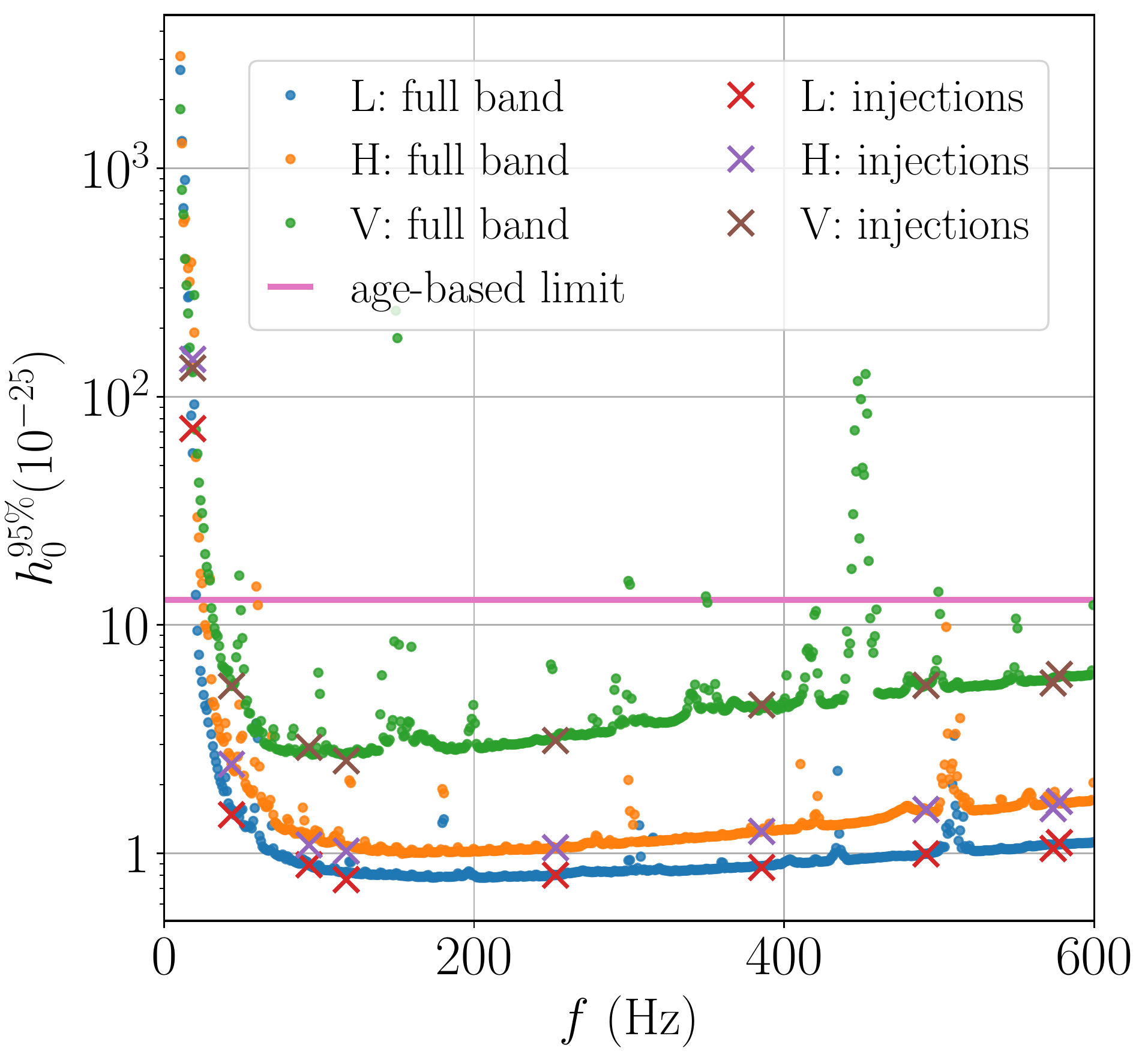}}
	}
	\subfigure[][G189.1+3.0]
	{
		\label{fig:ulBSD_G189_1}
		\scalebox{0.29}{\includegraphics{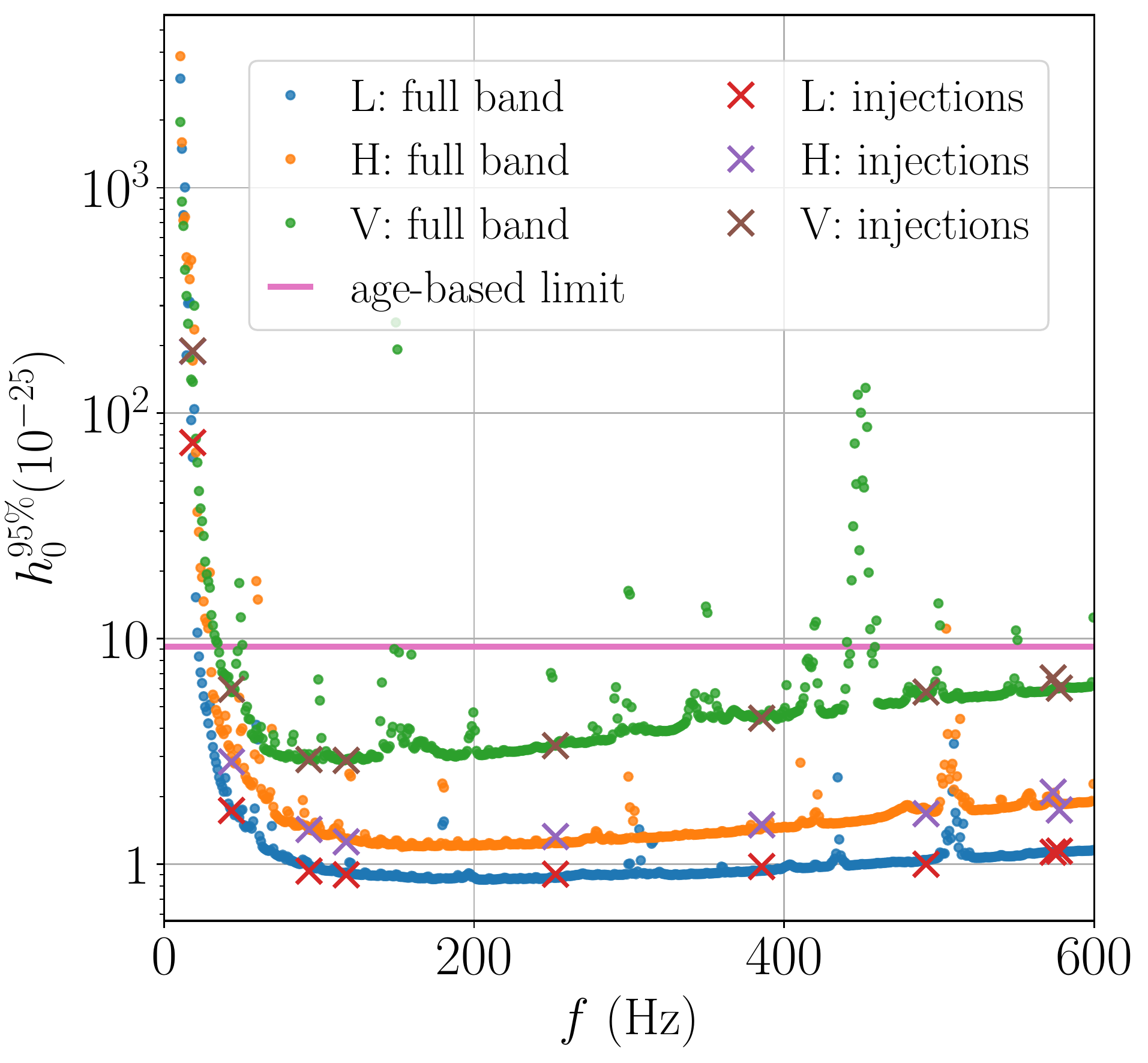}}
	}
 	\subfigure[][G266.2--1.2/Vela Jr.]
	{
		\label{fig:ulBSD_G266_2}
		\scalebox{0.29}{\includegraphics{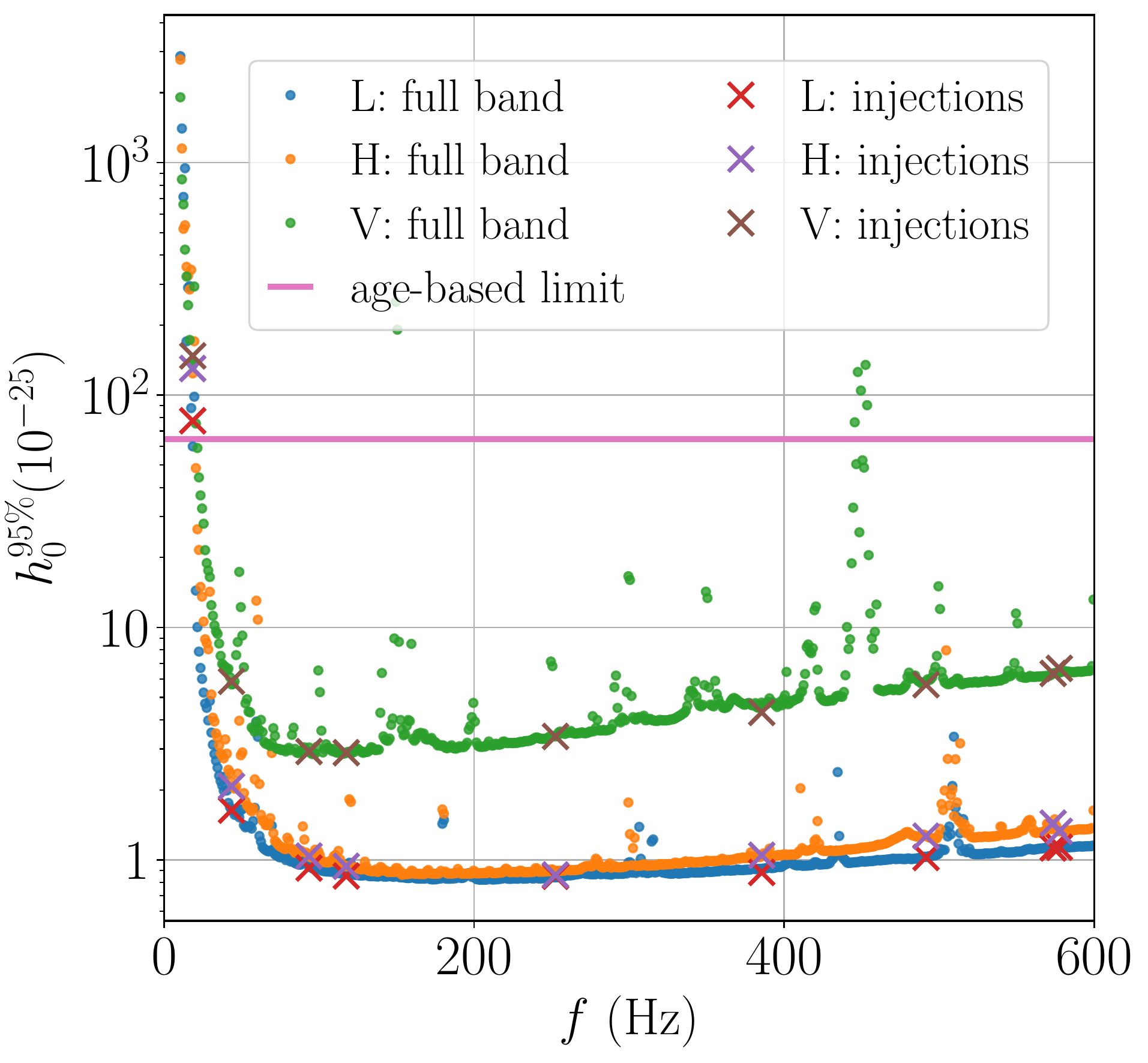}}
	}
	\subfigure[][G39.2--0.3]
	{
		\label{fig:ulBSD_G39_2}
		\scalebox{0.29}{\includegraphics{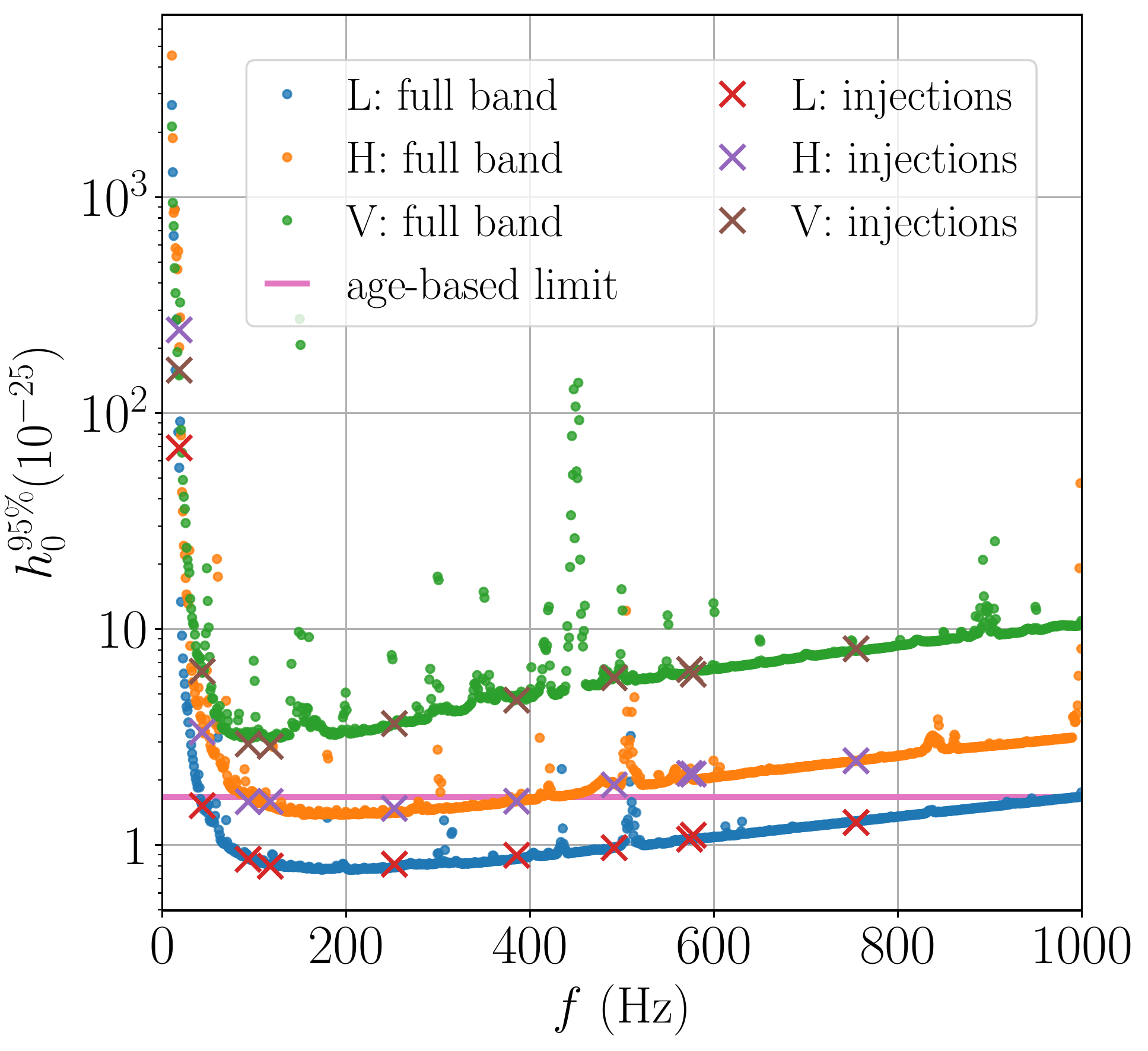}}
	}
	\subfigure[][G93.3+6.9]
	{
		\label{fig:ulBSD_G93_3}
		\scalebox{0.29}{\includegraphics{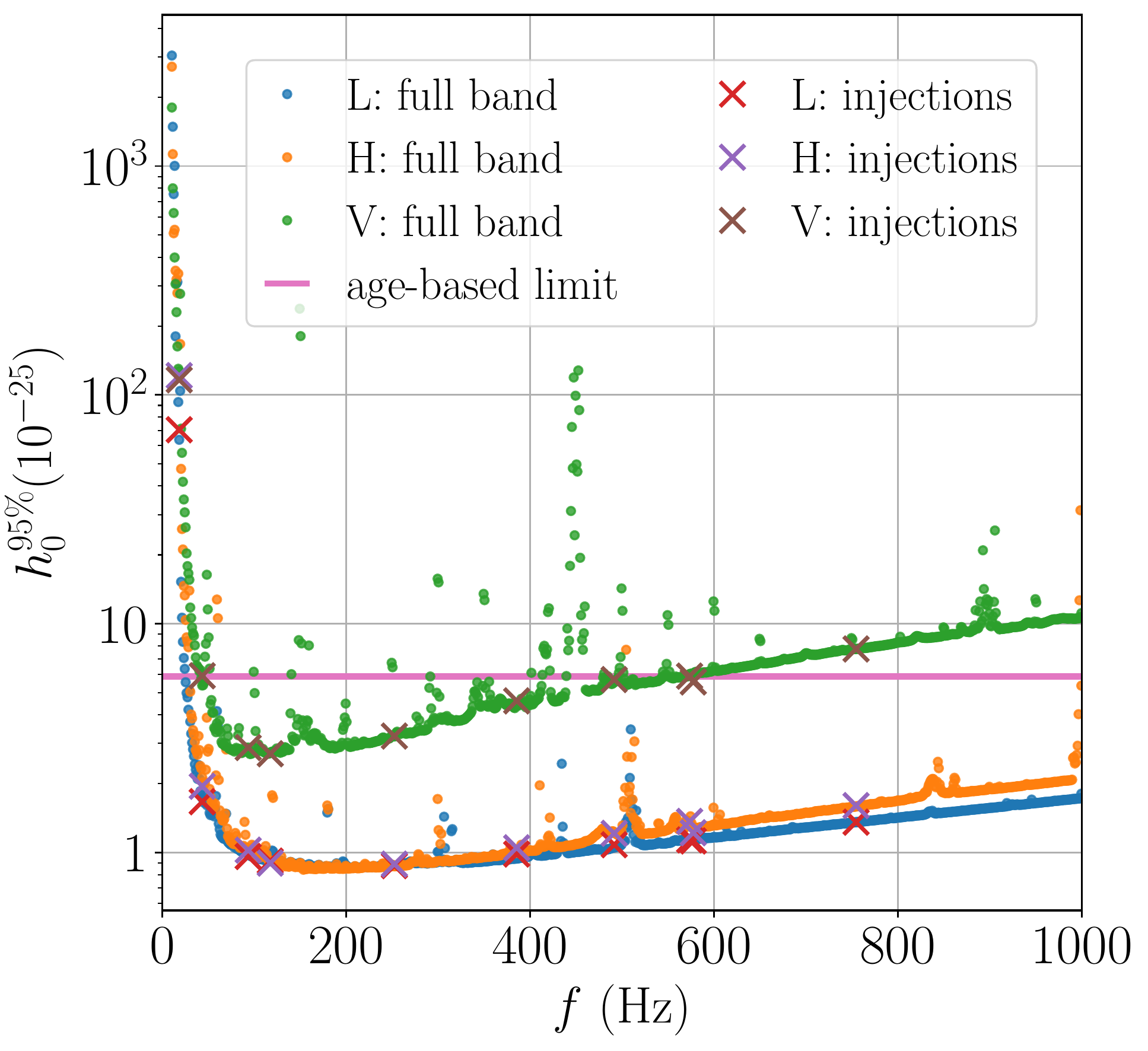}}
	}
	\subfigure[][G18.9--1.1]
	{
		\label{fig:ulBSD_G18_9}
		\scalebox{0.29}{\includegraphics{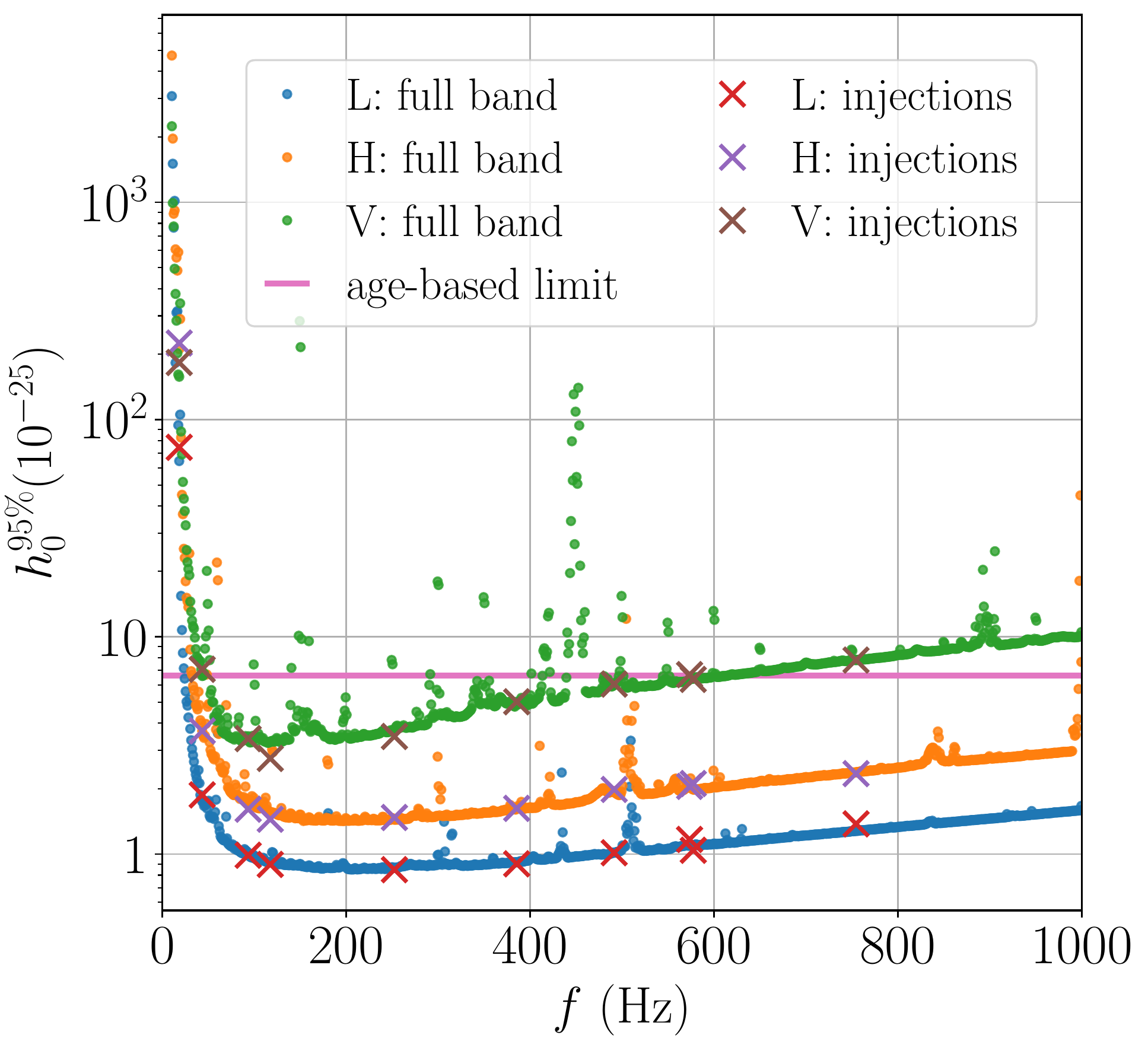}}
	}
	\subfigure[][G353.6--0.7]
	{
		\label{fig:ulBSD_G353_6}
		\scalebox{0.29}{\includegraphics{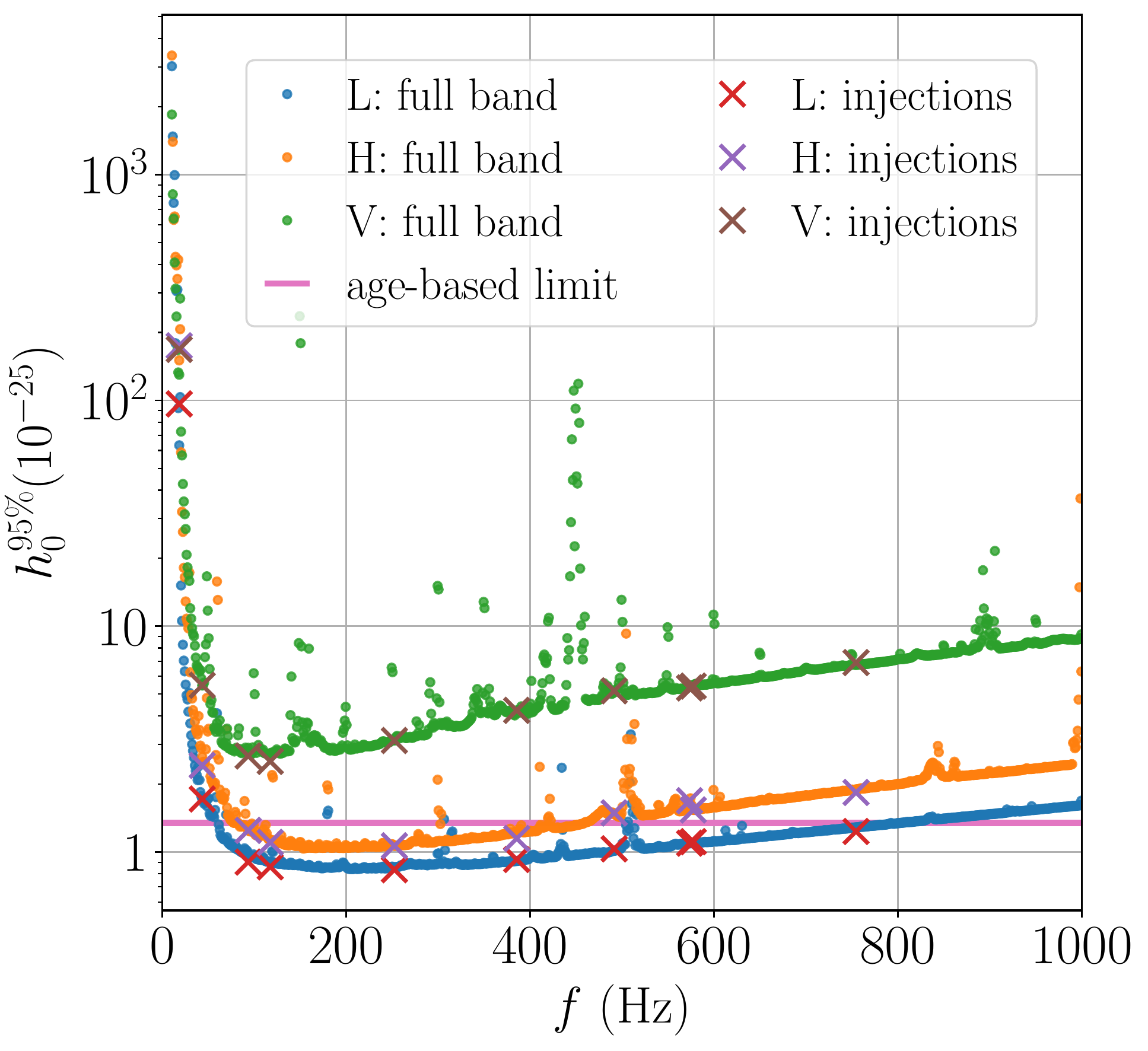}}
	}
	\caption{The sensitivity estimate $h_0^{95\%}$ obtained from the BSD search. The dotted curves represent the estimated $h_0^{95\%}$ in the full band of H, L and V detectors searched by the BSD pipeline (Table~\ref{tab:parabsd}). The crosses represent the frequentist strain upper limits at 95\% confidence level obtained empirically in the sample sub-bands of 1~Hz. Horizontal lines are the so-called indirect age-based limit as in Eq.~(\ref{eqn:maxstrain}). The limit is beaten across the full band also using Virgo data, except for the most disturbed regions, for G65.7+1.2, G189.1+3.0 and G266.2--1.2/Vela Jr. The remaining curves beat the limit on a limited parameter space and/or not for every detector.}
	\label{fig:BSDUL}
\end{figure*}

Surviving candidates are all compatible with noise fluctuations and no evidence of their presence is found in Virgo O3a and/or in the full LIGO O3 data. We compute the constraints on the strain amplitude using a well established method used in~\cite{DirectedBSD} and described in~\cite{PhysRevD.98.084058}. The sensitivity curve is obtained from the 95\% confidence level upper limits of 10 randomly selected frequency sub-bands of 1~Hz each for targets in the [10, 1000]~Hz frequency band, and 9 sub-bands for the remaining targets. The $h_0^{95\%}$ in the sub-bands is computed with the frequentist approach, i.e., injecting 50 signals with a given amplitude $h_0$ and computing the corresponding detection efficiency. The injections are done for each source, assuming the same sky position as the selected source for each injection. The spin-down and polarization parameters ($\cos \iota$ and $\psi$) are randomly chosen from their uniform distributions.
We repeat the injections in a given sub-band using 6–18 values of $h_0$ in the interval [1.3$\times 10^{-26}$, 3$\times 10^{-23}$]. The detection efficiency for a given amplitude $h_0$ is given by the fraction of injections recovered. The actual $h_0^{95\%}$ corresponding to a detection efficiency of 0.95 is derived from the sigmoidal fit of the detection efficiency curve versus the injected amplitude. 

Given that the sensitivity to $h_0$ is proportional to $\sqrt{S_n(f)}$, which is the noise amplitude spectral density, we compute the Normalized Upper Limit (NUL), $h_{\rm NUL}(f_i)={h_0^{95\%}(f_i)}/{\sqrt{S_n(f_i)}}$, in each of the randomly chosen sub-bands. We remark that it is the inverse of the more widely used ``sensitivity depth"~\citep{Behnke2015}. Since the NUL values should follow a linear trend, given by the dependence of the coherence time used in each 10 Hz band, we extrapolate the NUL values of the remaining bands with a linear fit of the NUL versus frequency.  In this way we can translate the NUL values, interpolated from the linear fit for each 1~Hz band, into the $h_0^{95\%}(f)$ curve.
The final $h_0^{95\%}(f)$ curve is then obtained for each detector, by multiplying the NUL values extrapolated from the linear fit in each 1-Hz band with the corresponding value of $\sqrt{S_n(f)}$ in that band, i.e., 
\begin{equation}
\label{eqn:BSDNUL}
h_0^{95\%}(f)=h_{\rm NUL}(f)\sqrt{S_n(f)}.
\end{equation}

The sensitivity plots are presented in Figure~\ref{fig:BSDUL} where we also report the indirect age-based limit from Eq.~(\ref{eqn:maxstrain}) (solid line) for each target. The best sensitivity is below the indirect age-based limit for all the sources. In particular for G65.7+1.2, G189.1+3.0 and G266.2–1.2/Vela Jr., this happens for the full frequency band analyzed, except for the most disturbed regions, and for all the detectors. 
The difference in sensitivity among the analyzed targets, is caused by the different antenna pattern response due to different sky locations of the sources, even when the same coherence time is used for multiple sources. 
We present different curves for each detector; the combined $h_0^{95\%}(f)$ result would correspond to the one for the less sensitive LIGO detector.
The best sensitivity at 95\% confidence level occurs at the Livingston detector at $h_0 \approx 7.8\times 10^{-26}$ near 200~Hz for G65.7+1.2 and at $h_0 \approx 7.7\times 10^{-26}$ for G39.2--0.3 in the same bucket region.

\subsection{Single-harmonic Viterbi constraints}
\begin{figure*}
	\centering
	\subfigure[][G18.9--1.1]
	{
		\label{fig:ul_viterbi_18911}
		\scalebox{0.25}{\includegraphics{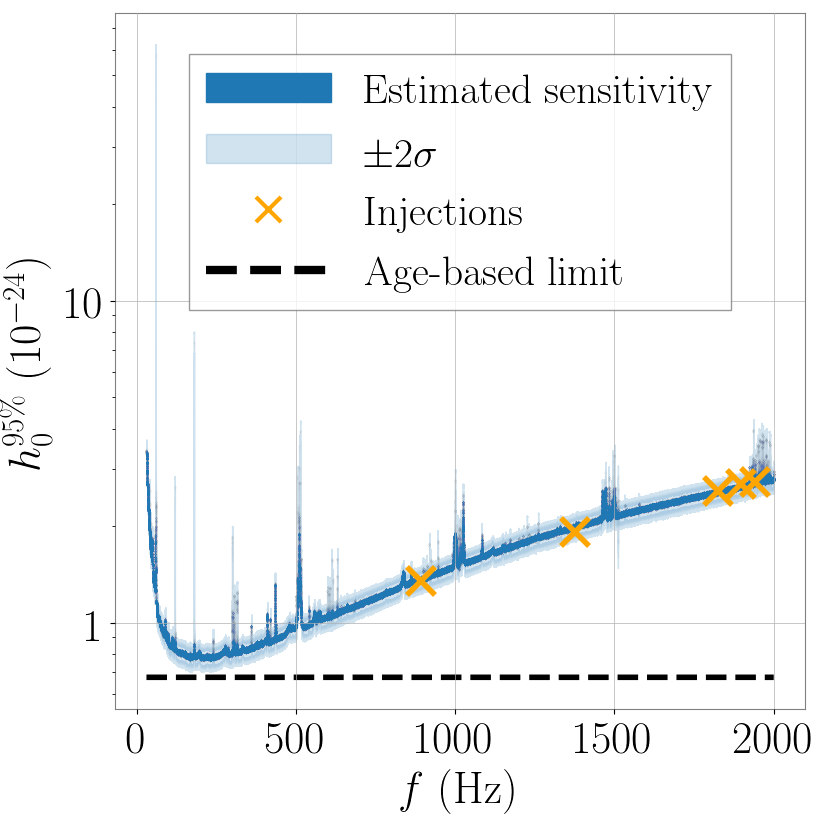}}
	}
	\subfigure[][G39.2--0.3]
	{
		\label{fig:ul_viterbi_39203}
		\scalebox{0.25}{\includegraphics{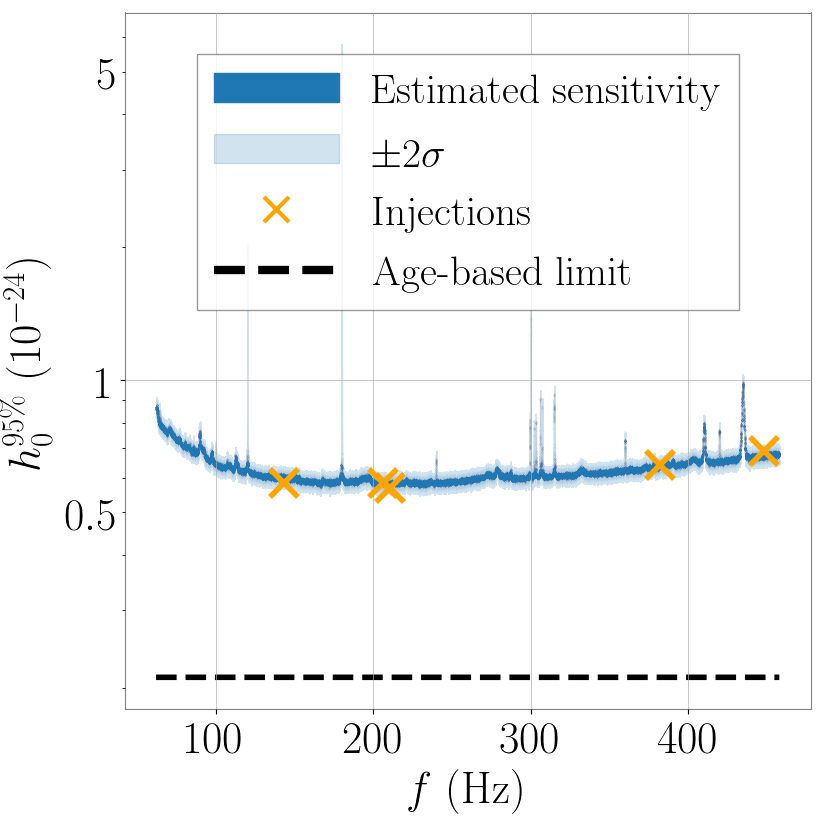}}
	}
	\subfigure[][G65.7+1.2]
	{
		\label{fig:ul_viterbi_65712}
		\scalebox{0.25}{\includegraphics{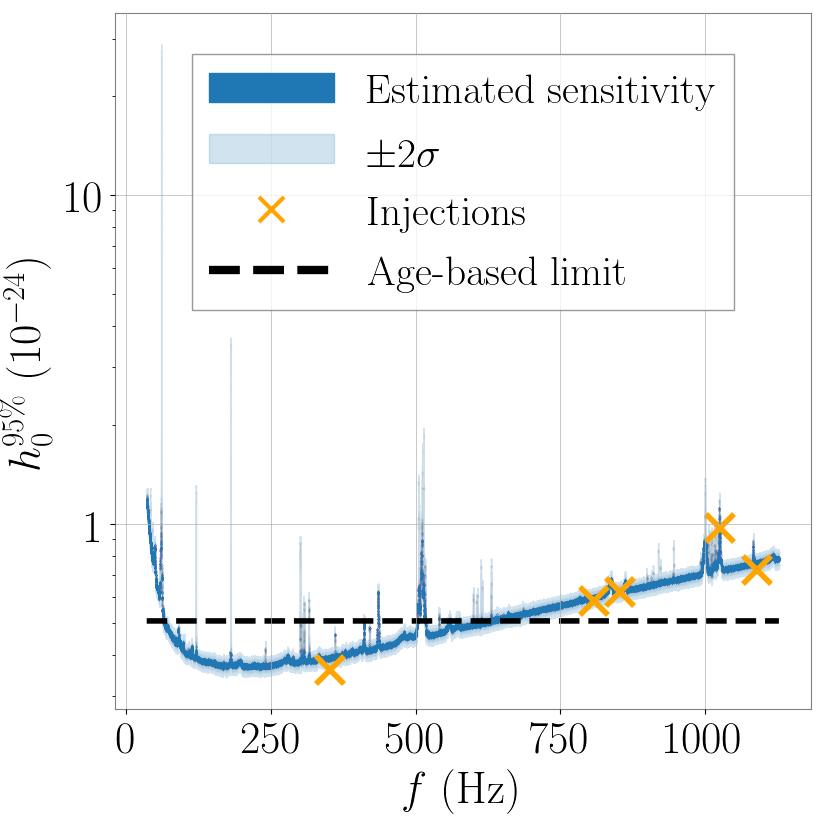}}
	}
	\subfigure[][G93.3+6.9]
	{
		\label{fig:ul_viterbi_93369}
		\scalebox{0.25}{\includegraphics{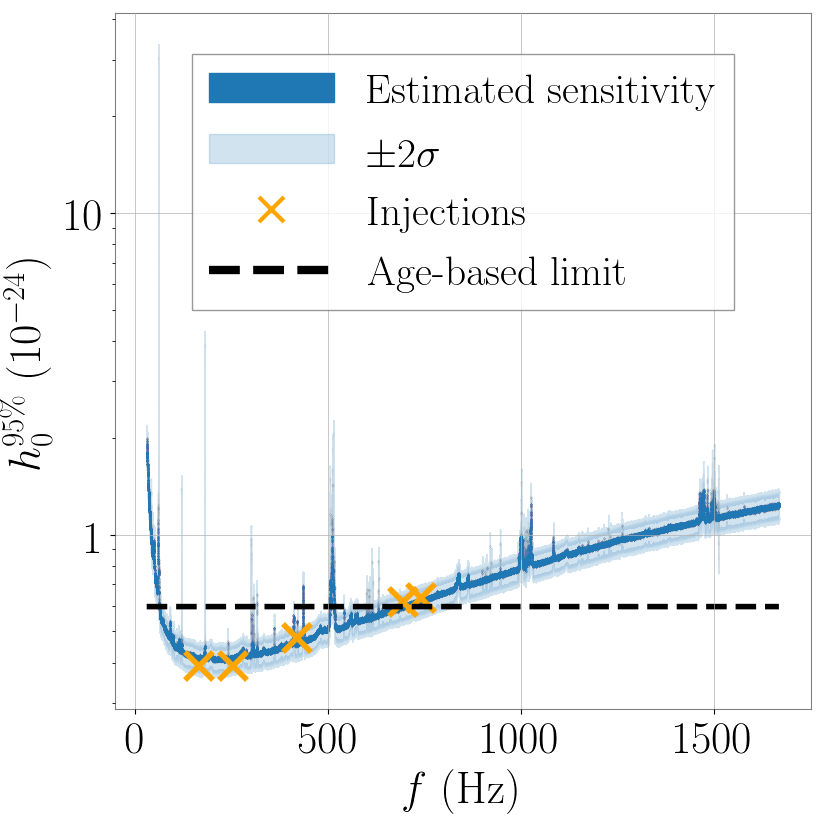}}
	}
	\subfigure[][G189.1+3.0]
	{
		\label{fig:ul_viterbi_189130}
		\scalebox{0.25}{\includegraphics{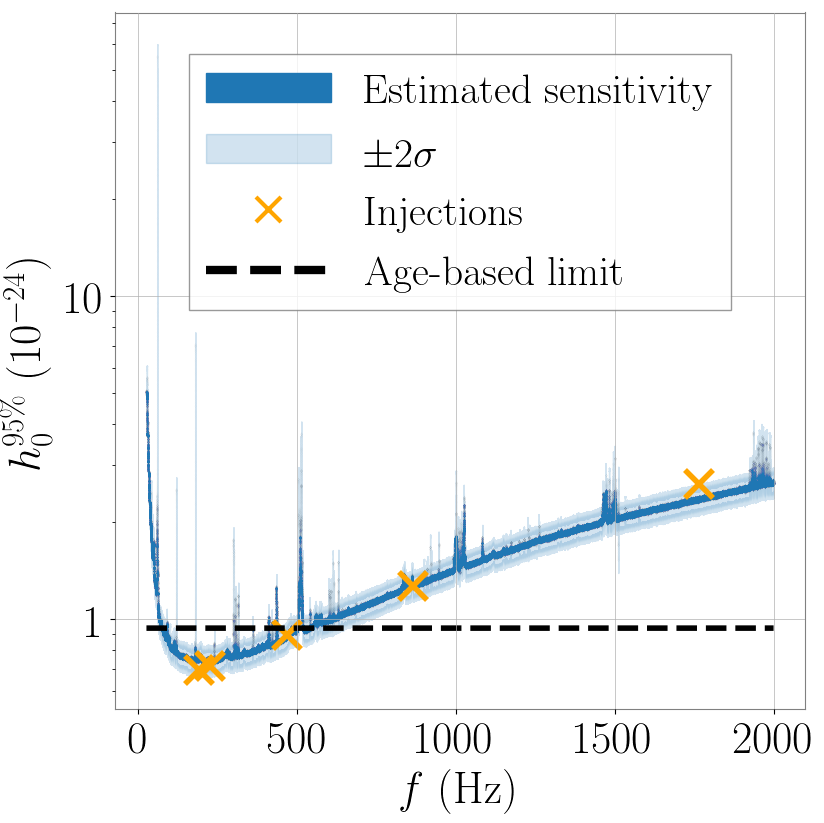}}
	}
	\subfigure[][G330.2+1.0]
	{
		\label{fig:ul_viterbi_330210}
		\scalebox{0.25}{\includegraphics{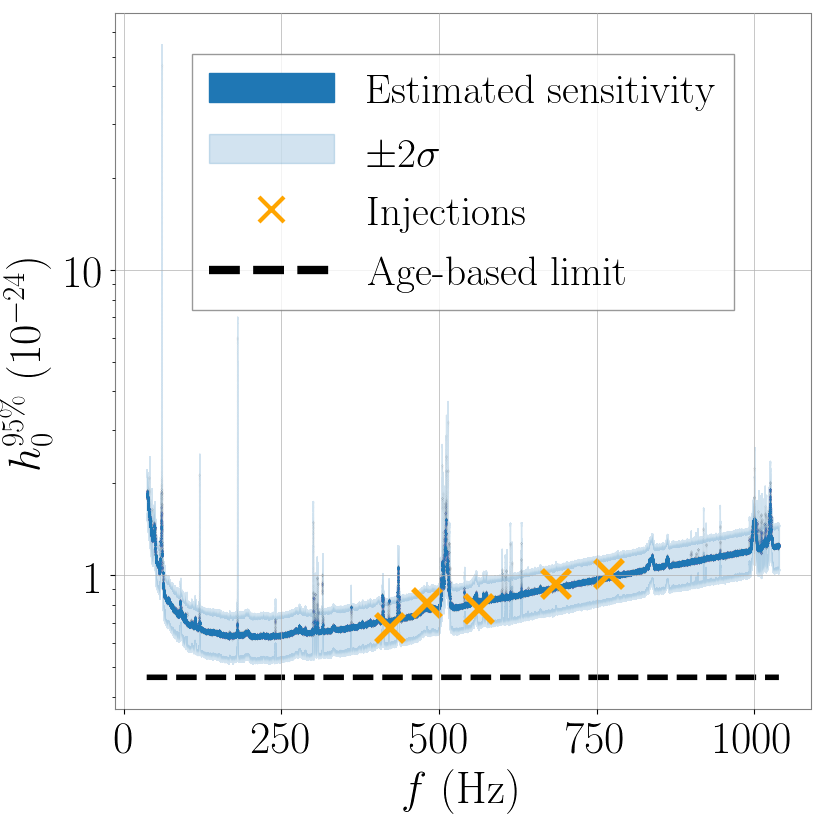}}
	}
	\subfigure[][G353.6--0.7]
	{
		\label{fig:ul_viterbi_353607}
		\scalebox{0.25}{\includegraphics{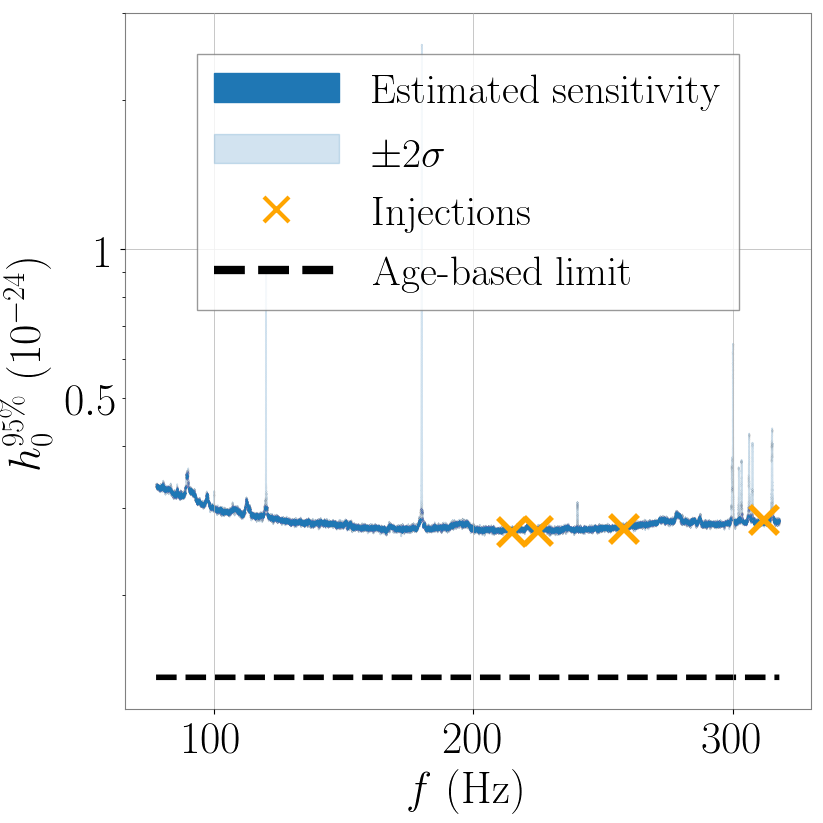}}
	}
	\subfigure[][G347.3--0.5]
	{
		\label{fig:ul_viterbi_347305}
		\scalebox{0.25}{\includegraphics{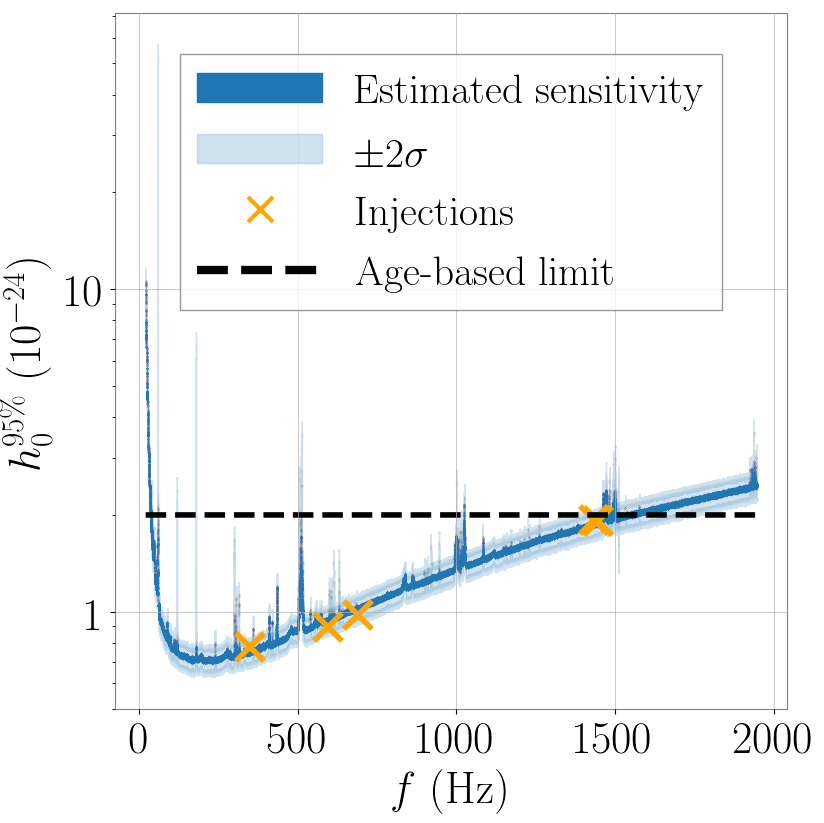}}
	}
 	\subfigure[][G1.9+0.3, G15.9+0.2, G111.7--2.1, \textbf{G266.2--1.2}, G291.0--0.1, G350.1--0.3, G354.4+0.0]
	{
		\label{fig:ul_viterbi_266212}
		\scalebox{0.25}{\includegraphics{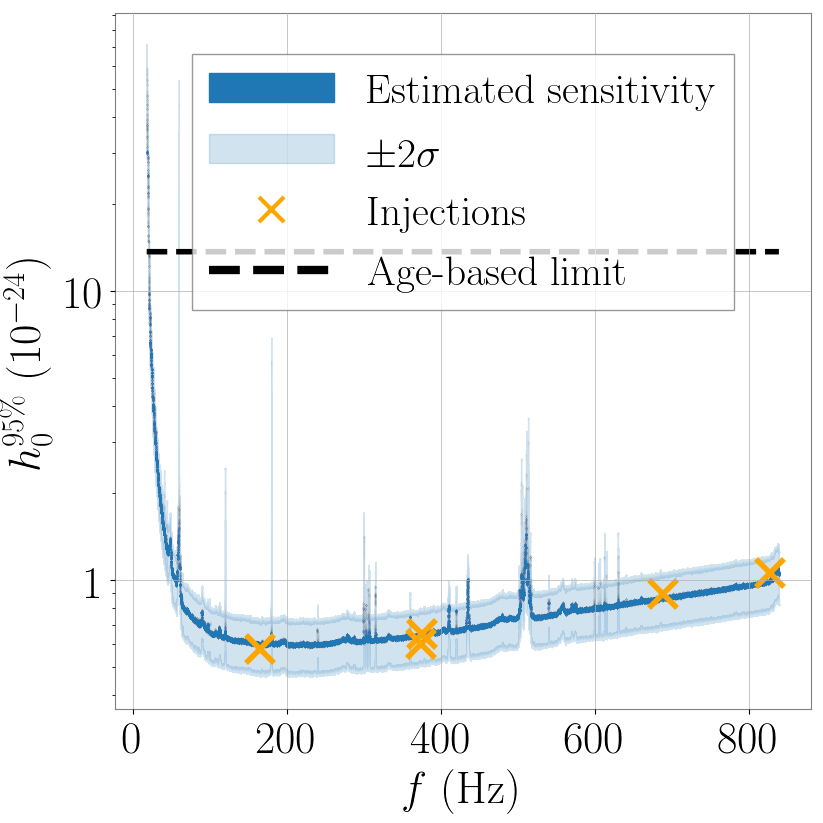}}
	}
	\caption{The sensitivity estimate $h_0^{95\%}$ obtained from the single-harmonic Viterbi search for each source. Multiple sources have $T_{\rm coh} = 1$~hr and have the same sensitivity; these sources are shown on one plot for a representative source, G266.2--1.2. The blue curves represent the estimated $h_0^{95\%}$ in the full band searched by the single-harmonic Viterbi pipeline (Table~\ref{tab:vit_targets}). The orange crosses represent the $h_0^{95\%}$ values obtained empirically in the sample sub-bands. The black dashed line is the age-based upper limit on the gravitational-wave strain from Eq.~(\ref{eqn:maxstrain}).}
	\label{fig:viterbi}
\end{figure*}

We report no evidence of CWs in the single-harmonic Viterbi search.
In this section, we estimate the sensitivity of this search across nine of the fifteen sources.
We estimate the sensitivity first using Eq.~(\ref{eqn:sensitivity}) and assume this is a reasonable representation of the key parameters determining the sensitivity, i.e. that between sources, the sensitivity of the search is predominantly determined by $T_{\rm coh}$.
So we determine the sensitivity for $T_{\rm coh} = 1$~hr using G266.2--1.2 and G347.3--0.5 and assume the variation in sky position for other targets with the same $T_{\rm coh}$ has a negligible effect on sensitivity.
This assumption has been validated through detailed simulations.
For each source we set limits on, we inject 100 simulated signals with fixed $h_0$, and randomly selected $f$ and $\dot{f}_0$ into five frequency sub-bands, selected at random from a set of bands with no known lines, and which returned $< 2$ unique paths with $\mathcal{L} > \mathcal{L}_{\rm th}$ in the original search.
We then apply the Viterbi algorithm to each injection.
We repeat this for 5--10 values of $h_0$. 
Each set of $N_I = 100$ injections forms a binomial distribution, with each injection and search acting as a Bernoulli trial with a probability of success (efficiency) $p$.
We infer the value of $p$ given $s$ successes for each $h_0$ given using the Wilson interval \citep{wilson1927probable}
\begin{equation}
p \approx \frac{s + \frac{1}{2}(1-\alpha_F/2)^2}{N_I + (1-\alpha_F/2)^2} \pm \frac{1-\alpha_F/2}{N_I+(1-\alpha_F/2)^2}\sqrt{\frac{s(N_I-s)}{N_I} + \frac{(1-\alpha_F/2)^2}{4}},
\end{equation}
where $\alpha_F$ is the false alarm probability. 
For each frequency band, we fit a sigmoid curve (as in \citealt{banagiri2019search}) to the set of $h_0$ and the corresponding $p$ using the Bayesian inference package Bilby \citep{Ashton2019} with a uniform prior over the sigmoid parameters.
We sample the posterior and, for each sample, determine the $h_0^{\rm 95\%}$ as the $h_0$ corresponding to $p = 95\%$.
We take the average $h_0^{95\%}$ of this population to be the 95\% frequentist confidence upper limit in that frequency band.
For each frequency band, we calculate $a = h_0^{95\%}/h_0^{\rm est}$ at the appropriate frequency, where $h_0^{\rm est}$ is estimated by Eq.~(\ref{eqn:sensitivity}).
Lastly, we find the mean $a$ across the five frequency bands and calculate the sensitivity across the full frequency band as $h_0^{95\%} = a h_0^{\rm est}$, plotted as the curves in Figure~\ref{fig:viterbi}.
We overplot the age-based limit from Eq.~(\ref{eqn:maxstrain}) (dashed line) for each target.
Our search is more sensitive than the age-based limit for all targets except G18.9-1.1, G39.2-0.3, G330.2+1.0, and G353.6-0.7, despite G353.6-0.7 having the smallest detectable strain in this search, $2.64 \times 10^{-25}$ at 172~Hz.
The targets with the poorest overall sensitivity (those with short $T_{\rm coh}$) place the tightest constraints relative to the age-based spin-down limit.

The constraints obtained in this search are for a random-walk signal model including spin down and spin wandering. 
The random-walk signal model (including spin down and spin wandering) and the range of $\dot{f}_0$ searched (up to $\dot{f}_0^{\rm max} = 3.9\times 10^{-8}$~Hz/s for $T_{\rm coh} = 1$~hr) mean the $h_0^{95\%}$ for this search is less stringent than for the other pipelines in this and other papers, which use a different signal model (e.g. Taylor expansion) and smaller range of $\dot{f}$.
For G65.7+1.2, one of the injections at just over 1000 Hz appears to be on a noise spike despite known noise features being filtered out, however, the scale factor obtained for that band is consistent with the other four bands tested.

\subsection{Dual-harmonic Viterbi constraints}
\begin{figure*}[!tbh]
	\centering
	\subfigure[][G65.7+1.2, G189.1+3.0, G353.6--0.7]
	{
		\label{fig:ul-12hr}
		\scalebox{0.38}{\includegraphics{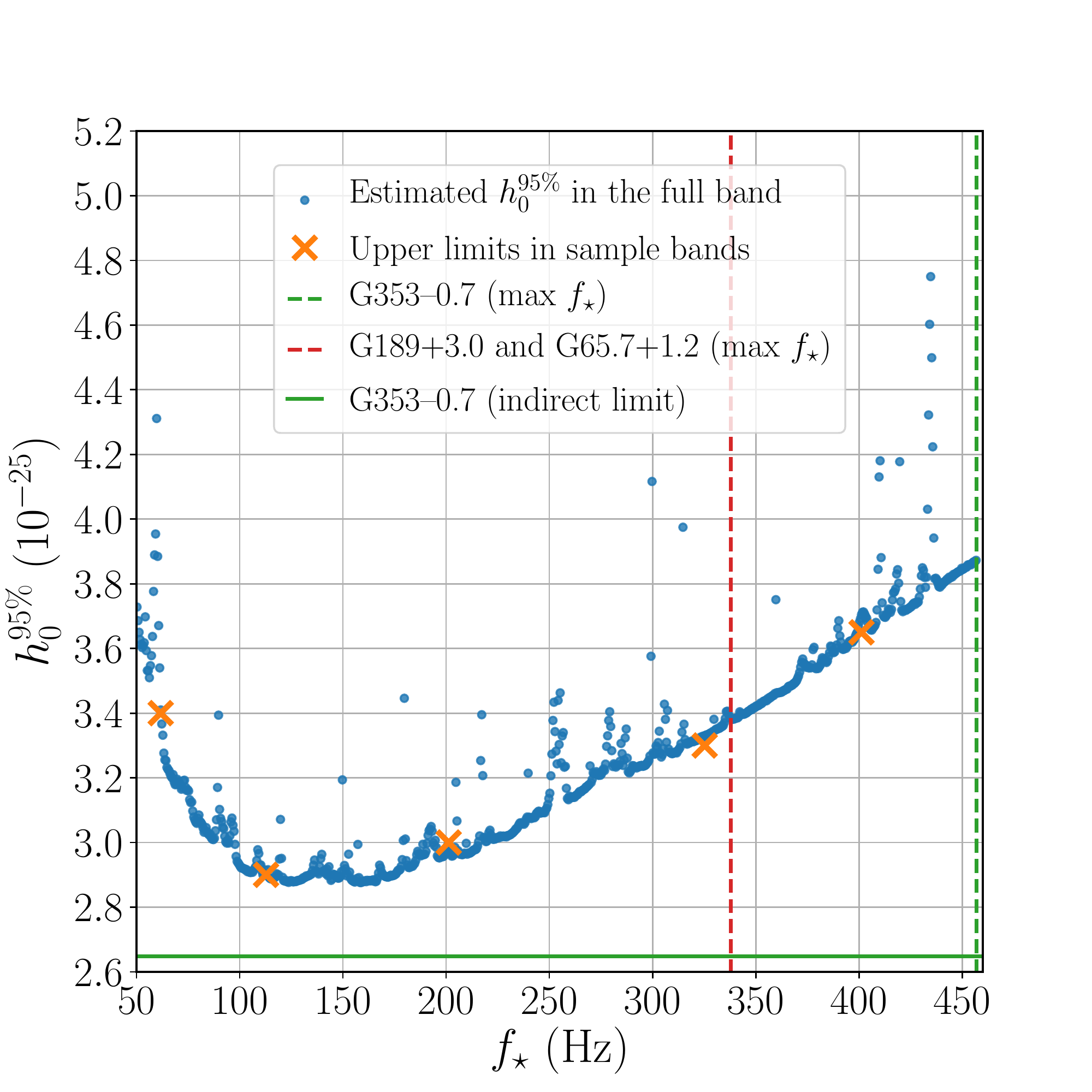}}
	}
	\subfigure[][G18.9--1.1, G39.2--0.3, G93.3+6.9, G266.2--1.2]
	{
		\label{fig:ul-9hr}
		\scalebox{0.38}{\includegraphics{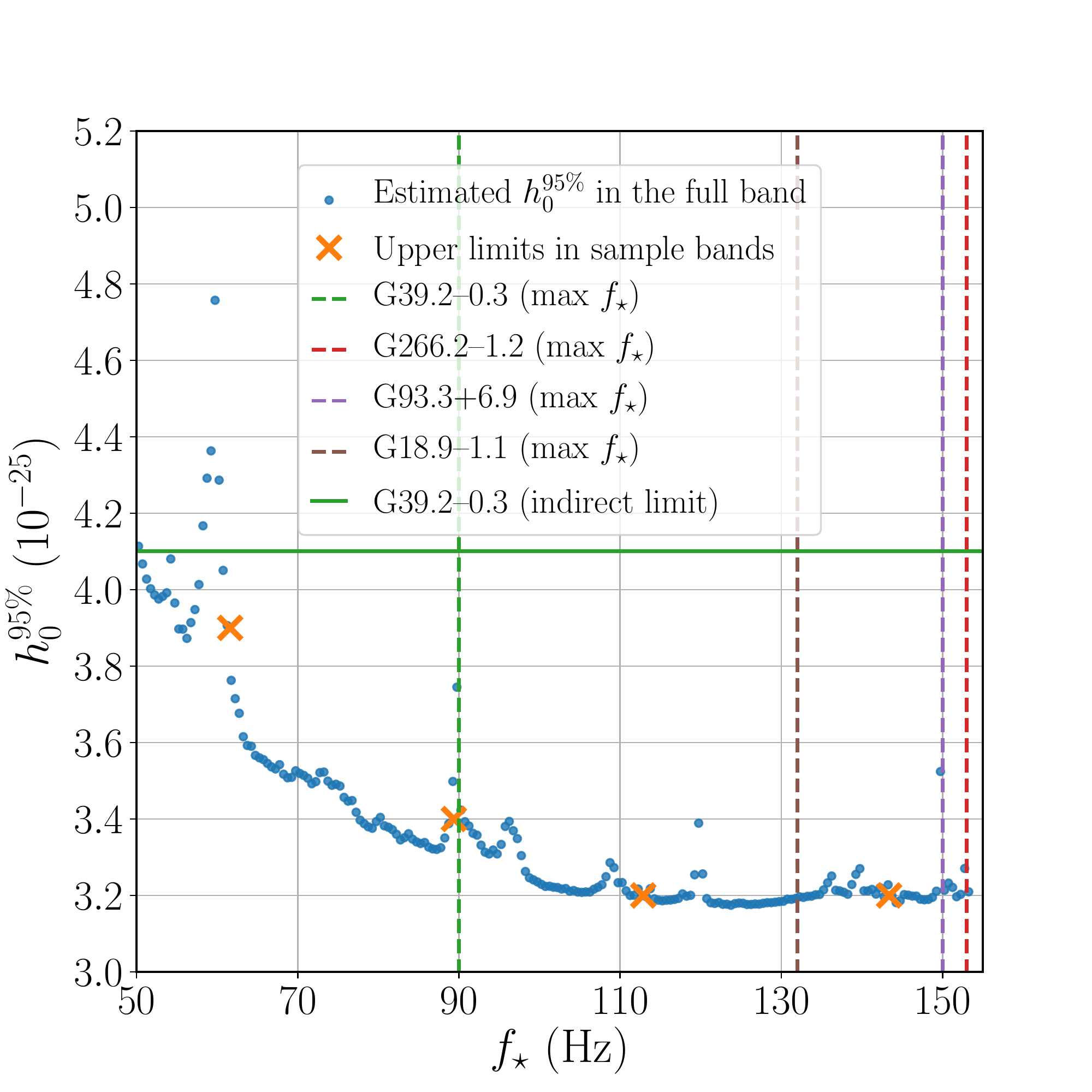}}
	}
	\caption{The estimated sensitivity $h_0^{95\%}$ obtained from the dual-harmonic Viterbi search as a function of $\fstar$ for (a) $\Tcoh=12$~hr and (b) $\Tcoh=9$~hr, assuming a specific scenario with source properties $\theta=45$~deg and $\cos\iota = 0$ (signals at both $\fstar$ and $2\fstar$ are linearly polarized). The blue dots represent the estimated $h_0^{95\%}$ in the full band searched by the dual-harmonic Viterbi pipeline (Table~\ref{tab:para}). The orange cross markers represent the $h_0^{95\%}$ strain upper limits obtained from injections in the randomly selected sample sub-bands. 
	These estimates are obtained from randomized sky positions and hence apply to all sources using the same $\Tcoh$. 
	The vertical dashed lines indicate the maximum $f_\star$ covered for each SNR.
	The horizontal lines indicate the age-based indirect strain limits $h_0^{\rm age}$ derived for the dual-harmonic signal model when $\theta=45$~deg. For G353--0.7, the $h_0^{95\%}$ obtained from the search has not beaten the indirect limit. For G39.2--0.3, the $h_0^{95\%}$ has beaten the indirect limit at most of the frequencies except for the noisy bands around 60~Hz. For all other sources, the $h_0^{\rm age}$ values are much larger than $h_0^{95\%}$ across the full band and thus are not shown in the figure. 
    }
	\label{fig:viterbi1f2fUL}
\end{figure*}

No evidence of CWs is found in the dual-harmonic Viterbi search. 
We empirically derive the sensitivity by estimating the signal strain $h_0^{95\%}$ in each frequency sub-band (as a function of $\fstar$), such that a signal with $h_0 \geq h_0^{95\%}$ can be detected on 95\% or more occasions.
Since this pipeline considers a signal model with both $\fstar$ and $2\fstar$ components, we use $\fstar$ instead of the GW frequency to avoid confusion.
Note that the sensitivity on the strain $h_0$ quoted in this pipeline is based on a different signal model from the other two pipelines [see Eqs.~(\ref{eqn:h2p})--(\ref{eqn:h1c})].
Here we assume the special scenario $\theta=45$~deg and $\cos\iota = 0$, i.e., signals at both $\fstar$ and $2\fstar$ are linearly polarized. 
In this scenario, tracking the two frequency bands simultaneously offers the most significant sensitivity improvement from searching a single band, compared to other choices of $\theta$ and $\cos\iota$~\citep{Sun2019}. 

Figure~\ref{fig:viterbi1f2fUL} shows $h_0^{95\%}$ in all sub-bands and a set of frequentist upper limits obtained through injections in a handful of randomly selected sample sub-bands (orange cross markers). 
The procedure to produce these results is as follows.
First, we derive the frequentist upper limit in one sample sub-band, starting from 112.5~Hz for $\fstar$ and 225~Hz for $2\fstar$. A set of 200 synthetic signals are injected into the O3a data at random sky positions in that sub-band with a fixed $h_0$. We use fixed $\theta=45$~deg and $\cos\iota = 0$. The other source parameters, including $f_\star$, $\dot{f}_\star$, the polarization angle, and the initial phase, are randomly drawn from their uniform distributions. The corresponding detection rate is calculated. This process is repeated with different $h_0$ values with step size $1 \times 10^{-26}$ and $2 \times 10^{-26}$ in the regions where the detection rate is roughly above and below 50\%, respectively. With all the injected $h_0$ values and the corresponding detection rates, $h_0^{95\%}$ is obtained through a sigmoidal fit. The $h_0^{95\%}$ value found in the sample sub-band is $2.9 \times 10^{-25}$ and $3.2 \times 10^{-25}$ for $\Tcoh=12$~hr and 9~hr, respectively.
Next, we use these values obtained in the sample sub-band (starting from 112.5~Hz for $\fstar$ and 225~Hz for $2\fstar$) to analytically calculate $h_0^{95\%}$ in the full frequency band (blue dots), using the scaling~\citep{Sun2019}
\begin{equation}
h_0^{95\%}(f) \propto \left(\frac{S_n(f)S_n(2f)}{S_n(f)+S_n(2f)}\right)^{1/2},
\label{eq:ULscaling}
\end{equation}
where $S_n$ is the effective power spectral density calculated from the harmonic mean of the two detectors over all the 30-min SFTs collected from September 1 to October 1, 2019 (GPS time 1251331218--1253923218).
Finally, in order to verify the analytical scaling, the simulation procedure in the first step is repeated in several other randomly selected sub-bands, indicated by the orange cross markers. The $h_0^{95\%}$ values obtained empirically in those sample sub-bands agree to $<1.5\%$ with the analytic sensitivity estimates. 
In the full frequency band searched, the best $h_0^{95\%}$ values for $\Tcoh=12$~hr and 9~hr are $2.88 \times 10^{-25}$ at $\fstar = 158.75$~Hz and $3.17 \times 10^{-25}$ at $\fstar = 123.75$~Hz, respectively.
These results are obtained from randomized sky positions and hence apply to all sources using the same $\Tcoh$. 
In the dual-harmonic Viterbi pipeline, the sensitivity is dominated by the length of $\Tcoh$ (for a fixed $\Tobs$) rather than the sky position of the source. 
Additional spot checks validate that the difference between the empirical $h_0^{95\%}$ values obtained from a fixed sky location and those from randomized sky positions is negligible.

Assuming that the star's rotational kinetic energy loss is all radiated in GWs, the age-based indirect strain limits $h_0^{\rm age}$ can be calculated for each source by fixing $\theta=45$~deg (consistent with the scenario presented in Figure~\ref{fig:viterbi1f2fUL}), setting $t_{\rm age}$ to the value in Table~\ref{tab:para}, and setting the distance to the minimum value in Table~\ref{tab:targets}. Note that the $h_0^{\rm age}$ value derived explicitly for the dual-harmonic model with $\theta=45$~deg is a factor of $\sim 2$ larger than the value calculated from Eq.~(\ref{eqn:maxstrain})~\citep{Zimmermann1979,wette2008searching}.
For five out of the seven sources, the indirect limits are much larger than the constraints obtained in this search across the full band. For G39.2--0.3, the $h_0^{95\%}$ has beaten the indirect limit at most of the frequencies except for the noisy bands around 60~Hz. For G353--0.7, the $h_0^{95\%}$ obtained from the search is close to $h_0^{\rm age}$ but has not reached it at any frequency. 
We emphasize that the sensitivity in the dual-harmonic Viterbi pipeline, and whether it beats the indirect limit, is not directly comparable to other methods due to the model difference.
\subsection{Astrophysical implications}
The sensitivity in terms of the GW strain amplitude can be converted into constraints on the fiducial ellipticity of the neutron star, $\epsilon$ \citep{Jaranowski1998}, and the $r$-mode amplitude parameter, $\alpha$ \citep{Owen2010}.
We first discuss the constraints obtained in the BSD and single-harmonic Viterbi pipelines.
For ellipticity, we assume the GW frequency $f$ (equivalent to $f_0$ in the single-harmonic Viterbi pipeline) is at $2\fstar$, which aligns with the model of a perpendicular biaxial rotor considered in both pipelines.
Given $h_0^{95\%}$ (derived with a uniform prior on the $\cos\iota$), we constrain the ellipticity of the neutron star in terms of the GW frequency $f=2f_\star$ via \citep{Jaranowski1998}
\begin{equation} \label{eqn:eps} \epsilon = 9.46\times 10^{-5} \left(\frac{h_0}{10^{-24}}\right) \left(\frac{D}{1{\rm~kpc}}\right) \left(\frac{100{\rm~Hz}}{f}\right)^2,
\end{equation}
assuming the moment of inertia with respect to the rotation axis ($I_{zz}$ for a perpendicular biaxial rotor) is $10^{38}$~kg\,m$^2$.
We can also convert $h_0^{95\%}$ to limits on the amplitude of $r$-mode oscillations via \citep{Owen2010}
\begin{equation}
\label{eqn:alpha}
\alpha \simeq 0.028\left(\frac{h_{0}}{10^{-24}}\right)\left(\frac{D}{1 \mathrm{~kpc}}\right)\left(\frac{100 \mathrm{~Hz}}{f}\right)^{3}.
\end{equation}

\begin{figure*}[!tbh]
	\centering
		\subfigure[]
	{
		\label{fig:eps_BSD_600}
		\scalebox{0.38}{\includegraphics{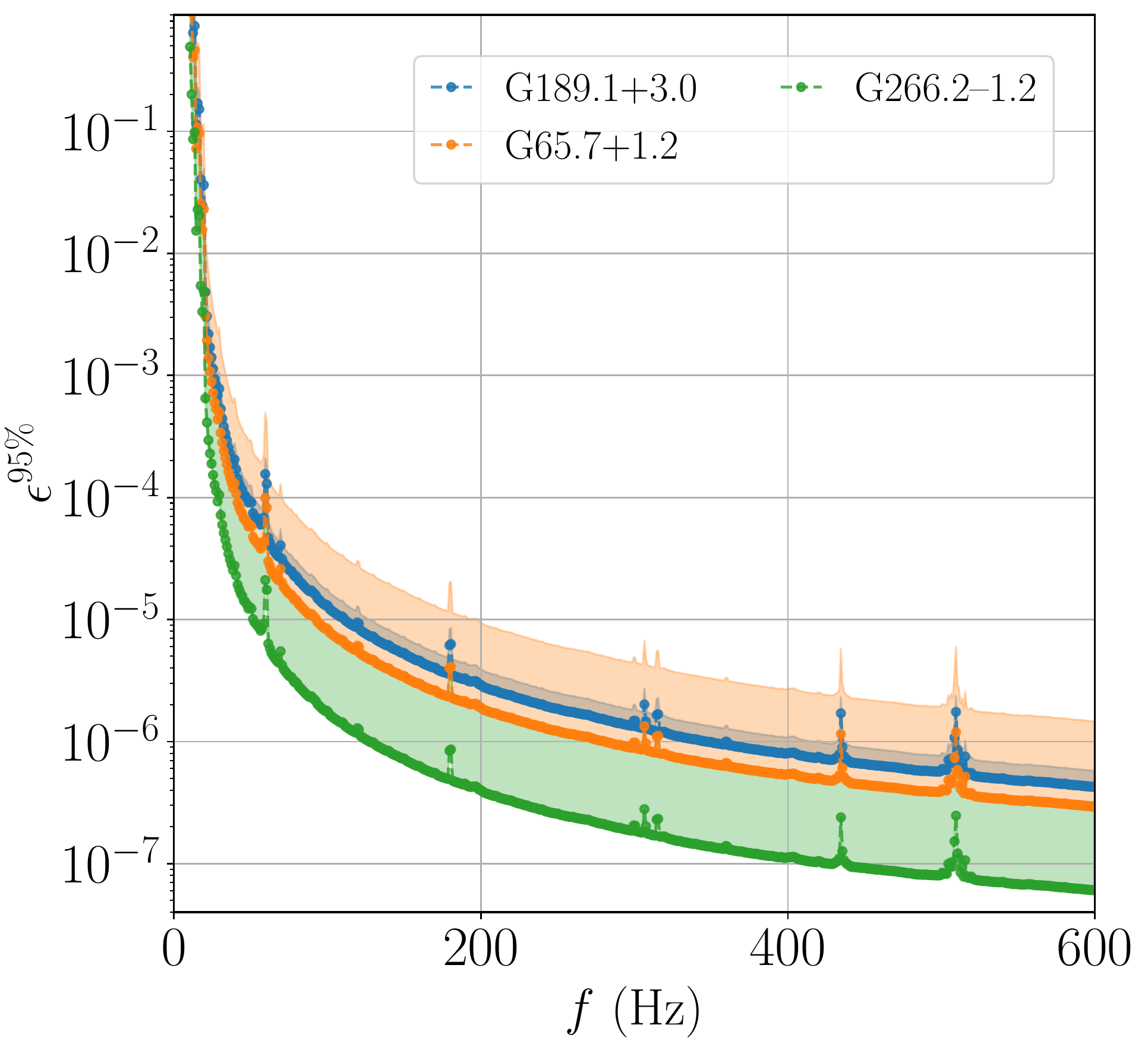}}
	}
		\subfigure[]
	{
		\label{fig:eps_BSD_1000}
		\scalebox{0.38}{\includegraphics{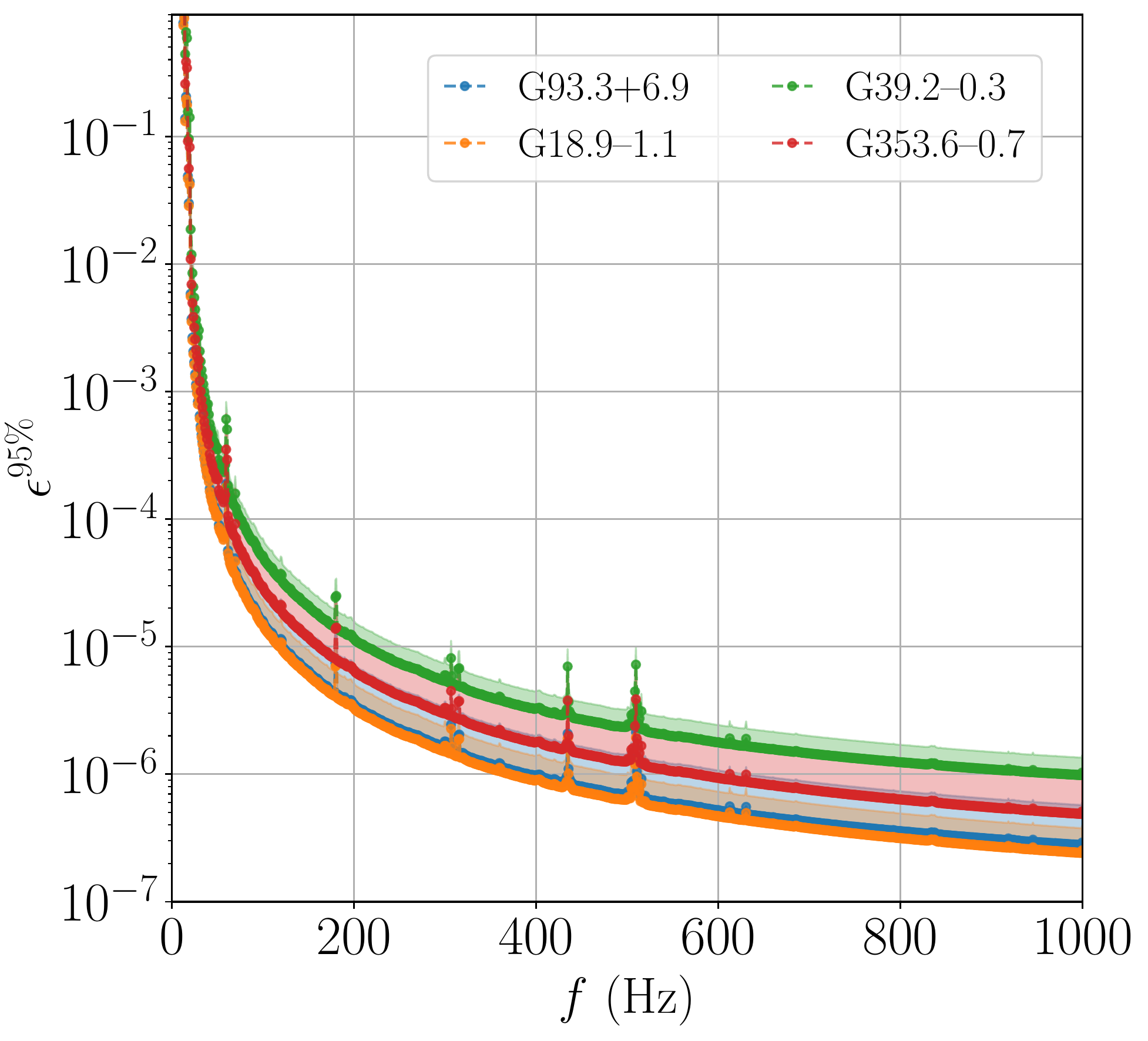}}
	}
    	\subfigure[]
	{
		\label{fig:alpha_BSD_600}
		\scalebox{0.38}{\includegraphics{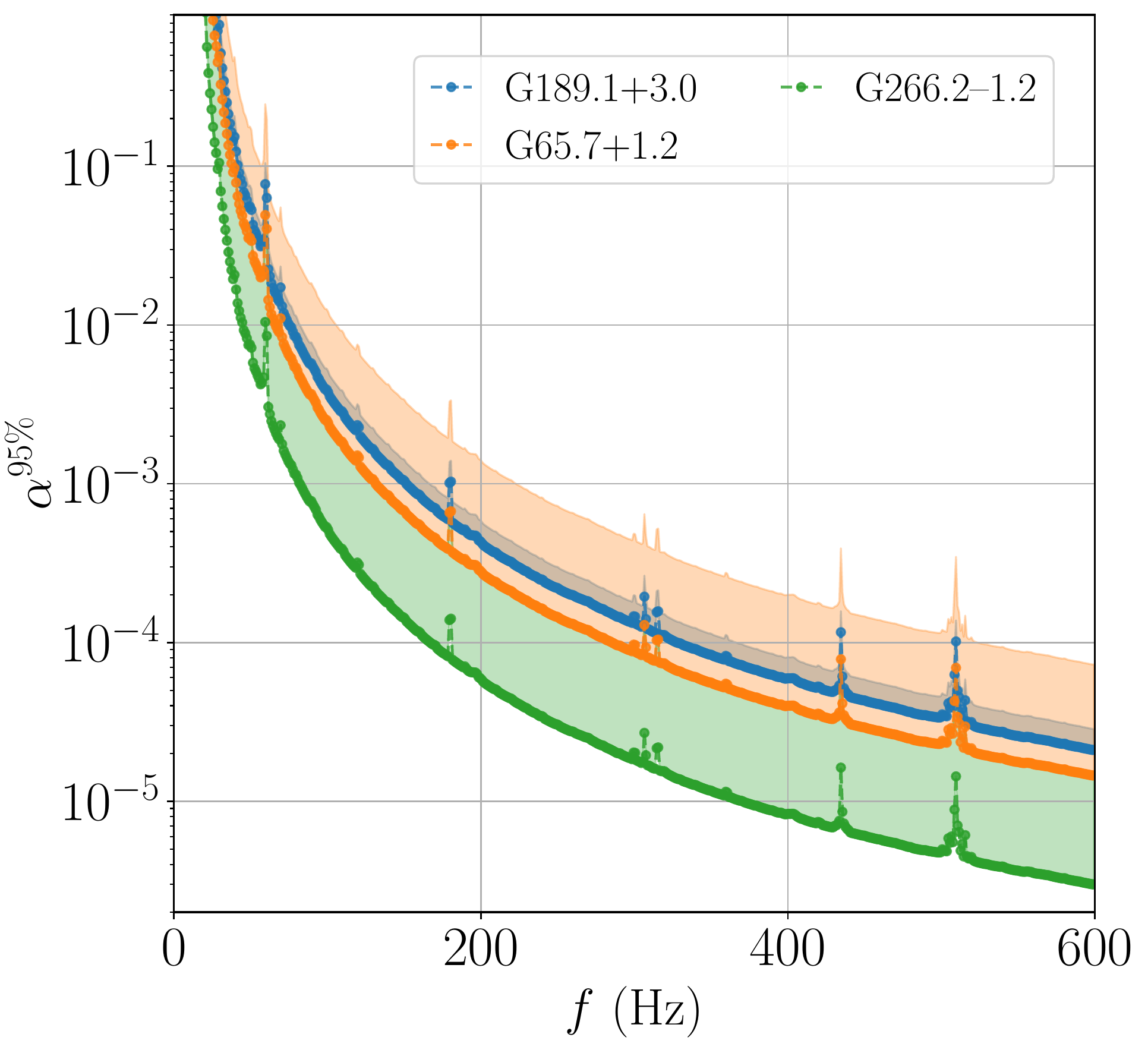}}
	}
		\subfigure[]
	{
		\label{fig:alpha_BSD_1000}
		\scalebox{0.38}{\includegraphics{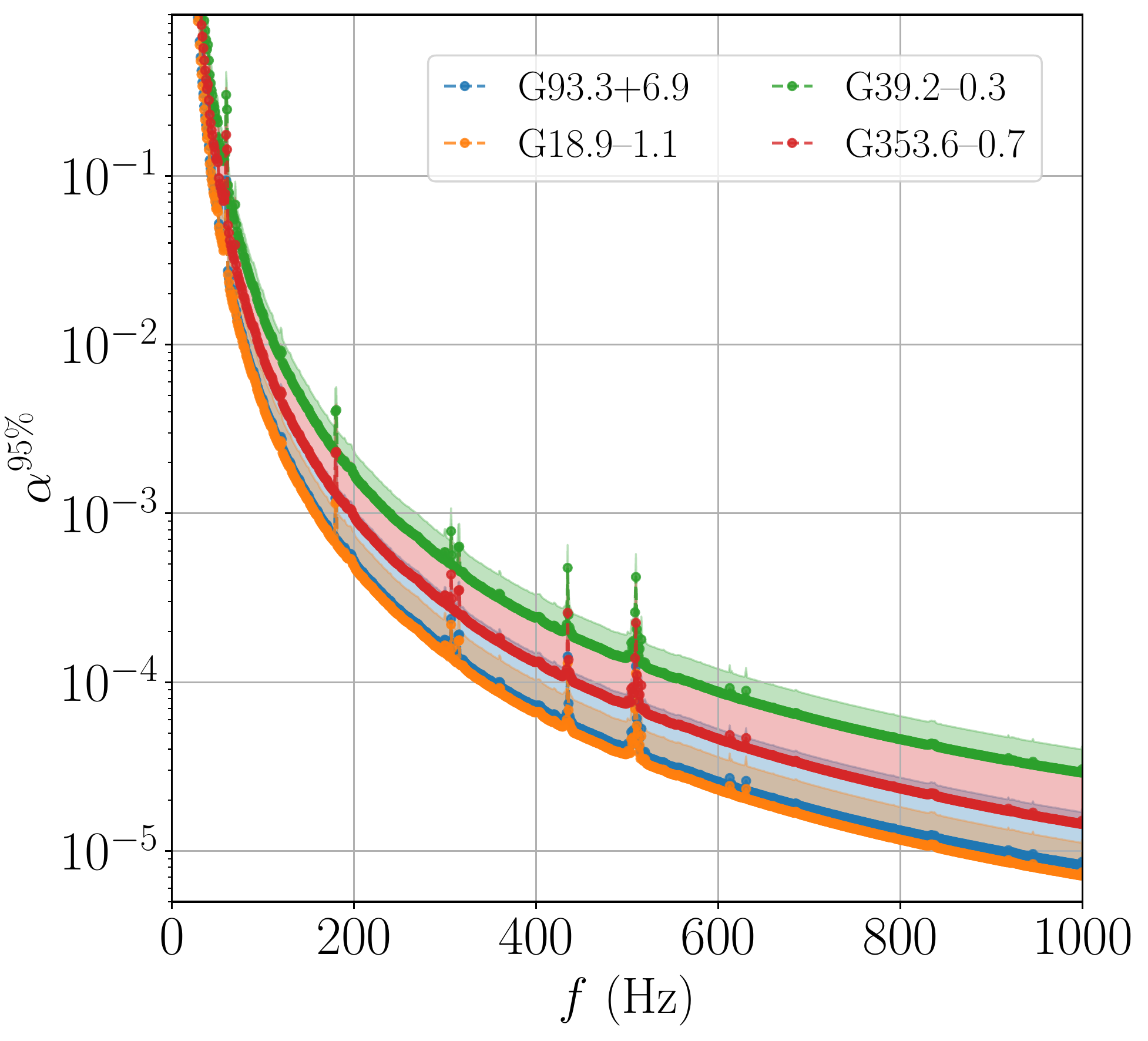}}
	}
	\caption{Constraints on the (a)--(b) neutron-star ellipticity $\epsilon^{95\%}$,  and (c)--(d) $r$-mode amplitude $\alpha^{95\%}$, from the BSD pipeline, converted from the $h_0^{95\%}$ values in Figure~\ref{fig:BSDUL}. Panels (a) and (c) report the results derived for G189.1+3.0, G65.7+1.2 and G266.2--1.2 (Vela Jr.), covering the [10, 600] Hz frequency band. Panels (b) and (d) report the results for G93.3+6.9, G18.9--1.1, G39.2--0.3 and G353.6--0.7, where the [10, 1000] Hz frequency band is investigated. Curves have been converted from $h_0^{95\%}$ derived for the L detector. Shaded regions correspond to the inferred ellipticity and $r$-mode amplitude using the full range of distances in Table~\ref{tab:targets}. The minimum distance is assumed for the solid dot curves.}
	\label{fig:BSDepsalpha}
\end{figure*}

\begin{figure*}
	\centering
	\subfigure[][Ellipticity constraints for targets with $f_{\rm max} < 1500 \, {\rm Hz}$]
	{
		\label{fig:v_ellipticities_low.png}
		\scalebox{0.40}{\includegraphics{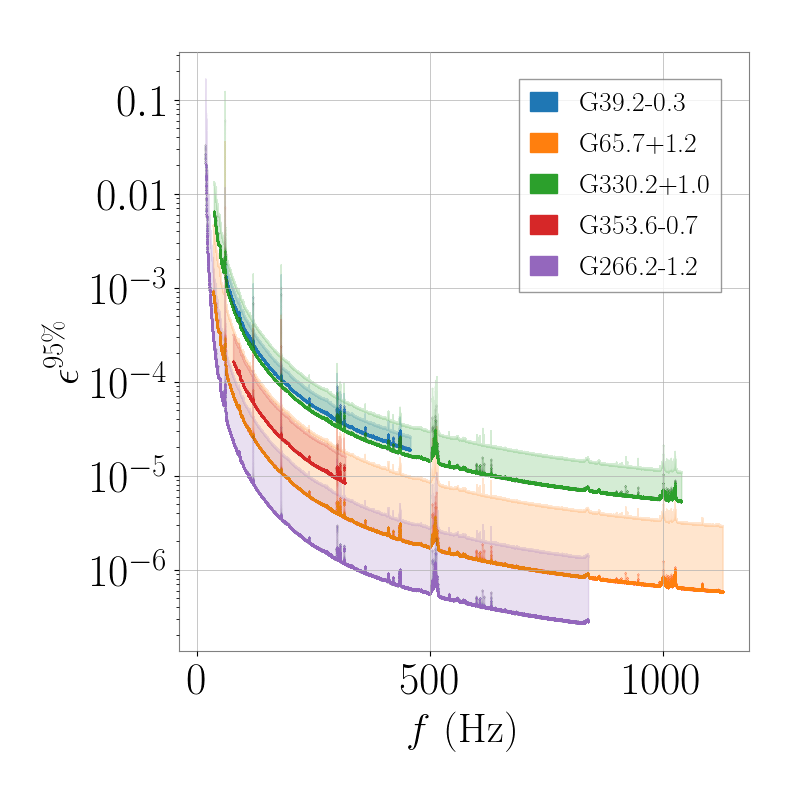}}
	}
	\subfigure[][Ellipticity constraints for targets with $f_{\rm max} > 1500 \, {\rm Hz}$]
	{
		\label{fig:v_ellipticities_high.png}
		\scalebox{0.40}{\includegraphics{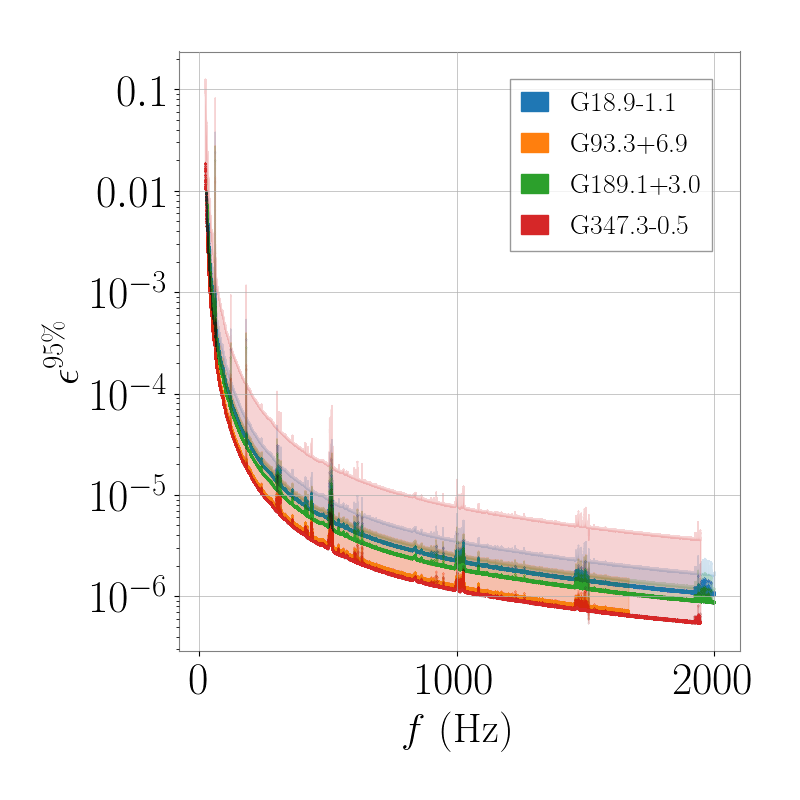}}
	}
	\subfigure[][r-mode constraints for targets with $f_{\rm max} < 1500 \, {\rm Hz}$]
	{
		\label{fig:v_rmode_low.png}
		\scalebox{0.40}{\includegraphics{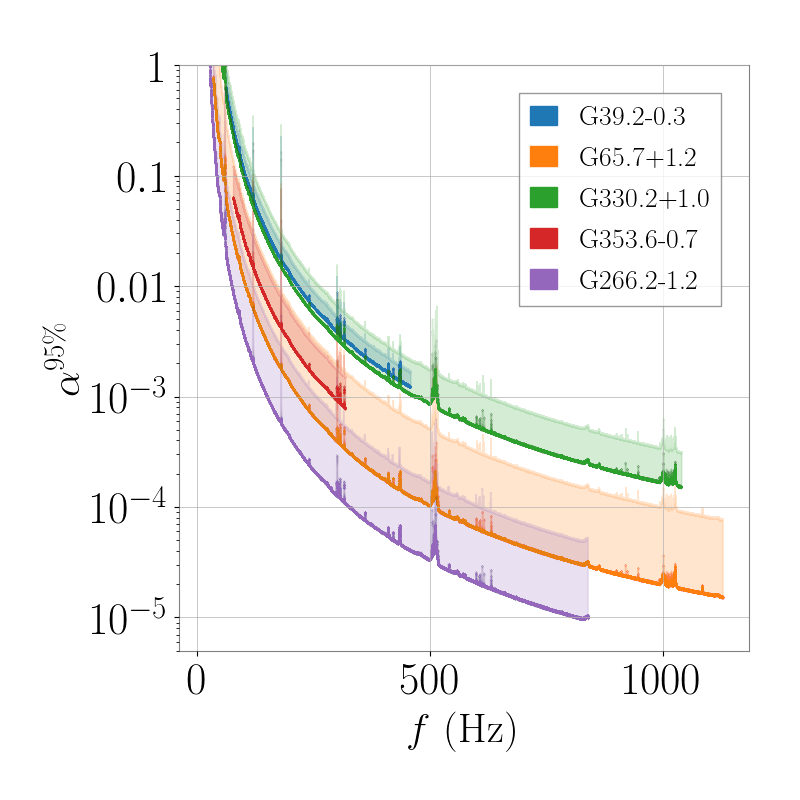}}
	}
	\subfigure[][r-mode constraints for targets with $f_{\rm max} > 1500 \, {\rm Hz}$]
	{
		\label{fig:v_rmode_high.png}
		\scalebox{0.40}{\includegraphics{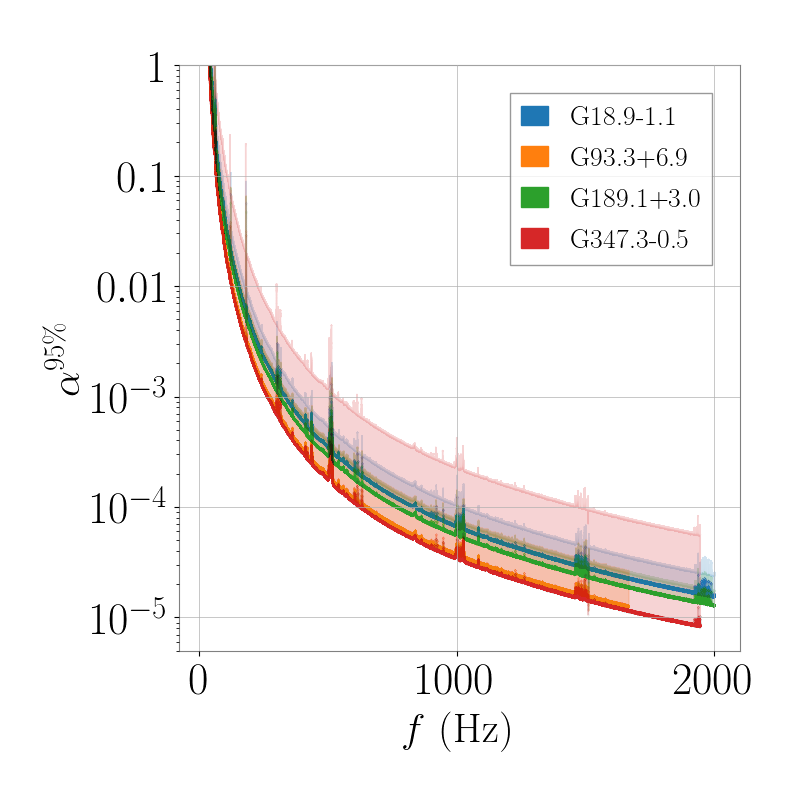}}
	}

	\caption{Constraints on the neutron star ellipticity ((a) and (b)) and $r$-mode amplitude ((c) and (d)) from the single-harmonic Viterbi pipeline converted from the $h_0^{95\%}$ values in Figure~\ref{fig:viterbi}. The results are plotted as a function of the GW frequency $f$. The solid line uses the closest distance estimate in Table~\ref{tab:targets} and the shaded area indicates the results across the full distance range. Panels (a) and (c) display the results for targets with $f_{\rm max} < 1500 \, {\rm Hz}$; panels (b) and (d) display results for targets with $f_{\rm max} > 1500 \, {\rm Hz}$.}
	\label{fig:viterbi_astro}
\end{figure*}

Figures~\ref{fig:BSDepsalpha} and \ref{fig:viterbi_astro} present the constraints on $\epsilon$ and $\alpha$ obtained from the BSD and single-harmonic Viterbi pipelines, respectively.  
The most stringent constraints, $\epsilon \lesssim 10^{-7}$ and $\alpha \lesssim 10^{-5}$, come from the BSD pipeline (see Figure~\ref{fig:BSDepsalpha}), where the values are converted using the $h_0^{95\%}$ curve from the L detector.
The results from the single-harmonic Viterbi pipeline, covering more targets and a wider parameter space, are presented in Figure~\ref{fig:viterbi_astro}.

\begin{figure*}[!tbh]
	\centering
	\subfigure[]
	{
		\label{fig:eps-12hr}
		\scalebox{0.34}{\includegraphics{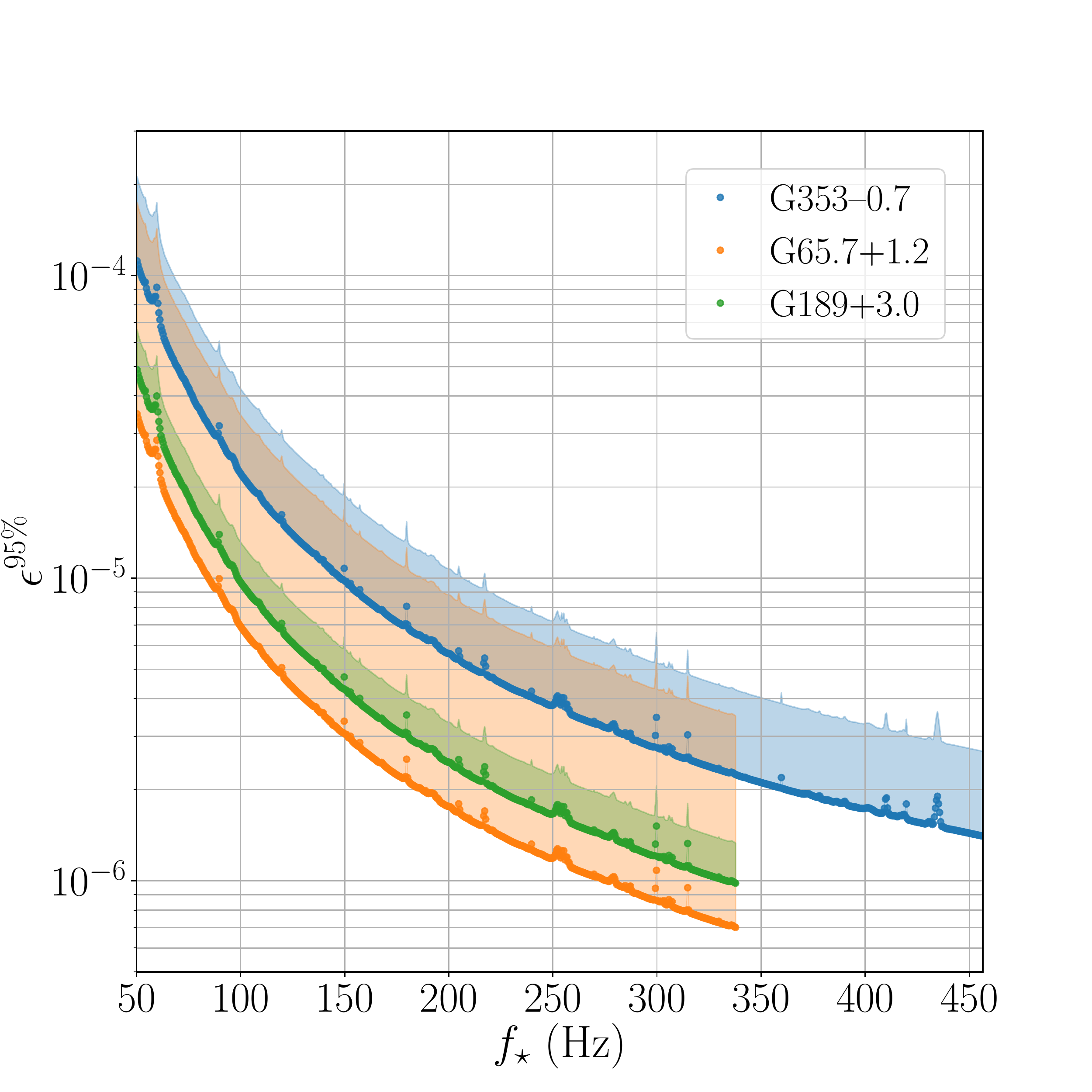}}
	}
	\subfigure[]
	{
		\label{fig:eps-9hr}
		\scalebox{0.34}{\includegraphics{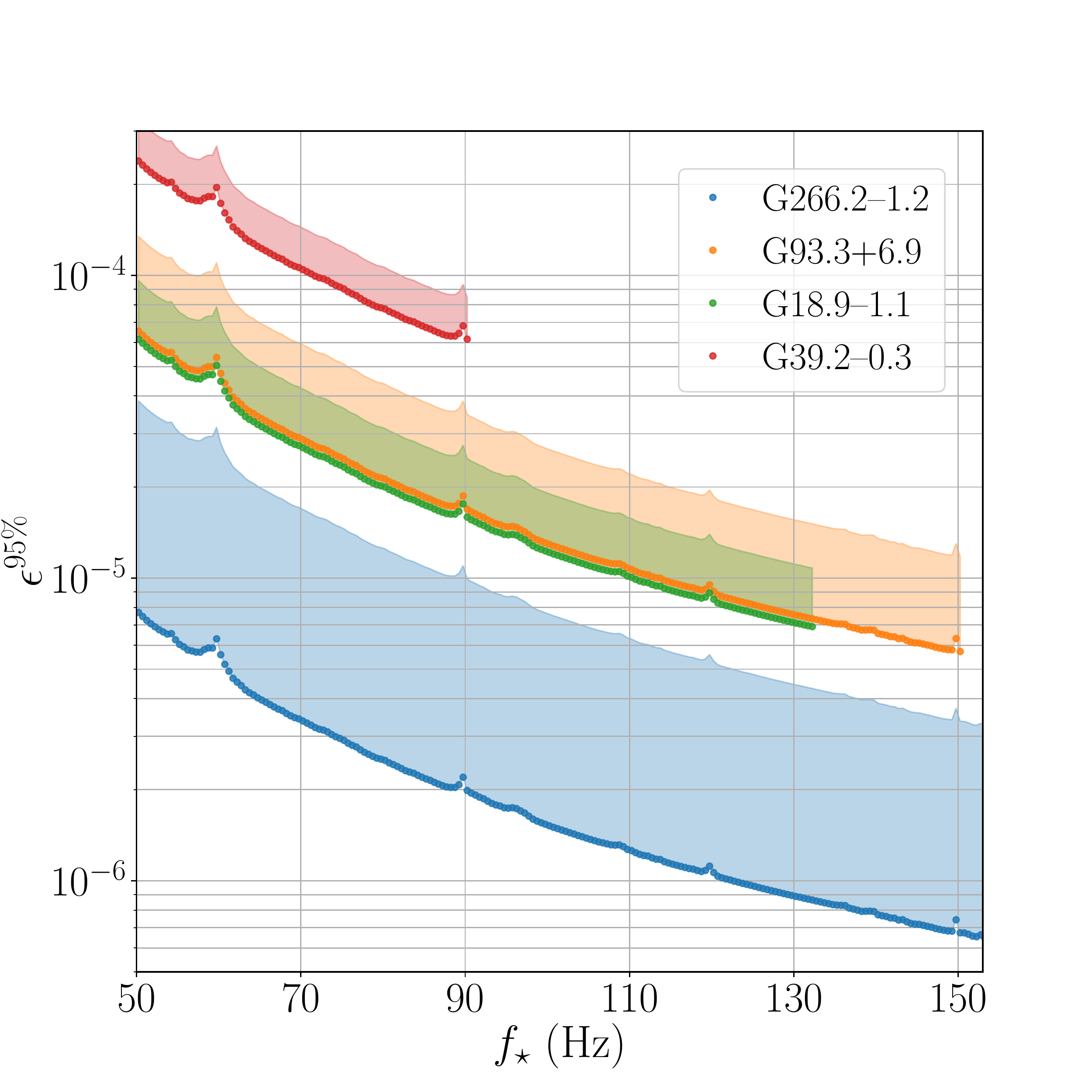}}
	}
	\caption{Constraints on the neutron star ellipticity with 95\% confidence from the dual-harmonic Viterbi search as a function of $\fstar$ for (a) $\Tcoh=12$~hr and (b) $\Tcoh=9$~hr, converted from the $h_0^{95\%}$ values in Figure~\ref{fig:viterbi1f2fUL}. The shaded region indicates the resulting ellipticity range calculated from the full estimated distance range for each SNR (see Table~\ref{tab:targets}), and the solid dots correspond to the minimum distance estimate.}
	\label{fig:viterbi1f2feps}
\end{figure*}

We also convert the $h_0^{95\%}$ obtained in the dual-harmonic Viterbi pipeline to the 95\% confidence constraint on the ellipticity of the star.
Since GW emission is at both $\fstar$ and $2\fstar$, Eq.~(\ref{eqn:eps}) can be written in terms of the spin frequency of the star $\fstar$,
\begin{equation} \label{eqn:eps_1f2f} \epsilon = 2.36\times 10^{-5} \left(\frac{h_0}{10^{-24}}\right) \left(\frac{D}{1{\rm~kpc}}\right) \left(\frac{100{\rm~Hz}}{\fstar}\right)^2.
\end{equation}
Figure~\ref{fig:viterbi1f2feps} shows the limits on $\epsilon$ as a function of $\fstar$.
Note that the results here are converted from the $h_0^{95\%}$ values obtained for a specific scenario with source properties $\theta=45$~deg and $\cos\iota = 0$, and hence the $\epsilon$ values in Figure~\ref{fig:viterbi1f2feps} are not directly comparable to the results obtained in other conventional searches where $\theta=90$~deg and the emission is only at $2\fstar$.
The signal model adopted in the dual-harmonic search cannot be interpreted as current quadrupole emission from an $r$-mode, so we do not infer $r$-mode amplitudes from $h_0^{95\%}$.

The strictest constraints on the intrinsic GW strain from the BSD pipeline are $h_0^{95\%}\approx 7.7 \times 10^{-26}$ for G39.2--0.3 and $h_0^{95\%}\approx 7.8 \times 10^{-26}$ for G65.7+1.2 near 200 Hz.
The results obtained by the Viterbi pipelines set the first constraints on CWs which allow for spin wandering in the signal model. Note that the authors of \citet{millhouse2020search} conducted a search for 13 out of the 15 sources in Advanced LIGO O2 data using the single-harmonic Viterbi method but did not derive the constraints from the search sensitivity.
Furthermore, the dual-harmonic Viterbi analysis provides the first results for these SNR sources derived considering two frequency harmonics simultaneously.
The best constraints on the star's ellipticity obtained at frequencies $f \gtrsim 100$~Hz reach $\epsilon < 10^{-6}$ for most of the sources, reaching below the rough theoretical upper limit for normal neutron stars~\citep{Johnson-McDaniel2013eps}, and reach as low as $\epsilon \approx 6 \times 10^{-8}$ for the closest source G266.2--1.2/Vela Jr., well below the theoretical limits. However, these limits are model dependent; the uncertainties on the star's geometry and composition, like the internal equation of state, the moment of inertia, and the magnetic field, play a significant role when deriving these limits. For example, \cite{Woan_2018} shows that an ellipticity of $\epsilon \sim 10^{-9}$ can be sustained by neutron stars with a buried magnetic field of $\sim 10^{11}$~G. 
The most stringent constraints on the $r$-mode amplitude obtained above $\sim 100$~Hz arrive at the theoretical prediction level of $\alpha \sim 10^{-3}$, expected
for the nonlinear saturation mechanisms~\citep{Bondarescu2009rmode}, and reach as low as $\alpha \sim 10^{-5}$ at higher frequencies.

\section{Conclusion}
\label{sec:conclusion}
In this work, we present the results of the search for CWs from neutron stars in 15 young SNRs by analyzing the data collected in the first half of O3.
No evidence of CWs is identified.
We present constraints on the GW strain, as well as the implied mass ellipticity and $r$-mode amplitude for each source.
The inferred upper limits on the latter quantities reach below the maximum values allowed on theoretical grounds.
The strictest constraints on the intrinsic GW strain from the BSD pipeline are $h_0^{95\%}\approx 7.7 \times 10^{-26}$ for G39.2--0.3 and $h_0^{95\%}\approx 7.8 \times 10^{-26}$ for G65.7+1.2, both near 200 Hz.
The Viterbi pipelines set the first constraints on the signal strain allowing for spin wandering. 
The dual-harmonic Viterbi analysis reports the first results for these SNR sources derived considering two frequency harmonics simultaneously.
All of the three pipelines are computationally efficient, costing $\sim 10^2$~core-hr for each source per pipeline (postprocessing excluded).
The significantly improved search efficiency compared to the existing studies for this type of source mainly comes from two factors: (1) we use a much shorter coherence time $T_{\rm coh}$ in this analysis compared to other studies \citep[e.g.][]{Abbott_2019,Lindblom_2020,Papa:2020vfz} and the computing cost scales roughly as $T_{\rm coh}^{4}$ to $T_{\rm coh}^{7}$ depending on the orders of the time derivatives of frequency searched; (2) advanced signal processing techniques are used in these three pipelines to further accelerate the search, e.g., the dynamic programming in Viterbi pipelines \citep{ViterbiOriginal,sun2018hidden} and the use of sub-sampled data and heterodyne correction in the BSD pipeline \citep{BSD}.

We briefly compare the constraints derived in this work to existing constraints in the literature.
The constraints from the two Viterbi-based pipelines are not directly comparable with other analyses because they assume different signal models.
The dual-harmonic Viterbi pipeline searches simultaneously for emission at $f_\star$ and $2f_\star$.
In addition, both Viterbi pipelines search a set of frequency random walks, unlike other pipelines that search a set of Taylor expansion coefficients.
A previous search for these sources in O1 reported $h_0^{95\%} \approx 2\times 10^{-25}$ for most of the sources and $h_0^{95\%} \approx 1\times 10^{-25}$ for one source in the most sensitive band ~\citep{Abbott_2019}, excluding results on Fomalhaut B \citep{jones2021search}.
The best limit on $h_0$ for G39.2--0.3 is $h_0^{95\%}\approx2\times 10^{-25}$, approximately 2.5 times higher than the results obtained here for the same source with the BSD pipeline.
\citet{Lindblom_2020} presented a similar search to \citet{Abbott_2019} using the O2 data and reported strain limits slightly above $1 \times 10^{-25} $ at 90\% confidence level, e.g., for G39.2--0.3, which is $\sim 1.4$ times higher than the results obtained here for the same source with the BSD pipeline.
Also, using the LIGO data in the full O1 run and a coherent integration duration longer than 10 days, \citet{Papa:2020vfz} set 90\% confidence limits of $h_0^{90\%} = 1.2\times10^{-25}$, $9.3\times10^{-26}$, and $8.8\times10^{-26}$ for Cas A, Vela Jr. and G347.3--0.5 near 172.5~Hz, respectively.
These limits for Cas A and G347.3--0.5 are lower than those obtained in the single-harmonic Viterbi search for these two SNRs, without any adjustments being made for the different signal models assumed. 
We do not search for these two SNRs using the BSD and dual-harmonic Viterbi pipelines. 
The results on Vela Jr. in this analysis from the BSD search slightly improve the previous constraints set by \citet{Papa:2020vfz}, despite the use of a coherent integration duration $\sim 20$ times shorter.
Future data collection and improved analysis methods will further extend the sensitivity of CW searches and increase the probability of a future discovery.


\section{Acknowledgments}
This material is based upon work supported by NSF’s LIGO Laboratory which is a major facility
fully funded by the National Science Foundation.
The authors also gratefully acknowledge the support of
the Science and Technology Facilities Council (STFC) of the
United Kingdom, the Max-Planck-Society (MPS), and the State of
Niedersachsen/Germany for support of the construction of Advanced LIGO 
and construction and operation of the GEO600 detector. 
Additional support for Advanced LIGO was provided by the Australian Research Council.
The authors gratefully acknowledge the Italian Istituto Nazionale di Fisica Nucleare (INFN),  
the French Centre National de la Recherche Scientifique (CNRS) and
the Netherlands Organization for Scientific Research, 
for the construction and operation of the Virgo detector
and the creation and support  of the EGO consortium. 
The authors also gratefully acknowledge research support from these agencies as well as by 
the Council of Scientific and Industrial Research of India, 
the Department of Science and Technology, India,
the Science \& Engineering Research Board (SERB), India,
the Ministry of Human Resource Development, India,
the Spanish Agencia Estatal de Investigaci\'on,
the Vicepresid\`encia i Conselleria d'Innovaci\'o, Recerca i Turisme and the Conselleria d'Educaci\'o i Universitat del Govern de les Illes Balears,
the Conselleria d'Innovaci\'o, Universitats, Ci\`encia i Societat Digital de la Generalitat Valenciana and
the CERCA Programme Generalitat de Catalunya, Spain,
the National Science Centre of Poland and the Foundation for Polish Science (FNP),
the Swiss National Science Foundation (SNSF),
the Russian Foundation for Basic Research, 
the Russian Science Foundation,
the European Commission,
the European Regional Development Funds (ERDF),
the Royal Society, 
the Scottish Funding Council, 
the Scottish Universities Physics Alliance, 
the Hungarian Scientific Research Fund (OTKA),
the French Lyon Institute of Origins (LIO),
the Belgian Fonds de la Recherche Scientifique (FRS-FNRS), 
Actions de Recherche Concertées (ARC) and
Fonds Wetenschappelijk Onderzoek – Vlaanderen (FWO), Belgium,
the Paris \^{I}le-de-France Region, 
the National Research, Development and Innovation Office Hungary (NKFIH), 
the National Research Foundation of Korea,
the Natural Science and Engineering Research Council Canada,
Canadian Foundation for Innovation (CFI),
the Brazilian Ministry of Science, Technology, and Innovations,
the International Center for Theoretical Physics South American Institute for Fundamental Research (ICTP-SAIFR), 
the Research Grants Council of Hong Kong,
the National Natural Science Foundation of China (NSFC),
the Leverhulme Trust, 
the Research Corporation, 
the Ministry of Science and Technology (MOST), Taiwan,
the United States Department of Energy,
and
the Kavli Foundation.
The authors gratefully acknowledge the support of the NSF, STFC, INFN and CNRS for provision of computational resources.

This work was supported by MEXT, JSPS Leading-edge Research Infrastructure Program, JSPS Grant-in-Aid for Specially Promoted Research 26000005, JSPS Grant-in-Aid for Scientific Research on Innovative Areas 2905: JP17H06358, JP17H06361 and JP17H06364, JSPS Core-to-Core Program A. Advanced Research Networks, JSPS Grant-in-Aid for Scientific Research (S) 17H06133, the joint research program of the Institute for Cosmic Ray Research, University of Tokyo, National Research Foundation (NRF) and Computing Infrastructure Project of KISTI-GSDC in Korea, Academia Sinica (AS), AS Grid Center (ASGC) and the Ministry of Science and Technology (MoST) in Taiwan under grants including AS-CDA-105-M06, Advanced Technology Center (ATC) of NAOJ, and Mechanical Engineering Center of KEK.

{\it We would like to thank all of the essential workers who put their health at risk during the COVID-19 pandemic, without whom we would not have been able to complete this work.}

\appendix

\section{Candidate follow up}
\label{sec:postprocessing}

Narrow-band noise features and the non-Gaussianity in the interferometric data can cause outliers with detection statistic above the threshold.
Consequently, each pipeline requires postprocessing of the results to eliminate candidates originating from noise artifacts. 
We follow up all the first-stage candidates identified in each pipeline with a hierarchy of predefined veto procedures as well as additional manual scrutiny. 
No candidate survives from any pipeline. 
In this section, we detail the postprocessing procedures. 

\subsection{Vetoes}
\label{sec:veto}
We first describe the predefined veto procedures in this section.   

\subsubsection{Known-line veto}
\label{subsec:line_veto}

Candidates caused by known instrumental lines are rejected in the first step in all three pipelines. For each candidate identified at a starting frequency $f_0$ at $t=0$, we veto the candidate if the band \mbox{$[f_0-\delta f, f_0+\delta f]$} intersects any known instrumental lines present in either the Hanford or Livingston interferometer, where $\delta f = 10^{-4} f_0$ is used to account for the Doppler shift due to the Earth's motion. Note that there is a subtle difference between the pipelines when applying the Doppler shift effect. The BSD and Viterbi pipelines apply $\delta f$ to the line frequency and the candidate frequency, respectively.
The three pipelines use a list of known instrumental lines in C01 data~\citep{linelist01}.
In the dual-harmonic Viterbi pipeline, candidates caused by instrumental lines in either of the two separate sub-bands, corresponding to the $f_\star$ and $2f_\star$ components, are rejected. 

\subsubsection{Interferometer veto}
\label{subsec:itf_veto}
In general, a candidate signal with an astrophysical origin should be present in the data of all detectors. If the candidate is louder in one detector than the other, it ought to be louder in the detector with better sensitivity for the source and in the frequency band considered. Each pipeline therefore applies a veto to the consistency of the candidate signal strength across each detector.

For the BSD algorithm, this means vetoing candidates with a weighted CR in the less sensitive detector more than three times higher than the corresponding weighted CR in the more sensitive one. This is applied using the CR computed from the statistical distribution of the FH number counts. This veto is repeated after the next veto step \ref{subsec:Doppler_veto}, using another statistic, namely the 5-vector statistic ~\citep{Astone_2010,PhysRevD.89.062008}, as the second round of consistency check. All the candidates with ${\rho_{\rm CR_{1}}}/{\sqrt{S_{n_1}}} > 3 {\rho_{\rm CR_{2}}}/{\sqrt{S_{n_2}}}$ are vetoed, where $\sqrt{S_{n_i}}$ is the noise amplitude spectral density in each $i$-th detector, and assuming that detector 1 is less sensitive than detector 2. This is an arbitrary and conservative choice, already used in \citet{allskyO2}, with a factor of 3 we do not need to consider an eventual weak dependence due to the different detectors orientation.

For the Viterbi algorithms, we repeat the full search over $\Tobs$ in each individual interferometer. The candidate is vetoed if the two criteria are both satisfied: a) searching data from a single interferometer yields $\mathcal{L} \geq \mathcal{L}_{\cup}$ in single-harmonic Viterbi ($S \geq S_{\cup}$ in dual-harmonic Viterbi), where $\mathcal{L}_{\cup}$ ($S_{\cup}$) is the original statistic obtained with both interferometers combined, while searching the other interferometer yields $\mathcal{L} < \mathcal{L}_{\cup}$ ($S < S_{\rm th}$); and b) the Viterbi path from the interferometer with $\mathcal{L} \geq \mathcal{L}_{\cup}$ ($S \geq S_{\cup}$) intersects the original path, i.e., the increased significance in a single detector occurs at the same frequency as the original candidate.

\subsubsection{Doppler-shift (and spin-down-shift) veto}
\label{subsec:Doppler_veto}
All three pipelines apply a Doppler correction to transform the observation in the detector frame to the source frame.
For a true astrophysical signal, this correction should increase the significance of any candidate in the data; for local noise, the significance should decrease or remain unaffected.
Both the BSD and the dual-harmonic Viterbi pipeline apply a veto based on this correction. This is not applied to the single-harmonic Viterbi pipeline because the $\Tcoh$ are short enough to track the Doppler shift.

For the BSD algorithm, we compute the significance of the candidate in terms of the CR and signal-to-noise ratio ($\rho_i$ where $i=\rm CR, snr$) with and without the Doppler and spin-down corrections applied to the time series. The Doppler and spin-down corrections are done using a heterodyne phase correction, where the assumed phase evolution of the signal, $\phi(f_0,\dot{f}_0)$, is fully described by the frequency and spin-down parameters of the candidate, given by $f_0$ and $\dot{f}_0$, respectively. The corrected time series is computed by multiplying the original (uncorrected) time series by the exponential factor $\exp{j\phi(f_0, \dot{f}_0)}$. An easy way to compute the statistical significance of a candidate, once the $\phi(f_0,\dot{f}_0)$ is assumed to be known, is to use the 5-vector statistics~\citep{Astone_2010,PhysRevD.89.062008}, originally developed for the search of known pulsars. Hence, for this step, we use a statistic based on the 5-vector method, whose main properties are described in Appendix~\ref{subsubsec:FollowupBSD}. We use $\rho_{\rm CR}$ and $\rho_{\rm snr}$ to check the nature of a candidate, but this time we compute them from the 5-vector statistic $\mathcal{S}$ rather than from the FH number count.  For this reason, the CR computed in this step is not directly comparable with the CR of the FH map used for the first level selection of candidates. 
If a candidate is from astrophysical origin, we expect, after testing the procedure with simulated signals injected in O3 data, that the significance will increase after the correction, and that it would increase proportionally to the fourth root of the coherence time.
This comparison is done using two different coherence times, $T_{\rm sid}$ and $T_{\rm 4 \, sid}=4T_{\rm sid}$, where $T_{\rm sid}=86164.0905$~s is the duration of a sidereal day.
We use the 5-vector $\rho_{i}$ CR and signal-to-noise ratio to check the change of significance. We keep the candidates if a larger 5-vector $\rho_{\rm CR,C}$ is obtained with the correction applied than the 5-vector $\rho_{\rm CR,NC}$ obtained without the correction applied (in the two cases using $T_{\rm sid}$ and $T_{\rm 4 \, sid}$).
We also veto those candidates which do not show an increased signal-to-noise ratio after the correction. Simulation studies show that the false dismissal probability of this veto is below 10\% if a tolerance of 5\% is used, e.g. we keep all those candidates with $\rho_{i,C}-0.95\rho_{i,NC}>0$ where $\rho_{i,\rm C}$ and $\rho_{i, \rm NC}$ refer the corrected and uncorrected case, respectively.

In the dual-harmonic Viterbi search, for each candidate remaining, we recompute $\mathcal{F}$-statistics over the same $\Tcoh$ as listed in Table~\ref{tab:para} with Doppler modulation correction turned off (DM-off), and repeat the search using the DM-off $\mathcal{F}$-statistics~\citep{DMoff}.
If the candidate is of astrophysical origin, it should become undetectable in the DM-off search, returning a score $S_{\rm DM-off} < S_{\rm th}$ and a Viterbi path different from the original one.  
This criterion does not apply to high signal-to-noise candidate, i.e., $S\gg S_{\rm th}$. However, after previous veto steps, no such high signal-to-noise candidate is left in this search.
Instead, if a candidate is caused by noise artifacts on Earth, its significance is expected to increase in the follow-up. Hence we veto a candidate if the DM-off follow-up yields $S_{\rm DM-off} \geq S_\cup$ and returns a new Viterbi path intersecting the band \mbox{$[f_0-\delta f, f_0+\delta f]$}.

\subsubsection{Sky-position veto}
If a candidate is of astrophysical origin, it should yield the highest detection statistic at the sky position of the source~\citep{Isi2020}.
For candidates surviving previous steps, the Viterbi pipelines conduct another follow-up step by shifting the sky position away from the true position of the SNR~\citep{jones2021search}. This off-target veto contains two separate parts: (a) shift right ascension by an offset $\delta_{\rm RA}$ while keeping declination fixed at the true location, and (b) shift declination by $\delta_{\rm DEC}$ while keeping right ascension fixed at the true location.
We use $\delta_{\rm RA}=3$~hr and $\delta_{\rm DEC}=30$~deg. These offset values are chosen based on a large number of Monte-Carlo simulations that pass veto safety check. For the sources with a declination angle in the range of $[-90, 0]$~deg and $(0, 90]$~deg, we set $\delta_{\rm DEC}$ to 30~deg and $-30$~deg, respectively.
The single-harmonic Viterbi pipeline conducts (a) only.
The dual-harmonic Viterbi pipeline conducts both (a) and (b).
For (a), we veto the candidate if the off-target search yields $\mathcal{L}_{\rm off-target} > \mathcal{L}_{\rm th}$ ($S_{\rm off-target} > S_\cup$) and returns a new Viterbi path intersecting the band \mbox{$[f_0-\delta f, f_0+\delta f]$}.  
For (b), we veto the candidate if the off-target search yields $S_{\rm off-target} > S_{\rm th}$ and returns a new Viterbi path intersecting the band \mbox{$[f_0-\delta f, f_0+\delta f]$}. Note that the veto criterion for (b) is more stringent than that for (a) in the dual-harmonic Viterbi pipeline, because this analysis is more sensitive to the mismatch along the direction of declination. Only the candidates surviving both (a) and (b) are kept.

\subsubsection{Cumulative-significance veto}
\label{subsec:Cumulative_veto}
The significance of a CW signal should be consistent over $T_{\rm obs}$, and in presence of stationary noise, there should be no sudden increase or decrease in the significance when more data are used to integrate the signal.  
The BSD algorithm uses the 5-vector statistics \citep{Astone_2010,PhysRevD.89.062008} to compute the cumulative signal-to-noise ratio and CR on a monthly base, increasing the amount of data used in each iteration by one month. We also compute this trend using an heterodyne-corrected time series with a phase correction $\phi(f_0,\dot{f}_0)$ obtained from the candidates parameters $(f_0,\dot{f}_0)$. The two trends, derived from the corrected and uncorrected time series, are then compared. The comparison is done looking at the plots of the CR, the signal-to-noise ratio and the 5-vector statistics $\mathcal{S}$ (defined in Eq. \ref{Eq:5vec_stat} of Appendix~\ref{subsubsec:FollowupBSD}) as a function of the number of months used to compute these quantities. We visually inspect these plots by comparing the trend of both curves in the corrected and uncorrected case. We veto the candidates if the plot of the corrected case has lower values than the corresponding uncorrected case for the entire duration of the run.  We also veto candidates when the CR cumulative curve of the most sensitive detector is well below the less sensitive one, which is a clear clue that the candidate is actually due to some noise present in the less sensitive detector. For other more complicated cases, we do not automatically veto the candidate but leave them for further investigation in the full O3 H and L data. This is a conservative choice, since vetoing all the candidates that simply present a sudden increase or decrease in the significance (in terms of either CR, signal-to-noise ratio and 5-vector statistic value) is not safe when the noise is not Gaussian.

\subsubsection{Sub-band veto}
If a sub-band is heavily contaminated by non-Gaussian noise, it can be challenging to distinguish noise from a candidate signal.
In the case of the single-harmonic Viterbi pipeline, this renders $\mathcal{L}_{\rm th}$ invalid because $\mathcal{L}_{\rm th}$ is calculated using the results of Gaussian noise simulations.
Furthermore, we do not expect multiple CW signals in a single sub-band.
Consequently, we veto any candidates in a sub-band if the sub-band has more than two unique Viterbi paths with $\mathcal{L} > \mathcal{L}_{\rm th}$.
Simulations in Gaussian noise found $<1\%$ of bands returned two unique paths with $\mathcal{L} > \mathcal{L}_{\rm th}$, justifying our assumption that a sub-band with multiple candidates is dominated by non-Gaussian noise.
The dual-harmonic Viterbi pipeline uses the Viterbi score as the detection statistic and only keeps the optimal path, and hence this step does not apply.

\subsubsection{Coherence-time veto}
\label{subsec:coh_veto}
Both implementations of the Viterbi algorithm use the $\mathcal{F}$-statistic computed over each $\Tcoh$ interval in Tables~\ref{tab:vit_targets} and \ref{tab:para}. 
In the original search, we select $\Tcoh$ assuming a range of $\fdot$. Candidates returned with relatively low $|\fdot|$ allow us to increase the coherent time in a follow-up search.
The sensitivity of the $\mathcal{F}$-statistic increases with longer $T_{\rm coh}$, as long as there is no power leakage over $T_{\rm coh}$ given the inferred $|\fdot|$; a more sensitive $\mathcal{F}$-statistic facilitates a more sensitive Viterbi search.
Hence the significance of the candidate should increase with longer $T_{\rm coh}$ if the candidate is a real astrophysical signal. This has been verified using simulations.

In practice, we first calculate the mean $\fdot$ value over the candidate path, then estimate the maximum $T_{\rm coh}$ capable of tracking the inferred spin down.
In the single-harmonic Viterbi pipeline, we conduct a follow-up search using $T_{\rm coh} = 4$~hr for all survivors. 
With the increased $T_{\rm coh}$, the ratio $\mathcal{L}/\mathcal{L}_{\rm th}$ should increase for a real signal, so we veto any candidates for which $\mathcal{L}(T_{\rm coh} = 4 \, {\rm hr})/\mathcal{L}_{\rm th}(T_{\rm coh} = 4 \, {\rm hr}) < \mathcal{L}/\mathcal{L}_{\rm th}$, where $\mathcal{L}/\mathcal{L}_{\rm th}$ are the values from the initial search.
Similarly, in the dual-harmonic Viterbi pipeline, we veto a candidate if a decreased Viterbi score is returned with the increased $T_{\rm coh}$.
No candidate survives in the dual-harmonic Viterbi pipeline after this step.

\subsection{Further verification}
\label{subsec:further_followup}
After the hierarchy of well-defined veto steps, we discuss the additional verification conducted in each pipeline.

\subsubsection{BSD follow up}
A total of 35 candidates identified by the BSD pipeline survive the vetoes described in Appendix~\ref{sec:veto}. 
This is consistent given the low CR threshold chosen in Section~\ref{sec:BSD}, indeed we are exposed to false alarm candidates and most of them could arise from noise fluctuations. The $\rho_{\rm CR,thr}$ chosen corresponds to the probability of picking, on average, more than one noise candidate. Indeed the CR threshold corresponding to the selection of only one false candidate over the total number of points in the parameter space would be $\rho_{\rm CR,thr}\sim5.7$, while for instance, in the search described in \citet{DirectedBSD}, the CR threshold used is 6.5.
In this section, we describe the extra steps taken to investigate and disqualify the surviving candidates (listed in Table~\ref{tab:BSD_survivors}).
We repeat the Doppler-shift (and spin-down-shift) veto (Appendix~\ref{subsec:Doppler_veto}), the cumulative-significance veto (Appendix~\ref{subsec:Cumulative_veto}), and the interferometer veto (Appendix~\ref{subsec:itf_veto}) using the full O3 (O3a and O3b) C01 data. We remind the reader that the CR computed by the FH is based on the number count associated to the pixel in the FH map where the candidate has been found, while the CR in the 5-vector is computed from the $\mathcal{S}$ statistic. In this step, we use the CR from the $\mathcal{S}$ statistic.
None of the candidates in Table~\ref{tab:BSD_survivors} survive the full-O3 vetos and as can be seen the original CR associated to the candidate in the FH map is below the $\rho_{\rm CR,thr}\sim5.7$, hence with a very low significance compatible with noise.

\begin{table}
	\centering
	\setlength{\tabcolsep}{3pt}
	\renewcommand\arraystretch{1.2}
	\begin{tabular}{lcrr}
		\hline
		Source & Original CR (FH) & ${f_0}$ (Hz) & $\dot{f}_{0}$ (Hz/s)\\
		\hline
		G189.1+3.0   & 5.40 &	 65.6458598 &	$-3.6202\times 10^{-10}$\\
                     & 4.80 &	 75.2342077	&   $-7.9666\times 10^{-10}$ \\
                     & 5.44 &	117.6621941	&   $-7.3779\times 10^{-10}$ \\
                     & 4.89 &	321.8914027	&   $-3.3898\times 10^{-9}$ \\
                     & 4.96 &	355.9654409	&   $-1.1674\times 10^{-9}$ \\
                     & 4.90 &   498.1299761	&   $-1.9622\times 10^{-9}$ \\
		G65.7+1.2    & 5.56 &	 97.5785468  &	$-8.5222\times 10^{-10}$ \\
                     & 4.92 &	120.7226480  &	$-1.0230\times 10^{-9}$ \\
		G266.2--1.2  & 5.07 &	 90.3241483	&   $ 4.0128\times 10^{-11}$ \\ 
                     & 5.48 &	101.8200802 &	$-6.1815\times 10^{-10}$ \\
                     & 5.18 &	139.6310303	&   $-1.1902\times 10^{-10}$ \\
                     & 5.07 &	142.8760455	&   $-2.7103\times 10^{-12}$ \\
                     & 4.87 &	217.9480484	&   $ 2.1858\times 10^{-10}$ \\
                     & 5.15 &	274.5506302	&   $ 1.1958\times 10^{-10}$ \\
		G93.3+6.9    & 5.61 &	208.8871413	&   $-9.6487\times 10^{-10}$ \\
                     & 5.25 &	758.8185308	&   $-4.4728\times 10^{-9}$ \\
                     & 4.83 &	807.0006637	&   $-2.6116\times 10^{-9}$ \\
                     & 4.86 &	851.3276865	&   $-3.9519\times 10^{-9}$  \\ 
                     & 5.49 &	858.9144800	&   $-3.9696\times 10^{-9}$ \\ 
		G18.9--1.1   & 5.24	&    65.0196186	&   $-4.7373\times 10^{-11}$ \\
                     & 5.18 &	 91.4561088	&   $-4.2390\times 10^{-10}$ \\ 
                     & 5.78 &	332.4543731	&   $-3.6398\times 10^{-10}$ \\
                     & 4.88 &	488.0994499	&   $-2.5631\times 10^{-9}$ \\
                     & 4.72 &	841.6011602	&   $-2.8949\times 10^{-9}$ \\
                     & 4.95 &	844.8167574	&   $-4.1613\times 10^{-9}$ \\
		G39.2--0.3   & 4.98 &	135.6997524	&   $-7.5073\times 10^{-10}$ \\
                     & 4.77 &	333.8910938	&   $-2.1219\times 10^{-9}$ \\
                     & 4.76 &	334.0656179	&   $-1.7914\times 10^{-9}$ \\
                     & 4.76 &	700.7408355	&   $-3.9014\times 10^{-9}$ \\ 
                     & 5.12 &	721.1474692	&   $-2.9249\times 10^{-9}$ \\
                     & 5.35 &	802.0985150	&   $-3.0294\times 10^{-9}$ \\ 
                     & 4.68 &	831.8034158	&   $-6.6128\times 10^{-10}$ \\
                     & 5.31 &	902.3896162	&   $-4.9226\times 10^{-9}$ \\
        G353.6--0.7  & 5.36 &	 64.6267590 &	$-9.1944\times 10^{-12}$ \\
                     & 4.99 &	754.3285519 &	$-4.0769\times 10^{-10}$\\	
		\hline
	\end{tabular}
	\caption{Surviving candidates from the BSD pipeline after vetos in Appendix~\ref{sec:veto}. The candidates investigated were excluded using the full O3 data in the BSD search. The columns list the source name, original mean CR from the FH map, the candidate frequency $f_0$, and the spin down $\dot{f}_0$ at the time of the coincidences. The initial set of candidates has been selected using $\rho_{\rm CR,thr}=4.7$ for G65.7+1.2, G189.1+3.0 and G266.2--1.2, $\rho_{\rm CR,thr}=4.6$ for G18.9--1.1 and G93.3+6.9, and $\rho_{\rm CR,thr}=4.5$ for G353.6--0.7 and G39.2--0.3.}
	\label{tab:BSD_survivors}
\end{table}

\subsubsection{Single-harmonic Viterbi follow up}
\begin{figure}[ht]
  \centering
  \includegraphics[width=0.5\linewidth]{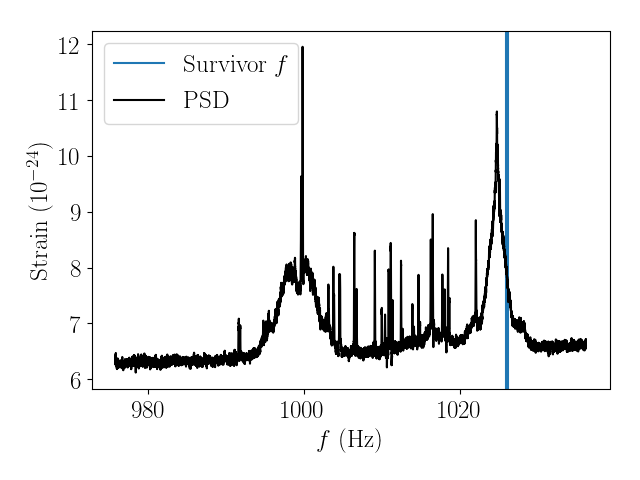}
  \caption{\label{fig:viterbi_survivor_psd} Power spectral density (black curve) versus frequency and the frequency of the last surviving candidate for source G93.3+6.9 (blue vertical line). The power spectral density is built using data collected from both Hanford and Livingston detecters over the full observing time of O3a.}
\end{figure}

One candidate identified by the single-harmonic Viterbi pipeline survives the veto process defined in Appendix~\ref{sec:veto}. 
The candidate is associated with G93.3+6.9 and has a frequency path with mean $f_0 = 1025.95 \, {\rm Hz}$ and $\dot{f}_0 = -2.13\times 10^{ -9 } \, {\rm Hz/s}$ and has a likelihood $\mathcal{L} =18154.0 $, within 5\% of $\mathcal{L}_{\rm th}$.
In Figure \ref{fig:viterbi_survivor_psd}, we plot the power spectral density (black curve) for both detectors combined, calculated over the full duration of O3a, and overplot the frequency of the surviving path (blue vertical line).
Visual inspection plainly indicates that it is associated with noise.
Though it does not lie on the peak of the noise disturbance, it is in the wings.

\subsubsection{Dual-harmonic Viterbi cross check}
All candidates identified in this search pipeline are rejected after the veto procedure described in Appendix~\ref{sec:veto}. 

Here we provide more details of the last set of candidates processed in the coherence-time veto (Appendix~\ref{subsec:coh_veto}). In Table~\ref{tab:viterbi1f2f_survivors}, the original Viterbi score, estimated $2f_\star$ at the beginning and the end of the observation ($2{f_\star}_{\rm start}$ and $2{f_\star}_{\rm end}$), and the mean $2\dot{f}_\star$ values [i.e., $(2{f_\star}_{\rm start} - 2{f_\star}_{\rm end})/\Tobs$] of each candidate are provided. 
Note that the pipeline returns estimated frequencies and spin-down rates corresponding to the $2f_\star$ component. We directly report the returned values in this section rather than converting them into the $f_\star$ component.
The increased $\Tcoh$ used for each candidate is listed in the sixth column. All of the new scores obtained by increasing $T_{\rm coh}$ fall below the original score. They are all considered vetoed according to the criteria set in Appendix~\ref{subsec:coh_veto}. 
To be more conservative, we further discuss the only two candidates with new scores above $S_{\rm th}$ (although decreased compared to the original score), marked by ``$*$" and ``$\dagger$" in the table. 
The one marked by ``$*$" can be confidently ruled out since it returns a completely different path. 
The follow-up search for the candidate marked by ``$\dagger$" using $\Tcoh=11$~hr yields a lower mean spin-down rate, $2\dot{f}_\star = -1.56 \times 10^{-10}$~Hz/s, which allows us to further increase $\Tcoh$. By searching the same sub-band using $\Tcoh=15$~hr, we find the optimal path overlaps with the original one, but with a further decreased significance, $S=4.97$. 
Hence it does not survive the further scrutiny. 

\begin{table}[!tbh]
	\centering
	\setlength{\tabcolsep}{3pt}
	\renewcommand\arraystretch{1.2}
	\begin{tabular}{llllllll}
		\hline
		Source & $S$ &$2{f_\star}_{\rm start}$ (Hz) & $2{f_\star}_{\rm end}$ (Hz)& Mean $2\dot{f}_\star$ ($10^{-10}$~Hz/s) & New $\Tcoh$ (hr) & New $S$ & Paths overlap?\\
		\hline
		G353.6--0.7  & 5.48   &	129.1071064815320 &	129.1048495370870 &	$-$1.4276& 15& 5.17& $\checkmark$\\
		& 5.68   &   130.0884606481900 &	130.0864004630040 &	$-$1.30314  & 15 & 5.49 *& $\times$\\
		& 5.52   &	256.5881944432800 &	256.5861342580990 &	$-$1.30314  & 15& 5.08 & $\checkmark$\\
		& 5.56 & 	416.7312615726270 &	416.7296180541120 &	$-$1.03959 & 15 & 4.85 & $\checkmark$\\
		& 5.52	& 453.4695254620340 &	453.4668981472240 &	$-$1.66187 & 15 &4.45& $\times$\\
		& 5.80    &	464.3793518511010 &	464.3771180548090 &	$-$1.41296  & 15& 4.33& $\times$\\
		& 5.82   & 578.1490972219270 &	578.1467592589690 &	$-$1.47885  & 15 & 4.71& $\times$\\
		& 5.84   & 636.2839467586970 &	636.2820833327750 &	$-$1.17869 & 15 & 4.26 & $\times$\\
		& 5.60   & 742.4446990731940 &	742.4426273139391 &	$-$1.31046  & 15& 5.21& $\times$\\
		& 5.82   & 870.4499074065170 &	870.4476504620770 &	$-$1.4276  & 15& 3.64& $\times$\\
		G189.1+3.0 & 5.66	&253.1534143519250 &	253.1509490741460 &	$-$1.55938 &  15& 4.34& $\times$\\
		G65.7+1.2  & 5.49   & 248.5372569447000 &	248.5349537039590 &	$-$1.45689 & 15& 4.72& $\times$\\
		&	5.77   & 269.5971180543740 &	269.5952662025260 &	$-$1.17137 & 15 &4.74& $\times$\\
		&	5.71   & 277.6064583321330 &	277.6043981469520 &	$-$1.30314 & 15 &5.12& $\times$\\
		&  5.69   & 321.1599189811650 &	321.1574537033920 &	$-$1.55938 & 15  &5.03& $\times$\\
		&	5.74  & 404.2299537032490 &	404.2277314810310 &	$-$1.40564 & 15& 4.34& $\times$\\
		& 	5.51  &486.9126851833790 &	486.9104398130130 &	$-$1.42028 & 15&3.94& $\times$\\
		\hline
		G266.2--1.2 & 5.76 &172.1456481481280 &	172.1412808641780 &	$-$2.76247 & 11  & 4.96 & $\times$\\
		& 	5.45   & 296.5154012345940 &	296.5116358024890 &	$-$2.38178 & 12& 3.74 & $\times$\\
		G93.3+6.9 &	5.51   &121.0981018518380 &	121.0943364197400 &	$-$2.38178 & 12 &4.45 & $\times$\\
		&	6.49 & 138.5975617282440 &	138.5931481479980 &	$-$2.79175 & 11 & 5.80 $\dagger$& $\checkmark$\\	
		G18.9--1.1 &	5.34   & 193.5356481480750 &	193.5317746912850 &	$-$2.45011 &12 & 4.64& $\checkmark$\\
		&5.71  &	219.8922530862280 &	219.8871604936370 &	$-$3.22125 & 12 &3.68& $\times$\\
		&	5.85   & 253.5746296295510 &	253.5707407406620 &	$-$2.45987 & 12 &4.06 & $\times$\\	
		G39.2--0.3 & 6.33   &	109.5362962962230 &	109.5325771604210 &	$-$2.35249 &12& 4.91&$\times$\\ 
		\hline
	\end{tabular}
	\caption{Final candidates from the dual-harmonic Viterbi pipeline and the coherence-time veto results (all vetoed). The first five columns list the source name, original score, estimated start and end 2$f_\star$, and the mean $2\dot{f}_\star$. Column 6 lists the new $\Tcoh$ used in the coherence-time veto. The last two columns shows the follow-up results: the new score obtained by increasing $\Tcoh$, and whether the new optimal path overlaps the original candidate path. The top and bottom halves of the table correspond to the searches using original $\Tcoh = 12$~hr ($S_{\rm th}=5.47$) and $\Tcoh = 9$~hr ($S_{\rm th}=5.33$), respectively. The GPS times for the start and end of the observation are 1238166353 and 1253975702, respectively. The two new scores marked by ``*" and ``$\dagger$" are above $S_{\rm th}$. (Note that the pipeline returns estimated frequencies and spin-down rates corresponding to the $2f_\star$ component.)}
	\label{tab:viterbi1f2f_survivors}
\end{table}

Next we describe additional verification conducted to ensure that we do not veto a weak signal at the final step accidentally. 
For all the final-stage candidates in Table~\ref{tab:viterbi1f2f_survivors}, we cross-check them by searching in the data collected over the second half of O3, with the same configuration as in the original search, in the sub-bands where these candidates are found. 
None of the optimal paths returned in O3b data overlaps the original path (taking into consideration the possible spin down during the shutdown time between two halves of observation).

\begin{figure*}[tbh]
	\centering
	\subfigure[]
	{
		\label{fig:df-12hr}
		\scalebox{0.38}{\includegraphics{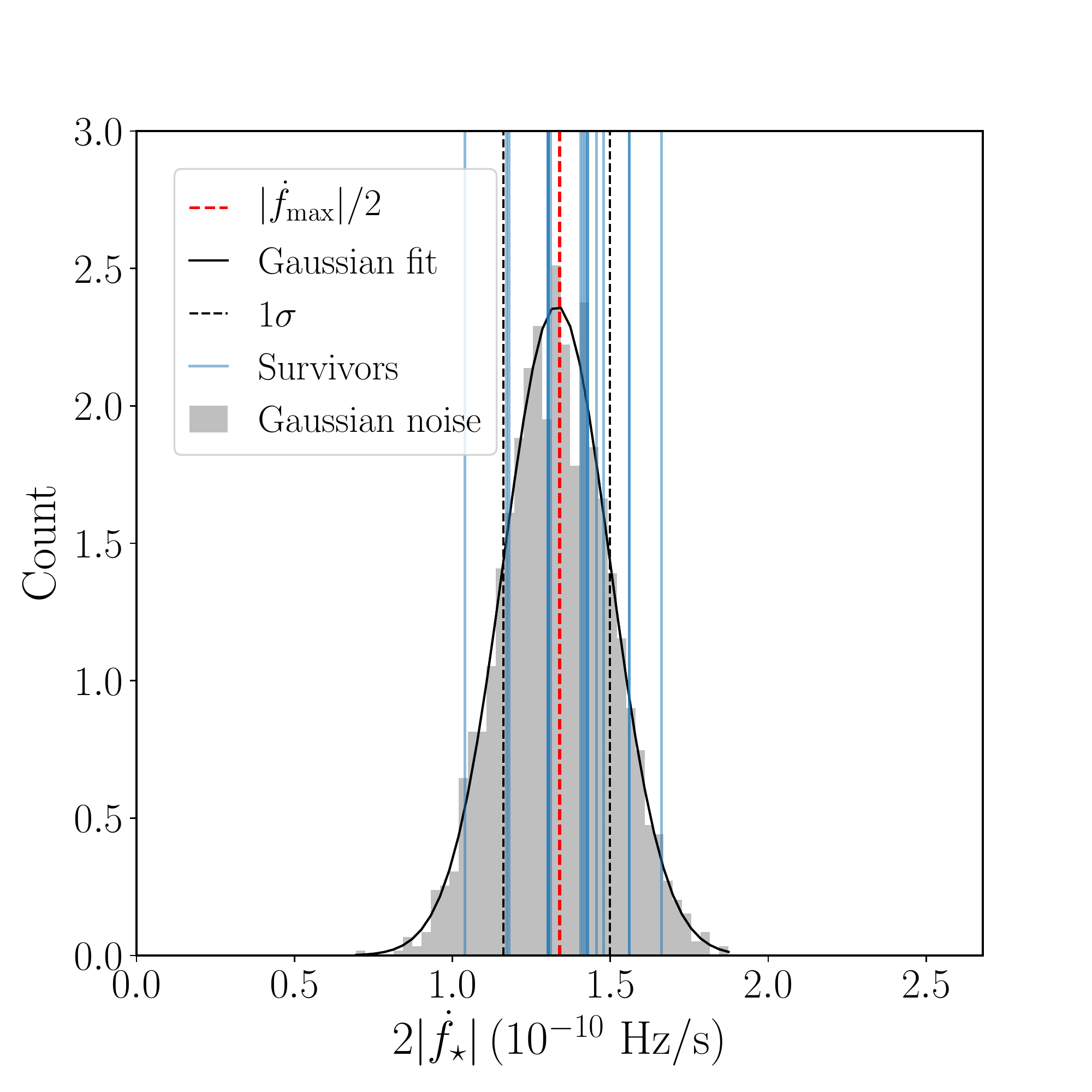}}
	}
	\subfigure[]
	{
		\label{fig:df-9hr}
		\scalebox{0.38}{\includegraphics{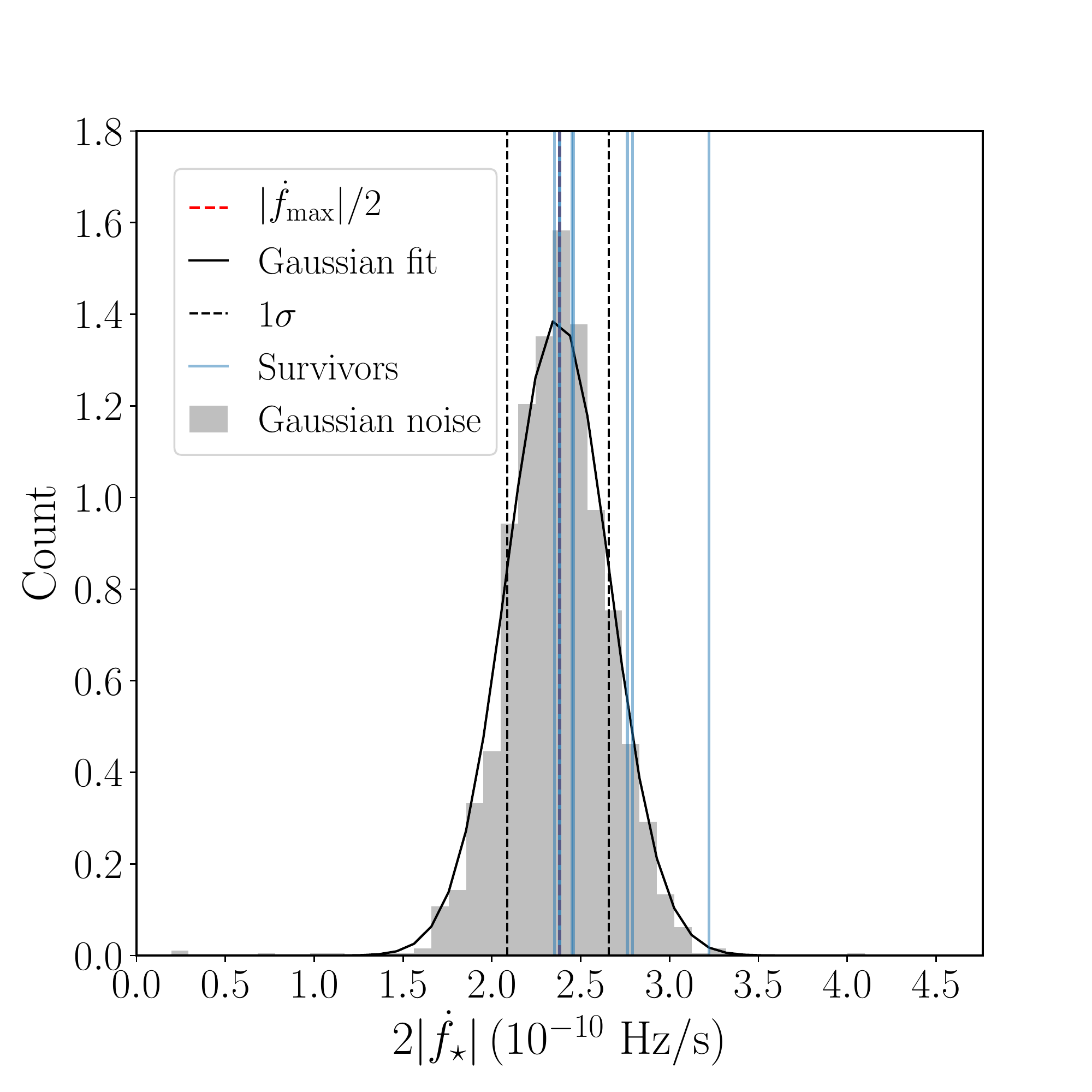}}
	}
	\caption{Noise-only distribution of the mean $2|\dot{f}_\star|$ (gray histogram) and the values obtained from the remaining candidates (blue vertical lines) for (a) $\Tcoh=12$~hr and (b) $\Tcoh=9$~hr in the dual-harmonic Viterbi search. The right edge of each panel is the maximum spin-down rate $|\fdotmax|=1/(2\Tcoh^2)$ for the $2f_\star$ component. The red dashed line indicates $|\fdotmax|/2$. The noise-only distribution is obtained from 2000 Gaussian noise realizations for each panel. The black solid curve indicates the Gaussian fit of the noise distribution. The two black dashed lines are the $\pm 1\sigma$ bounds. }
	\label{fig:df}
\end{figure*}

A further consistency verification is conducted for the candidates in Table~\ref{tab:viterbi1f2f_survivors}. 
All of them have low scores of $S \lesssim 1.2 S_{\rm th}$ (cf. $S \sim 5 S_{\rm th}$ vetoed at early steps). Hence we examine whether they are false alarms arising from noise given that the threshold chosen corresponds to 1\% false alarm probability.
Since the signal frequency is approximated by a negatively biased random walk, the mean $2|\dot{f}_\star|$ value of a path obtained from pure Gaussian noise over $\Tobs$ is expected to be around $|\fdotmax|/2$, where $|\fdotmax|=1/(2\Tcoh^2)$ is the maximum spin-down rate covered in the search for the $2f_\star$ component. 
This is because the method attempts to ``track" pure Gaussian noise with a transition probability $A_{q_{i-1} q_i} = A_{q_i q_i} = 1/2$ and $2|\dot{f}_\star|$ can take any value in the range of $[0, |\fdotmax|]$. 
Figure~\ref{fig:df} shows the distribution of the mean $2|\dot{f}_\star|$ obtained by tracking 2000 pure Gaussian noise realizations (gray histograms; fit with black curve) and the values from the remaining candidates (blue vertical lines).
The left and right edges of each panel are the minimum and maximum spin-down rates covered in the search, respectively.
Out of all 25 candidates, 18 lie within the interval of $[-\sigma, \sigma]$ (black dashed lines). 
For both $\Tcoh=12$~hr and $\Tcoh=9$~hr, all the candidate paths are consistent with pure noise.
Moreover, the total number of remaining candidates is consistent with the false alarm probability (1\% in each sub-band). 
Hence, these candidates with low scores are likely to be false alarms. This explains why they are not confidently rejected at early steps.

\section{Details on search pipelines}
\label{appendix:pipelines}
\subsection{BSD}
\label{appendix:BSD}
\subsubsection{Impact of the age and the second-order spin down}
\label{subsubsec:ParametersBSD}
As mentioned in Section~\ref{sec:BSD}, the age of the source sets the range of frequency and spin down/up we can investigate for a given target, and it does not directly affect the sensitivity of the pipeline as it happens e.g. for the coherence time. In the BSD search the first and second order spin-down/up (typically referred to as spin-down) ranges are defined by the age $t_{\rm age}$ and the braking index $n$ of the source as
\begin{equation}
\left\{\begin{array}{l}
-f / t_{\rm age} \leq \dot{f} \leq 0.1 f / t_{\rm age} \\
0 \mathrm{~Hz} ~\mathrm{s}^{-2} \leq \ddot{f} \leq n|\dot{f}|_{\max }^{2} / f=n f / t_{\rm age}^{2}
\end{array}\right. .
\end{equation}
Since we are not explicitly removing the frequency modulation due to the second order spin down, we will limit our search to those sources which second order spin-down range is constrained in a single second order spin-down bin $\delta\ddot{f}$. In practice, we require that $n f / t_{\rm age}^{2} \leq \delta \ddot{f}$. 
Given that the size of the second order spin-down bin is $\delta \ddot{f}={\delta f}{T_{\rm obs}^{-2}}$ as in \citep{Frasca_2005,PhysRevD.89.062008}, where $\delta f$ is the frequency bin size and proportional to $\sqrt{f}$, we can write the maximum frequency allowed for a given source (in the case of a spin down dominated by gravitational emission, hence $n=5$) as
\begin{equation}
\label{eqn:agefreq}
f\leq 4.85\times10^{-13}\left(\frac{t_{\rm age}}{T_{\rm obs}}\right)^4 \rm Hz.
\end{equation}
This means that, e.g., for a source of $t_{\rm age}=3$ kyr and $T_{\rm obs}$ equal to the O3a run length, the maximum frequency covered in the search, neglecting the second order spin-down modulation, is $\sim 600$ Hz.

\subsubsection{Follow-up based on the 5-vector statistic}
\label{subsubsec:FollowupBSD}
In this section, we briefly recap the 5-vector method and its statistic in order to describe the new follow-up veto steps (Appendices~\ref{subsec:Doppler_veto} and \ref{subsec:Cumulative_veto}) used in this search by the BSD pipeline. The 5-vector detection statistic is built exploiting the feature of the amplitude modulation we observe at the detector. This modulation is induced by the detector radiation pattern in response to a CW signal and, given that the interferometers are on Earth, also by the change of the received polarization, called sidereal modulation. 
The response of the detector to a passing CW signal, after removing the Doppler and spin-down frequency modulations, can be described as (see \cite{Astone_2010} for more details)
\begin{equation}
h(t)=H_{0}(\eta)\left[H_{+}(\psi, \eta) A_{+}(t)+H_{\times}(\psi, \eta) A_{\times}(t)\right],
\end{equation}
where $A_{+/ \times}(t)$ are the two sidereal responses to plus and cross polarizations and $H_{0}(\eta)$ is the maximum signal strain. The plus and cross amplitudes $H_{+/ \times}$ are given by
\begin{equation}
H_{+}=\frac{\cos (2 \psi)-i \eta \sin (2 \psi)}{\sqrt{1+\eta^{2}}},
\end{equation}
\begin{equation}
H_{\times}=\frac{\sin (2 \psi)-i \eta \cos (2 \psi)}{\sqrt{1+\eta^{2}}},
\end{equation}
and depend on the polarization angle $\psi$ and the parameter $\eta$, which denotes the degree of polarization of the CW ($\eta=0$ for a linearly polarized wave, $\eta=\pm 1$ for a circularly polarized wave).
It can be shown  (see e.g. \cite{Astone_2010}) that the frequency components of the signal at the detector are all encoded in the $A_{+/ \times}(t)$ functions and in particular that the signal is fully described by its Fourier components at the five angular frequencies centered at the intrinsic angular frequency of the source, $\omega_0,\omega_0 \pm \Omega, \omega_0 \pm 2 \Omega$, where $\Omega$ is the Earth’s sidereal angular frequency. These five-component complex vectors identify the so called 5-vector space onto which interferometric data can be projected. Let us call $\vec{X}$ and $\vec{A}_{+/ \times}$ the 5-vectors for the data and the plus/cross polarization signal templates, respectively.  The scalar products between $\vec{X}$ and $\vec{A}_{+/ \times}$ correspond to the matched filters between the data and the signal templates, and if opportunely normalized, these two quantities are the estimators of the signal plus and cross amplitudes
\begin{equation}
\widehat{H}_{+/ \times}=\frac{\vec{X} \cdot \vec{A}_{+} / x}{\left|\vec{A}_{+/ \times}\right|^{2}},
\end{equation}
from which a detection statistic can be derived as
\begin{equation}
\label{Eq:5vec_stat}
\mathcal{S} \equiv\left|\vec{A}_{+}\right|^{4}\left|\widehat{H}_{+}\right|^{2}+\left|\vec{A}_{\times}\right|^{4}\left|\widehat{H}_{\times}\right|^{2}.
\end{equation}

We can use the value of this detection statistic to compute the associated significance and the false alarm probability of a given candidate. To do so, we need to estimate also the noise background distribution, by repeating the calculation of $\mathcal{S}$ in an ``off-source" analysis (far from the signal frequency).
In the veto step described in Appendix~\ref{subsec:Doppler_veto}, we compute the statistical properties of the data (and noise) over chunks of data of duration $T_{\rm sid}$, by summing up the values of the statistic in each iteration. In this step, we compare the statistical features of the data twice: first using a time series corrected for the Doppler and the spin-down parameters provided by the candidate, and then using no correction. We expect that even if the correction is not precise, if the candidate is of astrophysical origin, the significance with respect to the uncorrected case will be higher. The same comparison is repeated using data of longer duration (4 times longer), which should correspond to an increase of the candidate significance proportional to the fourth square root of the coherence time used.

\subsection{Single-harmonic Viterbi}
\label{appendix:Viterbi}
In this section, we outline the methods used to determine the parameter space and detection thresholds used in the single-harmonic Viterbi pipeline.

First, we outline the process to determine the frequency range and $T_{\rm coh}$ for each source.
We determine the maximum and minimum spin-down rate $\dot{f}_0$ expected for the source assuming a typical signal model with $f = 2f_\star$,
\begin{equation}
-\frac{f}{(n_{\rm min}-1)t_{\rm age}} \leq \dot{f}_0 \leq - \frac{f}{(n_{\rm max}-1)t_{\rm age}},
\label{eqn:fdotbound}
\end{equation}
where $t_{\rm age} = f_\star/\left[(n-1)\dot{f}_\star\right]$, and $n$ is the braking index.
Because the braking index is unknown for each source, we use the most extreme plausible values $n_{\rm min} = 2$ and $n_{\rm max} = 7$.
Also, we neglect stochastic spin wandering when determining the maximum $\dot{f}$ because we expect the spin-down rate to be much faster than the rate of spin wandering, especially in young SNRs.
Using the maximum value of $\dot{f}_0^{\rm max}$ from equation (\ref{eqn:fdotbound}), we make our first estimate of the coherence time
\begin{equation}
  T_{\rm coh} = 2^{-1/2} \left|\dot{f}_0^{\rm max}\right|^{-1/2},
  \label{def:tcoh}
\end{equation}
and calculate the analytically estimated sensitivity $h_0^{\rm est}$ using Eq.~(\ref{eqn:sensitivity}). We also calculate the inferred $h_0^{\rm max}$ using Eq.~(\ref{eqn:maxstrain}).
The initial estimate of $T_{\rm och}$ is used only to find the frequency range, with the maximum frequency covered in the search, $f_{\rm max}$, set to the maximum frequency for which $h_0^{\rm max} > h_0^{\rm est}$.
We then recalculate $T_{\rm coh}$ using $f_{\rm max}$ to identify the $T_{\rm coh}$ necessary to track the $\dot{f}_0^{\rm max}$ implied by $f_{\rm max}$ in Eq.~(\ref{def:tcoh}).
If, after recalculating $T_{\rm coh}$, we have $T_{\rm coh} < 1$~hr, we recalculate $\dot{f}_{\rm max}$ (by inverting Eq.~(\ref{def:tcoh})) using $T_{\rm coh} = 1$~hr and insert the new $\dot{f}_{\rm max}$ into Eq.~(\ref{eqn:fdotbound}) to obtain a new ${f}_{\rm max}$.
We do not search using $T_{\rm coh} < 1$~hr because to do so would require reproducing and cleaning short Fourier transforms (SFTs) explicitly for this search, instead of using the same standard SFTs as other pipelines. 
In addition, using coherent time shorter than an hour would significantly degrade the sensitivity. 
If, after recalculating $T_{\rm coh}$, we have $T_{\rm coh} \geq 1$~hr, we do not need to recalculate $f_{\rm max}$ and $\dot{f}_0^{\rm max}$.
Finally, we determine the minimum search frequency $f_{\rm min}$ which satisfies $h_0^{\rm max} > h_0^{\rm est}$.
The values $T_{\rm coh}$, $f_{\rm min}$, $f_{\rm max}$, and $\dot{f}_0^{\rm max}$ define the parameter space for the search and are summarized in Table \ref{tab:vit_targets}.

Next, we outline the process of setting the detection threshold for each source.
In each sub-band, the Viterbi algorithm obtains $N_Q$ frequency paths (ending in $N_Q$ different frequency bins), each with a log-likelihood $\mathcal{L}$.
We set a log-likelihood threshold to determine which, if any, of the $N = N_Q N_{\rm band}$ paths warrant further analysis. 
We require a false alarm probability $\alpha_{N} = 0.01$ for each source across all sub-bands.
This is equivalent to requiring a false alarm probability per sub-band of
\begin{equation}
\alpha_F = 1-(1-\alpha_{N})^{1/N}.
\label{eqn:fapn}
\end{equation}
The likelihood threshold $\mathcal{L}_{\rm th}$ is then determined by solving
\begin{equation}
\alpha_F =  \int_{\mathcal{L}_{\rm th}}^\infty {\rm d}\mathcal{L} \, p(\mathcal{L}).
\label{eqn:threshold}
\end{equation}
The threshold $\mathcal{L}_{\rm th}$ is unique to each source.
We follow up any path with $\mathcal{L}>\mathcal{L}_{\rm th}$. 

\begin{figure*}[!tbh]
	\centering
	\subfigure[]
	{
		\label{fig:mean}
		\scalebox{0.38}{\includegraphics{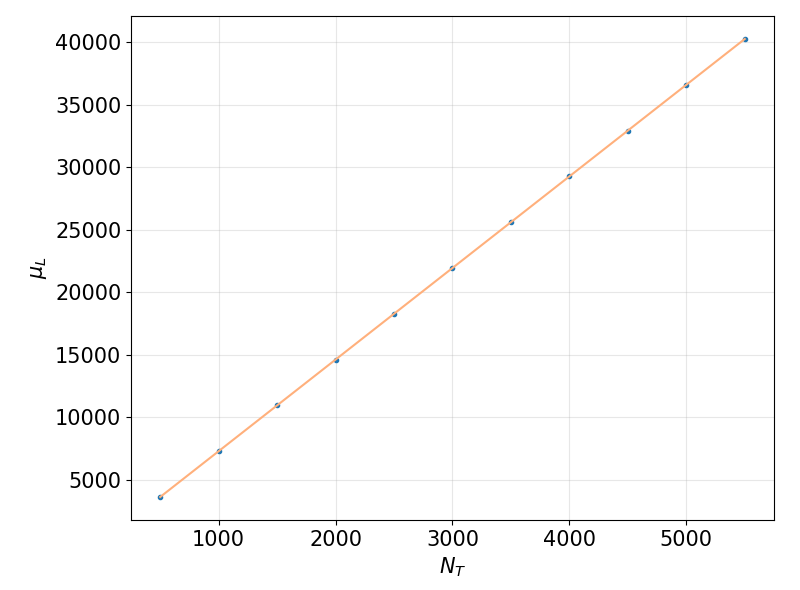}}
	}
	\subfigure[]
	{
		\label{fig:std}
		\scalebox{0.38}{\includegraphics{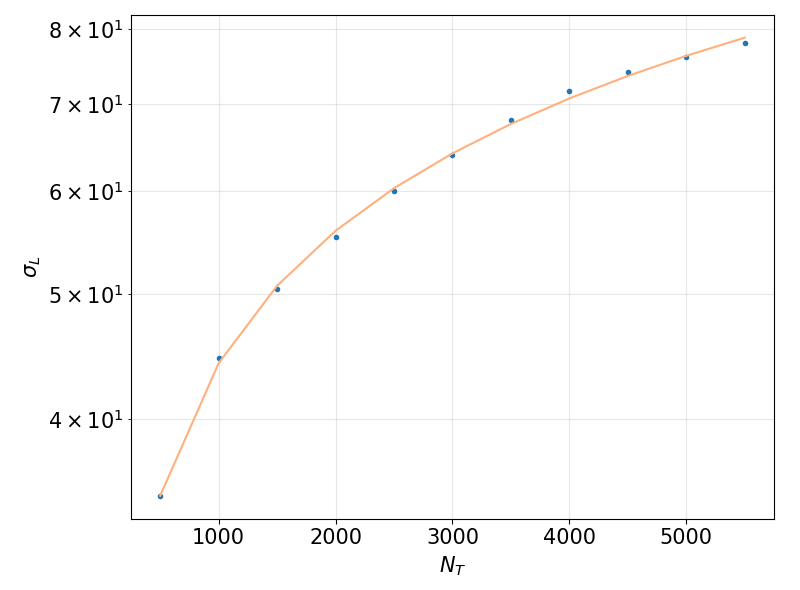}}
	}
  \caption{\label{fig:viterbi_dist_stats} (a) Mean $\mu_\mathcal{L}$ and (b) standard deviation $\sigma_\mathcal{L}$ of the log-likelihoods of the Viterbi paths as a function of $N_T$. Blue dots are values obtained from 100 trials of simulations. The orange curves are the linear and power-law fits describing the mean and standard deviation dependence on $N_T$. An analogous calibration is presented in \citep{millhouse2020search}.}
\end{figure*}

While the distribution of the log-likelhoods is unknown (see \citet{suvorova2016hidden} for details), \citet{millhouse2020search} demonstrated that the mean $\mu_L$ and standard deviation $\sigma_L$ depend only on $N_T$ and scale according to linear and power-law relationships, respectively.
We determine the form of these relationships by simulating 100 sub-bands of Gaussian noise for 11 different $N_T$ values in the range $500 \leq N_T \leq 5500$ and conduct a search on each band.
From the log-likelihoods for each Viterbi path we calculate $\mu_\mathcal{L}$, $\sigma_\mathcal{L}$, and scaling relations of the log-likelihood distribution for each $N_T$.
Figure \ref{fig:viterbi_dist_stats} displays $\mu_\mathcal{L}$ and $\sigma_\mathcal{L}$ from simulations (blue dots) and from the scaling relations (orange curves). 

For each source, we use the scaling relations from Figure \ref{fig:viterbi_dist_stats} to define a Gaussian log-likelihood distribution $p(\mathcal{L})$ for $N_T = T_{\rm obs}/T_{\rm coh}$ and solve $\alpha_F = \int_{\mathcal{L}_{\rm th}}^\infty p(\mathcal{L})$ to obtain $\mathcal{L}_{\rm th}$ (Table \ref{tab:vit_thresholds}).
We follow up all unique frequency paths with $\mathcal{L} > \mathcal{L}_{\rm th}$ using the procedure described in Appendix~\ref{sec:postprocessing}.

\begin{table}
\centering
\setlength{\tabcolsep}{3pt}
\renewcommand\arraystretch{1.2}
\begin{tabular}{ll}
\hline
Source & $\mathcal{L}_{\rm\ \text{th}}$\\
 \hline
G1.9+0.3 & 32552\\
G15.9+0.2 & 32575\\
G18.9--1.1 & 17417\\
G39.2--0.3 & 11683\\
G65.7+1.2 & 7160\\
G93.3+6.9 & 17174\\
G111.7--2.1 & 32568\\
G189.1+3.0 & 24187\\
G266.2--1.2 & 32579\\
G291.0--0.1 & 32586\\
G330.2+1.0 & 30130\\
G347.3--0.5 & 32589\\
G350.1--0.3 & 32577\\
G353.6--0.7 & 3334\\
G354.4+0.0 & 32553\\
  \hline
\end{tabular}
	\caption{\label{tab:vit_thresholds} The threshhold $\mathcal{L}_{\rm th}$ for each SNR in the single-harmonic Viterbi search.}
\end{table}

\subsection{Dual-harmonic Viterbi}
\label{appendix:Viterbi2f}

The HMM formulation in the dual-harmonic search is essentially the same as described in Section~\ref{sec:viterbiregular}, with modifications detailed in Section~\ref{sec:viterbi2f}. 
Here we briefly review the dual-harmonic formulation \citep{Sun2019} and describe the settings used in this analysis. 

We select a coherent time interval, $\Tcoh$, and assume that
\begin{equation}
\label{eqn:int_T}
\left|\int_t^{t+\Tcoh}dt' \dot{\fstar}(t')\right| < \Delta f
\end{equation}
is always satisfied, where $\Delta f= 1/(4 \Tcoh)$ is the frequency bin size in the $\mathcal{F}_1$ output. 
We use $2\Delta f = 1/(2 \Tcoh)$ as the frequency bin size when computing $\mathcal{F}_2$ such that the signal is expected to move at most one bin in the outputs of both $\mathcal{F}_1$ and $\mathcal{F}_2$ over each discrete time step, i.e., from one coherent time interval to next. 
The log emission probability computed over each $\Tcoh$ interval is \citep{Jaranowski1998,Sun2019}
\begin{eqnarray}
\label{eqn:emi_prob_matrix}
\ln L_{o(t_k) q(t_k)} &=& \ln P [o(t_k)|f_i \leq \fstar(t_k) \leq f_i+\Delta f]\\ 
\label{eqn:matrix_propto}
&= & \mathcal{F}_1(f_i) + \mathcal{F}_2(2f_i),
\end{eqnarray}
where $f_i$ is the frequency value in the $i$-th bin.
In this analysis, we assume that the frequency evolution in these young sources is dominated by the secular spin down of the star, and hence the transition probability matrix $A_{q_j q_i}$ becomes \citep{Sun2019}
\begin{equation}
\label{eqn:trans_matrix_sd}
A_{q_{i-1} q_i} = A_{q_i q_i} = \frac{1}{2},
\end{equation}
with all other entries being zero.
A uniform prior $\Pi[q(t_0)] = N_Q^{-1}$ is used.


\bibliography{o3asnrsearch}{}
\bibliographystyle{aasjournal}

\end{document}